\documentclass[aps,pre,preprint,groupedaddress]{revtex4-2}
\usepackage{xcolor}
\usepackage{graphicx}
\usepackage{epstopdf}
\usepackage{amsmath}
\usepackage{comment}
\usepackage{siunitx}
\usepackage[version=4]{mhchem}
\begin{document}

%\preprint{}

\title{A quantitative variational phase field framework}


\author{Arnab Mukherjee$^{1,2}$} \email[]{arnab.mukherjee@northwestern.edu}
\author{James A. Warren$^2$}
\author{Peter W. Voorhees$^{1,3,4}$}

%\thanks{}
\affiliation{$^1$Center for Hierarchical Materials Design, Northwestern University, Evanston, Illinois 60208, USA}
\affiliation{$^2$Material Measurement Laboratory, National Institute of Standards and Technology, Gaithersburg, Maryland 20899, USA}
\affiliation{$^3$Department of Materials Science and Engineering, Northwestern University, Evanston, Illinois 60208, USA}
\affiliation{$^4$Department of Engineering Sciences and Applied Mathematics, Northwestern University, Evanston, Illinois 60208, USA}


\date{\today}

\begin{abstract}
The finite solid-liquid interface width in phase field models results in non-equilibrium effects, including solute trapping. Prior phase field modeling has shown that this extra degree of freedom, when compared to sharp-interface models,  results in solute trapping that is well captured when realistic parameters, such as interface width,  are employed. However, increasing the interface width, which is desirable for computational reasons,  leads to artificially enhanced trapping thus making it difficult to model departure from equilibrium quantitatively. In the present work, we develop a variational phase field model that guarantees a temporal decrease in the free energy with independent kinetic equations for the solid and liquid phases. Separate kinetic equations for the phase concentrations obviate the assumption of point wise equality of diffusion potentials, as is done in previous works. Non-equilibrium effects such as solute trapping, drag and interface kinetics can be introduced in a controlled manner in the present model. In addition, the model parameters can be tuned to obtain ``experimentally-relevant" trapping while using significantly larger interface widths than prior efforts. A comparison with these other phase field models suggests that interface width of about three to twenty-five times larger than current best-in-class models can be employed depending upon the material system at hand leading to a speed-up by a factor of $W^{(d+2)}$, where $W$ and $d$ denote the interface width and spatial dimension, respectively. Finally the capacity to model non-equilibrium phenomena is demonstrated by simulating oscillatory instability leading to the formation of solute bands.
\end{abstract}

%\keywords{}

\maketitle
\section{Introduction}\label{intro}
Rapid solidification refers to the use of large cooling rates or undercoolings resulting in growth speeds of solid-liquid front typically of the order of a \SI{}{\cm\per\second} or larger. Rapid solidification processing encompasses a wide range of industrially-relevant manufacturing processes such as laser-induced surface melting \cite{cui2021additive}, spray forming \cite{wolf2020single}, welding \cite{oliveira2020revisiting} and glass formation \cite{kobold2017dendrite} amongst others. With the recent surge in additive manufacturing processes, there has been a renewed interest in experimental \cite{ghosh2018single, chen2021formation, enrique2021effect}, theoretical development and numerical modeling \cite{keller2017application, ghosh2017primary, karayagiz2020finite, chadwick2021development, wang2019investigation} of rapid solidification of metallic alloys.

Rapid solidification is characterized by the breakdown of the local interfacial equilibrium assumption. Owing to the high growth speeds, the solute atoms do not have enough time to diffuse through the characteristic solid-liquid interface width $\delta$ which is of the order of inter-atomic spacing. As a result solute is incorporated in the growing solid phase at a concentration significantly different from that predicted by the phase diagram. This non-equilibrium or incomplete solute partitioning at the solid-liquid interface is referred to as ``solute trapping." Cognizance of solute trapping holds the key in understanding the growth morphology \cite{kurz1994rapid, merchant1990morphological}, formation of secondary phases \cite{chen2021formation} and micro-segregation pattern of the resulting microstructure \cite{enrique2021effect}.

Many of the insights into the phenomenon of solute trapping have followed from analyses modeling the liquid-solid interface using either a sharp-interface  \cite{aziz1982model, aziz1988continuous, aziz1994transition, aagren1989simplified, jackson2004analytical} or a semi-sharp interface or continuum model \cite{baker1971solidification, hillert1977solute}. Typically, the sharp interface models employ a solution to the diffusion equation to describe the bulk phases which are subjected to external and interfacial conditions. The interfacial conditions are described by response functions, one each for the interfacial concentration and temperature \cite{baker1970interfacial, aziz1988continuous}. While most sharp-interface theories differ in the details of the interface response functions, most of them predict a velocity-dependent partition coefficient $k_v$, which increases monotonically from the equilibrium partition coefficient $k_e$ to $1$ as velocity increases from zero and tends to infinity, and an interface temperature that varies with velocity. The continuum models go a step further to incorporate the variation of the chemical potential and diffusivity through the interface which leads to non-monotonic dependence of $k_v$ on the velocity \cite{baker1970interfacial}. Free energy dissipated by solute diffusion through the interface during interface motion leads to  solute drag  \cite{hillert1977solute, aziz1988continuous}. While the sharp-interface and continuum models provide a thermodynamically-consistent framework to study trapping, tracking the morphologically complex interface makes implementing such models in higher dimensions an arduous task.

Phase field models have enjoyed great success in simulating microstructure evolution in myriad problems in solidification \cite{boettinger2002phase, tourret2022phase} and solid-state phase transformation \cite{chen2002phase, moelans2008introduction, steinbach2013phase}. Unlike the sharp-interface models, the efficacy of the phase field method lies in its treatment of the interface. An indicator variable $\phi$, known as the phase field is introduced to demarcate the regions of different phases, say, solid and liquid by assigning constant values (say $\phi =0,1$) in the bulk phases. The phase field varies smoothly between the bulk values in a thin transition region known as the diffuse interface width $W$. While the indicator function circumvents the need for interface tracking, the diffuse interface thickness introduces an additional length scale in the formulation. $W$ can be treated as either (i) a physical entity $\delta$, in which case the value is typically of the order of Angstrom or a few nanometers, or (ii) a mere mathematical construct, such that $W$ can be enlarged by orders of magnitude than the physical solid-liquid interface thickness. The value of $W$ is restricted by the condition to resolve the smallest microstructural feature. Ascribing the latter notion to $W$, microstructures of the order of few \SI{}{\micro\meter} can be simulated within reasonable times. However, the computation results are now dependent on $W$ and an additional step of assessing the $W$ dependence on results needs to be performed. This can be done numerically by performing convergence tests or analytically by the method of matched asymptotic analysis.

The finite interface width in phase field models introduces a number of non-equilibrium effects including solute trapping \cite{karma2001phase, almgren1999second}. Depending upon the solidification conditions the solute trapping might be physically realistic or an undesirable artifact. It is computationally beneficial to use as large an interface width as possible, while preserving the required physical behavior of the model. For low speed directional solidification, solute trapping is negligible and can be eliminated using a judicious choice of interpolation functions \cite{karma1998quantitative, almgren1999second} and incorporation of an ``antitrapping current" \cite{karma2001phase, ohno2009quantitative}, the form of which is selected based on the thin-interface analysis to recover local equilibrium conditions. Since the antitrapping current does not contribute to the free energy dissipation such models are termed as non-variational models. So, while solute trapping can and does occur during rapid solidification,  phase field models with large values of $W$ chosen for computational expedience will yield artificially enhanced trapping \cite{zhang2013diffuse, mullis2010solute}.

Ahmad and co-workers \cite{ahmad1998solute} studied solute trapping in context of the Wheeler-Boettinger-McFadden (WBM) phase field model \cite{wheeler1992phase}. In the WBM model, the interface is can be interpreted as a mixture of solid and liquid with the same composition. Thus, a single variable $c$ tracks the solute concentration in both phases. Selecting $W$ of the order of 0.1 \SI{}{\nano\meter}, the variation of the partition coefficient with velocity was found to be in good agreement with the sharp-interface continuous growth model (CGM). In addition, the interface temperature agrees well with CGM with partial solute drag. Trapping has also been studied for the Kim-Kim-Suzuki (KKS) phase field model \cite{kim1999phase}. Unlike the WBM model, the KKS phase field model treats the interface as a mixture of solid and liquid with distinct concentrations $c_s$ and $c_l$ respectively. However, the phase concentrations are mutually dependent via the constraint of equal  diffusion potential at each spatial point. The imposition of pointwise equality of diffusion potential, however, does not preclude solute trapping and the far-field diffusion potentials in the solid and liquid exhibit a jump that was shown to be consistent with the high velocity asymptotics of the WBM model. Similar conclusions have been deduced by a number of studies  employing different versions of the above models on different material systems \cite{danilov2006phase, conti1997solute, kim2001phase, mullis2010solute, glasner2001solute, wheeler1993phase, boettinger1999simulation}. A common feature of the above models is that the partition coefficient was found to be sensitive to the chosen interface width, diffusivity and free energy interpolation functions. The interpolation functions (say $h(\phi)$), which connect bulk quantities through the interface, are usually constructed based on the restriction that $h(0) = 0$, $h(1) = 1$ and $h^{\prime}(\phi =0,1)=0$. Since the interpolation functions are not associated with any inherent length scale, $W$ is the only free parameter in the model to tailor partitioning. It is evident that additional degrees of freedom are needed if certain control over solute trapping and drag is to be introduced in the phase field models. Moreover, the choice of $W$ puts a severe limitation on the computation time, especially in higher dimensions. The recent advances have therefore focused on addressing the issue of  introducing additional degrees of freedom such that interface thickness can be increased, while still maintaining experimentally/physically-relevant trapping.

A finite interface dissipation model was recently introduced by Steinbach et al. \cite{steinbach2012phase}. Similar to the KKS model, distinct phase concentrations are introduced. However, rather than assuming local equilibrium condition, separate kinetic equations for the phase concentrations $c_s$ and $c_l$ were derived and the rate of exchange of atoms across phases is determined by a kinetic parameter termed as the ``interface permeability'' which provides the necessary degree of freedom. The solute trapping in the model is controlled through a combination of values of the interface permeability and interface thickness. However, the model is not analytically tractable making it difficult to select appropriate model parameters. In addition, it is unclear whether the interface permeability can be tuned to additionally control solute drag or interface temperature. Pinomaa and Provatas \cite{pinomaa2019quantitative} have extended the thin-interface asymptotics to map the dilute alloy model of Echebarria-Folch-Karma-Plapp (EFKP) \cite{echebarria2004quantitative} to the sharp-interface CGM. A desired solute partitioning is attained by tuning the ``anti-trapping current."  A recent model by Ji et al. \cite{ji2022microstructural} controls the partitioning by introducing an additional free parameter in the diffusivity interpolation function to increase the diffusivity at the interface. For different choices of the interface width, this additional  parameter can be selected to give rise to same trapping behaviour. Both the above models are valid for dilute alloys and one-sided diffusivity i.e. the diffusion in the solid phase is zero.

A new class of phase field models termed as ``non-diagonal" have also been introduced independently by several authors \cite{brener2012kinetic, boussinot2014achieving, wang2013application, fang2013recovering, ohno2016variational, ohno2017variational}. Diagonality refers to the fact that in the previous phase field models such as WBM and KKS, the temporal derivative of concentration and the phase field depends only on the functional derivative of the free energy with respect to concentration (driving force for diffusion) and phase field (driving force for phase transformation) respectively. In other words, the diffusion flux is proportional to driving force for diffusion and rate of interface motion is coupled to the driving force for phase transformation. The dissipative flux in the bulk and the interface is coupled in a single diffusion flux term. However, the ``non-diagonal" models introduce a kinetic cross-coupling between solute diffusion and phase-transformation. This cross-coupling term introduces an additional degree of freedom such that the bulk diffusion and interfacial solute redistribution can be controlled independently. Furthermore, the dynamic cross-coupling restores thermodynamic consistency in phase field models with the cross-coupling terms naturally emerging in the kinetic equations of concentration and phase field respectively. The non-diagonal models have opened the possibility to model close-to-equilibrium solidification processes quantitatively by successfully eliminating all the undesirable interface kinetic effects in systems with finite diffusion in solid \cite{wang2020modeling, wang2021quantitative, ohno2017variational}. The full potential of non-diagonal models, however, have not been fully explored for non-equilibrium solidification problems. A recent phase field model has been proposed by Ohno and co-workers \cite{ohno2017variational} where the kinetic cross-coupling terms and their forms naturally arise in the model derivation. However, the model is only applicable to a low speed solidification regime. A local equilibrium condition is imposed to relate the phase concentrations. In addition, there are no degrees of freedom to control non-equilibrium effects. With this motivation in mind, we develop a variational phase field model that guarantees a decrease in free energy with sufficient degrees of freedom to incorporate non-equilibrium effects relevant to rapid solidification processing in a controlled manner. A key feature of the present model is that the phase concentrations are treated as mutually independent rather than relating them via the local equilibrium condition. Separate kinetic equations with interfacial source terms are derived for the phase concentrations which can be tuned to control solute trapping and drag. Unlike the KKS model and its derivatives, separate kinetic equations alleviate the need for the iterative solution of non-linear system of equations to relate the phase concentrations which can be computationally time consuming depending upon the alloy model and complexity of the system (for instance multi-phase multi-component systems). A number of previous phase field models are shown to be special cases of this new formulation. 

The article is organized as follows: the model framework derived from variational principles is presented in Section \ref{model}. We perform a detailed asymptotic analysis to derive the expression for the diffusion potential jump. The equations are then specialized  for a dilute alloy in Section \ref{dilute}. It is shown in Section \ref{pfm} that a number of previous phase field models can be derived as special cases from the present model. Numerical results are presented in Section \ref{results} providing insights on tailoring a desired level for solute trapping and drag. The model parameters can be tuned readily from available experimental datasets, atomistic simulations or sharp-interface theories to capture ``correct" level of partitioning. The model is applied to \ce{Si}--\SI{9}{\%}\ce{As} and \ce{Ni}--\SI{5}{\%}\ce{Cu} and the efficacy of the model is demonstrated by numerical studies on the partition coefficient and interface temperature. A comparative study with the previously developed phase field model suggests that interface width can be enlarged by a factor of three to twenty-five times in the velocity regime of \SI{0.01}{\meter\per\second} to \SI{1}{\meter\per\second}. For a finite-difference algorithm on a uniform grid spacing this upscaling expedites the computation by a factor $W^{(d+2)}$, where, $d$ is the dimension. Preliminary results on an oscillatory instability leading to the formation of solute bands are also presented. Finally, the paper is concluded by a summary of the main findings in Section \ref{conclusion}.

\section{Model}\label{model}
We consider a binary two-phase system (say, solid and liquid) such that the phase field variable $\phi$ assumes a value of $1$ in the solid and $0$ in the liquid. The interface, where $\phi$ varies smoothly between the bulk values is treated as a mixture of phases, each with a distinct composition $c_s$ and $c_l$. The overall solute concentration is defined as
\begin{equation}\label{model1}
    c = c_s g(\phi) + c_l\{ 1 - g(\phi) \},
\end{equation}
where $g(\phi)$ is a smoothed step-function satisfying the constraints $g(\phi=0) = 0$, $g(\phi=1) = 1$ and $g^{\prime}(\phi = 0,1) = 0$. The free energy functional is written as a contribution arising from the usual interfacial and bulk free energy density and taking in account the constraint Eq.~(\ref{model1}) as
\begin{equation}\label{model2}
    \overline{\mathcal{F}} = \mathcal{F} +  \int_V \lambda \Big[c - \{ c_s g(\phi) + c_l(1-g(\phi))\}\Big]\mathrm{d}V ,
\end{equation}
where
\begin{equation}\label{model1.1}
  \mathcal{F} =  \int_V \Big[H f_{\mathrm{dw}}(\phi) + \frac{\sigma}{2}|\nabla\phi|^2 + f_s(c_s) g(\phi) + f_l(c_l)\{1-g(\phi)\}\Big] \mathrm{d}V .
\end{equation}
$f_{\mathrm{dw}}(\phi)$ is the double well potential with a quartic form $\phi^2(1-\phi)^2$ with minima at $\phi = 0$ and $1$. $H$ is the barrier height. $\sigma$ is the gradient energy coefficient penalizing sharp gradients and imparting smoothness to the phase field profile. $f_s(c_s)$ and $f_l(c_l)$ are the bulk free densities of the solid and liquid respectively. $\lambda$ is the Lagrange multiplier  to ensure that constraint Eq.~(\ref{model1}) is obeyed. To select suitable choices of the evolution equations of the field variables, we consider the temporal change of free energy functional as
\begin{equation}\label{model3}
    \frac{\mathrm{d}\overline{\mathcal{F}}}{\mathrm{d}t} = \int_V \Bigg[ \frac{\delta \overline{\mathcal{F}}}{\delta c_s}\frac{\partial c_s}{\partial t} +  \frac{\delta \overline{\mathcal{F}}}{\delta c_l}\frac{\partial c_l}{\partial t} + \frac{\delta \overline{\mathcal{F}}}{\delta \phi}\frac{\partial \phi}{\partial t} + \frac{\delta \overline{\mathcal{F}}}{\delta c}\frac{\partial c}{\partial t} + \frac{\delta \overline{\mathcal{F}}}{\delta \lambda}\frac{\partial \lambda}{\partial t}\Bigg] \mathrm{d}V .
\end{equation}
Introducing the individual variation with respect to $c_s$, $c_l$, $\phi$, $c$ and $\lambda$ form Eqs.~(\ref{model2}) and (\ref{model1.1}) we obtain
\begin{equation}\label{model4}
    \frac{\mathrm{d}\overline{\mathcal{F}}}{\mathrm{d}t} = \int_V \Bigg\{\Bigg[ \frac{\delta \mathcal{F}}{\delta c_s} - \lambda g(\phi)\Bigg]\frac{\partial c_s}{\partial t} + \Bigg[ \frac{\delta \mathcal{F}}{\delta c_l} - \lambda \{1-g(\phi)\}\Bigg]\frac{\partial c_l}{\partial t} + \Bigg[ \frac{\delta \mathcal{F}}{\delta \phi} - \lambda (c_s - c_l)g^{\prime}(\phi)\Bigg]\frac{\partial \phi}{\partial t} + \lambda \frac{\partial c}{\partial t}\Bigg\} \mathrm{d}V .
\end{equation}
The conservation of mass is captured through the solute concentration $c$ continuity equation as
\begin{equation}\label{model5}
    \frac{\partial c}{\partial t} = - \nabla \cdot J .
\end{equation}
The net flux $J$ is assumed to be a simple interpolation of  the solid and liquid flux as
\begin{equation}\label{model6}
    J = J_s g(\phi) + J_l \{ 1 - g(\phi)\} .
\end{equation}
The temporal evolution of variables $c_s$ and $c_l$ can be postulated by noting that these quantities need only be conserved in their respective bulk phases when $c = c_s$ as $\phi = 1$ and $c = c_l$ as $\phi = 0$. In the interfacial region, a source term is permitted. A possible choice can be written as 
\begin{equation}\label{model7}
    \frac{\partial c_s}{\partial t} = - \nabla \cdot J_s + S_s(c_s,c_l,\phi)g^{\prime}(\phi)\frac{\partial \phi}{\partial t} ,
\end{equation}
and
\begin{equation}\label{model8}
    \frac{\partial c_l}{\partial t} = -\nabla \cdot J_l + S_l(c_s,c_l,\phi)g^{\prime}(\phi)\frac{\partial \phi}{\partial t}.
\end{equation}
The source terms in Eqs.~(\ref{model7}) and (\ref{model8}) are proportional to $\frac{\partial \phi}{\partial t}$ to take into account that the solute redistribution is dependent on the interface velocity. The functions $S_s$ and $S_l$ in the source terms introduce a degree of freedom, which in principle can be tailored to map the phase field model onto a desired set of sharp-interface relations. We will see later that the source functions $S_s$ and $S_l$ contribute to the free energy dissipation. However, the free energy dissipation does not depend explicitly on the forms or the sign of the source functions. The exact form of the interfacial source terms $S_s$ and $S_l$ will be deduced later based on the asymptotic analysis of the model. Substituting Eqs.~(\ref{model5}), (\ref{model6}), (\ref{model7}) and (\ref{model8}) in Eq.~(\ref{model4}), applying the divergence theorem and neglecting the surface integrals, we have
\begin{eqnarray}\label{model9}
    \frac{\mathrm{d}\overline{\mathcal{F}}}{\mathrm{d}t} = \int_V \Bigg[ \nabla\Big(\frac{\delta \mathcal{F}}{\delta c_s}\Big) - \lambda g^{\prime}(\phi)\nabla\phi\Bigg]\cdot J_s + \Bigg[ \nabla\Big(\frac{\delta \mathcal{F}}{\delta c_l}\Big) + \lambda g^{\prime}(\phi)\nabla\phi\Bigg]\cdot J_l \nonumber \\
     + \Bigg[ \frac{\delta \mathcal{F}}{\delta \phi} - \lambda (\tilde{c_s} - \tilde{c_l}) g^{\prime}(\phi) + \Big(S_s\frac{\delta \mathcal{F}}{\delta c_s} + S_l\frac{\delta \mathcal{F}}{\delta c_l}\Big)g^{\prime}(\phi) \Bigg]\frac{\partial \phi}{\partial t} ,
\end{eqnarray}
where $\tilde{c_s} - \tilde{c_l} = c_s - c_l + S_s g(\phi) + S_l\{1-g(\phi)\}$. We note that the terms associated with Lagrange multiplier in the square brackets that are dotted to the fluxes $J_s$ and $J_l$ are directed normal to the interface (because of the presence of the operator $\nabla\phi$). Therefore, decomposing the gradient operators and fluxes in the normal and tangential directions with respect to the interface as $\nabla(\delta \mathcal{F}/\delta c_{s,(l)}) = \nabla_n(\delta \mathcal{F}/\delta c_{s,(l)}) + \nabla_t(\delta \mathcal{F}/\delta c_{s,(l)})$ and $J_{s,(l)} = J_{s,(l)}^n + J_{s,(l)}^t$ where the subscripts $\mathrm{n}$ and $\mathrm{t}$ denote the normal and tangential directions with respect to the interface, respectively. The normal and tangential derivatives are expressed in terms of projection operators as $\nabla_n = (\mathbf{n}\otimes \mathbf{n})\nabla$ and $\nabla_t = (\mathbf{1}-\mathbf{n} \otimes \mathbf{n})\nabla$ respectively. $\mathbf{n}$ denotes the interface normal defined by $\mathbf{n} = -\frac{\nabla\phi}{|\nabla\phi|}$ and $\mathbf{1}$ is a unit tensor allows  Eq.~(\ref{model9}) to be re-written as
\begin{align}
    \frac{\mathrm{d}\overline{\mathcal{F}}}{\mathrm{d}t} = \int_V &\Bigg[ \nabla_n\Big(\frac{\delta \mathcal{F}}{\delta c_s}\Big) - \lambda g^{\prime}(\phi)\nabla\phi\Bigg]\cdot J_s^n + \nabla_t\Big(\frac{\delta \mathcal{F}}{\delta c_s}\Big)\cdot J_s^t  \nonumber \\
     + & \Bigg[\nabla_n\Big(\frac{\delta \mathcal{F}}{\delta c_l}\Big) + \lambda g^{\prime}(\phi)\nabla\phi\Bigg]\cdot J_l^n + \nabla_t\Big(\frac{\delta \mathcal{F}}{\delta c_l}\Big)\cdot J_l^t  \nonumber \\
     + & \Bigg[ \frac{\delta \mathcal{F}}{\delta \phi} - \lambda (\tilde{c_s} - \tilde{c_l}) g^{\prime}(\phi) + \Big(S_s\frac{\delta \mathcal{F}}{\delta c_s} + S_l\frac{\delta \mathcal{F}}{\delta c_l}\Big)g^{\prime}(\phi) \Bigg]\frac{\partial \phi}{\partial t},
\end{align}
To guarantee a temporal decrease in the free energy we choose the forms of fluxes as
\begin{equation}\label{model10}
    J_s^n = -{M_s^n(\phi)}\Bigg[ \nabla_n\Big(\frac{\delta \mathcal{F}}{\delta c_s}\Big) - \lambda g^{\prime}(\phi)\nabla\phi\Bigg],
\end{equation}
\begin{equation}\label{model10.2}
    J_s^t = -M_s^t(\phi)\nabla_t\Big(\frac{\delta \mathcal{F}}{\delta c_s}\Big),
\end{equation}
\begin{equation}\label{model11}
    J_l^n = -{M_l^n(\phi)}\Bigg[ \nabla_n\Big(\frac{\delta \mathcal{F}}{\delta c_l}\Big) + \lambda g^{\prime}(\phi)\nabla\phi\Bigg],
\end{equation}
\begin{equation}
    J_l^t = -M_l^t(\phi)\nabla_t\Big(\frac{\delta \mathcal{F}}{\delta c_l}\Big),
\end{equation}
and,
\begin{equation}\label{model12}
    \frac{\partial \phi}{\partial t} = -M_{\phi}\Bigg[ \frac{\delta \mathcal{F}}{\delta \phi} - \lambda (\tilde{c_s} - \tilde{c_l}) g^{\prime}(\phi) + \Big(S_s\frac{\delta \mathcal{F}}{\delta c_s} + S_l\frac{\delta \mathcal{F}}{\delta c_l}\Big)g^{\prime}(\phi) \Bigg].
\end{equation}
such that $\frac{\mathrm{d}\overline{\mathcal{F}}}{\mathrm{d}t} \leq 0$
as long as $M_s^n(\phi)$, $M_s^t(\phi)$, $M_l^n(\phi)$, $M_l^t(\phi)$ and $M_{\phi} > 0$. $M_{s(l)}^n(\phi)$ and $M_{s(l)}^t(\phi)$ are the atomic mobilities of the component in the solid (liquid) phase in the normal and tangential directions respectively. It is to be noted that the free energy dissipation does not depend on the form or the sign of source functions $S_s$ and $S_l$. The normal and tangential decomposition of the fluxes provides an additional degree of freedom in choosing different interpolation functions of the mobility in the normal and tangential directions. We will see later that this additional degree of freedom offers independent control over two thin-interface effects namely Kapitza jump and surface diffusion. 

To evaluate the Lagrange multiplier, we employ the overall solute conservation Eq.~(\ref{model5}) at each point in the domain. Substituting Eq.~(\ref{model1}), (\ref{model6}) and the kinetic Eqs.~(\ref{model7}) and (\ref{model8}) in Eq.~(\ref{model5}) we obtain
\begin{equation}\label{model14}
    (\tilde{c_s} - \tilde{c_l})\frac{\partial \phi}{\partial t} = -(J_s^n - J_l^n)\cdot \nabla\phi .
\end{equation}
The above relation (which is a Stefan-like condition) relates the three dissipative fluxes at each spatial point in the interface. Further, substituting $J_s^n$ and $J_l^n$ from Eqs.~(\ref{model10}) and (\ref{model11}) we have
\begin{align}\label{model15}
    \lambda g^{\prime}(\phi) = \frac{1}{\overline{M}(\phi)}\Bigg[& -\frac{(\tilde{c_s} - \tilde{c_l})}{|\nabla\phi|^2}\frac{\partial \phi}{\partial t} \nonumber \\
    & + \frac{\Bigg\{ M_s^n(\phi)\nabla_n\Big(\frac{\delta \mathcal{F}}{\delta c_s}\Big) - M_l^n(\phi)\nabla_n\Big(\frac{\delta \mathcal{F}}{\delta c_l}\Big)\Bigg\}}{|\nabla\phi|} \cdot \frac{\nabla\phi}{|\nabla\phi|}\Bigg],
\end{align}
where,
\begin{equation}
    \overline{M}(\phi) = M_s^n(\phi) + M_l^n(\phi).
\end{equation}
Substituting Eq.~(\ref{model15}) in Eqs.~(\ref{model10})-(\ref{model12}) we obtain 
\begin{align}\label{model16}
   J_s = & -\frac{M_s^n(\phi) M_l^n(\phi)}{\overline{M}(\phi)}\nabla_n \Bigg(\frac{\partial f_s}{\partial c_s}g(\phi) + \frac{\partial f_l}{\partial c_l}\{1-g(\phi)\}\Bigg)  \nonumber \\
   &- M_s^t(\phi)g(\phi) \nabla_t\frac{\partial f_s}{\partial c_s} -\frac{(\tilde{c_s} - \tilde{c_l})M_s^n(\phi)}{\overline{M}(\phi)}\frac{1}{|\nabla\phi|}\frac{\nabla\phi}{|\nabla\phi|}\frac{\partial \phi}{\partial t}
 \end{align}
\begin{align}\label{model17}
  J_l = & -\frac{M_s^n(\phi) M_l^n(\phi)}{\overline{M}(\phi)}\nabla_n \Bigg(\frac{\partial f_s}{\partial c_s}g(\phi) + \frac{\partial f_l}{\partial c_l}\{1-g(\phi)\}\Bigg)  \nonumber \\
  & - M_l^t(\phi)\{1-g(\phi)\} \nabla_t\frac{\partial f_l}{\partial c_l}+ \frac{(\tilde{c_s} - \tilde{c_l})M_l^n(\phi)}{\overline{M}(\phi)}\frac{1}{|\nabla\phi|}\frac{\nabla\phi}{|\nabla\phi|}\frac{\partial \phi}{\partial t}
 \end{align}
where, the normal and tangential components have been added to obtain $J_s$ and $J_l$ and
\begin{align}\label{model18}
    \Bigg[ \frac{1}{M_{\phi}}  + \frac{(\tilde{c_s} - \tilde{c_l})^2}{\overline{M}(\phi)}\frac{1}{|\nabla\phi|^2}\Bigg]&\frac{\partial\phi}{\partial t}  =   \nonumber \\
    & \frac{M_s^n(\phi)}{\overline{M}(\phi)}\Bigg[  \sigma\nabla^2\phi - Hf_{\mathrm{dw}}^{\prime}(\phi) - \Big( f_s - f_l -(c_s-c_l)\frac{\partial f_s}{\partial c_s} \nonumber \\
    & \hspace{1.5cm}-S_l\{1-g(\phi)\}(\frac{\partial f_s}{\partial c_s} - \frac{\partial f_l}{\partial c_l}) \Big) g^{\prime}(\phi) \Bigg] + \nonumber \\
    & \frac{M_l^n(\phi)}{\overline{M}(\phi)}\Bigg[ \sigma\nabla^2\phi - Hf_{\mathrm{dw}}^{\prime}(\phi) 
    - \Big( f_s - f_l -(c_s-c_l)\frac{\partial f_l}{\partial c_l}\nonumber \\
    & \hspace{1.5cm}-S_s g(\phi)(\frac{\partial f_l}{\partial c_l} - \frac{\partial f_s}{\partial c_s}) \Big) g^{\prime}(\phi) \Bigg] +\nonumber \\
    & \frac{(\tilde{c_s} - \tilde{c_l})}{\overline{M}(\phi)}\frac{1}{|\nabla\phi|} \Bigg(M_s^n(\phi)g(\phi) \nabla\frac{\partial f_s}{\partial c_s}  - M_l^n(\phi)\{1-g(\phi)\} \nabla\frac{\partial f_l}{\partial c_l}\Bigg) \cdot \frac{\nabla\phi}{|\nabla\phi|}.
\end{align}
Inserting Eqs.~(\ref{model16}) and (\ref{model17}) in Eqs.~(\ref{model6}) and (\ref{model7}) the individual kinetic equations for the phase concentrations can be derived as
\begin{align}\label{eq18.1}
    \frac{\partial c_s}{\partial t} =  \nabla \cdot \Bigg[& \frac{M_s^n(\phi) M_l^n(\phi)}{\overline{M}(\phi)}\nabla_n \Bigg(\frac{\partial f_s}{\partial c_s}g(\phi) + \frac{\partial f_l}{\partial c_l}\{1-g(\phi)\}\Bigg) \nonumber \\
    & + M_s^t(\phi)g(\phi) \nabla_t \frac{\partial f_s}{\partial c_s} \nonumber \\
    & + \frac{(\tilde{c_s} - \tilde{c_l})}{|\nabla\phi|}\frac{M_s^n (\phi)}{\overline{M}(\phi)}\frac{\nabla\phi}{|\nabla\phi|}\frac{\partial \phi}{\partial t}\Bigg] + S_s g^{\prime}(\phi)\frac{\partial\phi}{\partial t},
\end{align}
and,
\begin{align}\label{eq18.2}
    \frac{\partial c_l}{\partial t} =  \nabla \cdot \Bigg[& \frac{M_s^n(\phi) M_l^n(\phi)}{\overline{M}(\phi)}\nabla_n \Bigg(\frac{\partial f_s}{\partial c_s}g(\phi) + \frac{\partial f_l}{\partial c_l}\{1-g(\phi)\}\Bigg) \nonumber \\
    & + M_l^t(\phi)(1-g(\phi)) \nabla_t \frac{\partial f_l}{\partial c_l} \nonumber \\
    & - \frac{(\tilde{c_s} - \tilde{c_l})}{|\nabla\phi|}\frac{M_l^n (\phi)}{\overline{M}(\phi)}\frac{\nabla\phi}{|\nabla\phi|}\frac{\partial \phi}{\partial t}\Bigg] + S_l g^{\prime}(\phi)\frac{\partial\phi}{\partial t}.
\end{align}
Eqs.~(\ref{model18}), (\ref{eq18.1}) and (\ref{eq18.2}) constitute the equations of motions for the phase field model. The phase concentrations $c_s$ and $c_l$ are mutually independent. This independence is in contrast to models like KKS, where the assumption of pointwise equality of diffusion potential $\mu_s = \frac{\partial f_s}{\partial c_s} = \mu_l = \frac{\partial f_l}{\partial c_l}$ is the second equation that must be solved at each point in space and time to relate the phase concentrations. Rather, the kinetic Eqs.~(\ref{eq18.1}) and (\ref{eq18.2}) can be solved to evaluate the respective phase concentrations. At the same time, the present model retains the desirable feature of the KKS model that at equilibrium, it decouples the bulk and the interfacial contributions of the free energy. This implies that the equilibrium phase field profile is given by a hyperbolic tangent function with interface width $W$ given by
\begin{equation}\label{width}
    W = \sqrt{\frac{\sigma}{H}}.
\end{equation}
The interfacial energy can be expressed as
\begin{equation}\label{energy}
    \gamma = IWH,
\end{equation}
where $I$ is a constant which depends on the chosen double-well potential. The details of the equilibrium properties of the model is provided in the supplementary materials for the sake of brevity. 
Using Eqs.~(\ref{model16}) and (\ref{model17}) in Eq.~(\ref{model5}), the overall solute conservation equation is
\begin{align}\label{model19}
    \frac{\partial c}{\partial t} = \nabla \cdot \Bigg[& q_n(\phi)\nabla_n\Bigg(\frac{\partial f_s}{\partial c_s}g(\phi) + \frac{\partial f_l}{\partial c_l}\{1-g(\phi)\}\Bigg) \nonumber \\
    &+M_s^t(\phi)g(\phi)^2 \nabla_t \frac{\partial f_s}{\partial c_s} + M_l^t(\phi)\{1-g(\phi)\}^2 \nabla_t\frac{\partial f_l}{\partial c_l} \nonumber \\
    &+\frac{(\tilde{c_s} - \tilde{c_l})}{|\nabla\phi|}a(\phi)\frac{\nabla\phi}{|\nabla\phi|}\frac{\partial \phi}{\partial t}\Bigg],
\end{align}
where,
\begin{equation}\label{normal}
    q_n(\phi) = \frac{M_s^n(\phi)M_l^n(\phi)}{\overline{M}(\phi)}
\end{equation}
and,
\begin{equation}\label{antitrapping}
    a(\phi) = \frac{M_s^n(\phi)g(\phi) - M_l^n(\phi)\{1-g(\phi)\}}{\overline{M}(\phi)}
\end{equation}
 Several important features of the governing equations are noteworthy. First, the term proportional to $\frac{\partial\phi}{\partial t}$ (last term) in Eq.~(\ref{model19})  is directly related to solute trapping as it is dependent on the velocity (through $\frac{\partial\phi}{\partial t}$), the local concentration difference between solid and liquid and directed normal to the interface (through the operator $\frac{\nabla\phi}{|\nabla\phi|}$). Thus, this is an interface dissipation flux, having the mathematical characteristics of an anti-trapping term that has been employed in previous works \cite{karma2001phase, echebarria2004quantitative}. The anti-trapping term is split into two parts in the respective phase concentrations in Eqs.~(\ref{eq18.1}) and (\ref{eq18.2}). The corresponding conjugate term proportional to $(M_s^n (\phi)\nabla \frac{\partial f_s}{\partial c_s} - M_l^n(\phi) \nabla \frac{\partial f_l}{\partial c_l}) \cdot \mathbf{n}$ naturally arises during the evolution of the phase field Eq.~(\ref{model18}). Secondly, the diffusion flux in Eq.~(\ref{model19}) is decomposed into normal and tangential direction. The normal and tangential derivatives are associated with diffusion potential jump and surface diffusion respectively. The two interpolation functions can be chosen to eliminate two important thin interface effects: namely the Kapitza resistance, or jump in the diffusion potentials at a stationary interface when there is a mass flux through the interface,   and surface diffusion as  reported in Ref.\cite{nicoli2011tensorial} and as will be discussed next.
 
\subsection{Choice of interpolation functions}\label{interpolation_discussion}
The normal and tangential derivatives offer independent control over the diffusion potential jump due to solute transport across a stationary interface and surface diffusion respectively. It was shown by Nicoli et al.\cite{nicoli2011tensorial} that both these thin interface effects can be remedied by a choice of inverse and linear interpolation of the mobilities in the normal and tangential direction respectively. The rationale behind such choices also follow from the asymptotic analysis discussed in the next Section \ref{asymptotic}. While the mobilities were chosen to be tensorial in the former work, the decomposition of the derivatives arise naturally in the model formulation in the present work. We note that tensorial mobilities have also been employed to capture surface diffusion in phase field models \cite{gugenberger2008comparison, ahmed2013phase}. Following the above arguments we choose $M_s^n(\phi) = M_s/h(\phi)$ and $M_l^n(\phi) = M_l/\{1-h(\phi)\}$ with $h(\phi)$  as an anti-symmetric function which can be chosen to be the usual cubic $h(\phi) = \phi^2(3-2\phi)$, or the quintic potential $h(\phi) = \phi^3(6\phi^2 - 15\phi + 10)$. Substituting these forms in Eq.~(\ref{normal}) we obtain
\begin{equation}\label{inverse}
    \frac{1}{q_n(\phi)} = \frac{h(\phi)}{M_s} + \frac{1-h(\phi)}{M_l}.
\end{equation}
The mobilities in the tangential direction are also chosen as $M_s^t(\phi) = M_s/h(\phi)$ and $M_l^t(\phi) = M_l/\{1-h(\phi)\}$. In addition we also select $g(\phi) = h(\phi)$ to interpolate phase concentrations, free energies and fluxes in Eqs.~(\ref{model1}), (\ref{model1.1}) and (\ref{model6}). The above choice of interpolation functions naturally leads to a linear interpolation of the diffusion potentials of solid and liquid phase in the tangential direction as $M_sh(\phi)\nabla_t(\partial f_s/\partial c_s) + M_l\{1-h(\phi)\}\nabla_t(\partial f_l/\partial c_l)$. The form of the anti-trapping term in Eq.~(\ref{antitrapping}) can be re-written as
\begin{equation}\label{antitrapping2}
    a(\phi) = \frac{(M_s - M_l)h(\phi)\{1-h(\phi)\}}{M_s\{1-h(\phi)\} + M_lh(\phi)}.
\end{equation}
It is worth noting since $h(\phi)$ is used to interpolate the phase concentrations in Eq.~(\ref{model1}), the interface stretching term that modifies mass conservation is automatically negated. Thus, all the thin-interface effects can be eliminated by a single $h(\phi)$ function. 
 
\subsection{Asymptotic analysis and determining the source terms $S_s$ and $S_l$}\label{asymptotic}
Typically the source terms should be determined by performing an asymptotic analysis of the model. However, thin-interface asymptotic analysis (i.e. up to second order) of the model remains a mathematically challenging task. To make the problem mathematically amenable and for the purpose of estimating the choice of $S_s$ and $S_l$, we {\sl temporarily} assume pointwise equality of the diffusion potential i.e. $\mu_s = \mu_l$ in Eqs.~(\ref{model18}), (\ref{eq18.1}), (\ref{eq18.2}) and (\ref{model19}). This allows for substantial progress in the thin-interface asymptotics, and a choice of source term can be deduced. The asymptotic analysis builds upon and follows closely the expansion scheme of Refs. \cite{almgren1999second, echebarria2004quantitative, ohno2017variational, provatas2011phase}. The lowest order solutions are identical to the previous works and differences arise only at higher order expressions, namely, first order jump in diffusion potential, first order correction to Gibbs Thomson condition and first order correction to Stefan's condition. We restrict the discussion only to these new findings here and the details of the asymptotic analyses (up to first order for the model $\mu_s \neq \mu_l$ and up to second order for the model $\mu_s = \mu_l$) are presented in detail in the supplementary materials along with numerical comparisons of the solution obtained for both the cases (to ascertain the range of the validity of the assumption). We emphasize that $\mu_s = \mu_l$ is {\sl not} a constraint when solving the dynamic equations. 

The diffusion potential jump up to first order in the asymptotic expansion can be obtained as
\begin{align}\label{math1}
    \mu^- - \mu^+ = & \frac{WV}{D_l\partial c_{l,0}/\partial\mu_0}(c_{l,0}-c_{s,0})(F_1^- - F_1^+)  + \frac{WV}{D_l\partial c_{l,0}/\partial\mu_0}(F_2^- - F_2^+) \nonumber \\
    & + \frac{W}{D_l\partial c_{l,0}/\partial\mu_0} q_n^-\frac{\partial \mu_0}{\partial r}\Big|^-(G^- - G^+),
\end{align}
where,
\begin{equation}\label{math2}
    F_1^{\pm} = \int_{0}^{\pm\infty}\Big[\overline{p}(\phi_0) - \overline{p}(\phi_0|^{\pm})\Big]\mathrm{d}\eta,
\end{equation}
\begin{equation}\label{math3}
    \overline{p}(\phi_0) = \frac{g(\phi_0)-1-a(\phi_0)}{q_n(\phi_0)/M_{l,0}},
\end{equation}
\begin{equation}\label{math4}
    F_2^{\pm} = \int_0^{\pm\infty}\Big[S_sg(\phi_0) + S_l\{1-g(\phi_0)\}\Big]\frac{a(\phi_0)}{q_n(\phi_0)/M_{l,0}}
\end{equation}
and
\begin{equation}\label{integral3}
     G^{\pm} = M_{l,0}\int_0^{\pm\infty}\Bigg[ \frac{1}{q_n(\phi_0)} - \frac{1}{q_n^{\pm}}\Bigg]\mathrm{d}\eta.
 \end{equation}
 $M_{l,0}$ is the lowest order liquid mobility defined as $M_{l,0} = D_l\partial c_{l,0}/\partial\mu $.
 
The superscripts $+(-)$ denote that the quantities have been evaluated in the liquid (solid) side of the interface. The subscript $0$ represent the zeroth (lowest) order solution and $\phi_0$ represents the equilibrium phase field profile which is of the form of hyperbolic tangent and $\eta$ denotes the co-ordinate system normal to the interface.

The source term in the phase field model provides an additional degree of freedom to recover a desired sharp-interface relation. From the analysis it is evident from Eq.~(\ref{math1}) that the choice of source terms $S_s = S_l = -A(c_s - c_l)$ is sufficient to impart a degree of control over solute trapping in the phase field model. With the above choice, the diffusion potential jump up to first order in asymptotics can be simplified as,
 \begin{equation}\label{jump}
     \mu^- - \mu^+ = \frac{WV}{D_l\partial c_{l,0}/\partial\mu}(c_{l,0}-c_{s,0})(F^- - F^+) + \frac{W}{D_l\partial c_{l,0}/\partial \mu}q_n^-\frac{\partial\mu_s}{\partial r}\Big|^-(G^- - G^+),
 \end{equation}
 where,
 \begin{equation}\label{integral1}
     F^{\pm} = \int_0^{\pm\infty}\Big[\tilde{p}(\phi_0) - \tilde{p}(\phi_0|^{\pm})\Big]\mathrm{d}\eta,
 \end{equation}
 \begin{equation}\label{integral2}
     \tilde{p}(\phi_0) = \frac{g(\phi_0)-1-(1-A)a(\phi_0)}{q_n(\phi_0)/M_{l,0}}
 \end{equation}

The second term in the diffusion potential jump Eq.~(\ref{jump}) is associated with a stationary interface and arises due to solute flux across the interface. Since we are unaware of any computational or experimental evidence of the importance of this term in rapid solidification processing, we choose to eliminate the term by satisfying the condition $G^- = G^+$. The inverse interpolation function given by Eq.~(\ref{inverse}) naturally satisfies the constraint.
 A numerical comparison of the two models (shown in the supplementary material) suggest that the steady state diffusion potential in both the models are identical. The agreement between the predictions of the two models is better (within $10\%$) at higher values of $A$ for the entire velocity range considered. Even with lower values of $A$ the agreement is good until velocities of about \SI{0.1}{\meter\per\second}. A good agreement between both the models motivates us to utilize the same interpolation functions to negate the Kapitza jump for the case $\mu_s \neq \mu_l$.
 
 Once the second term in Eq.~(\ref{jump}) is eliminated, solute trapping can be controlled by the first term. The first term is the diffusion potential jump due to interface velocity and the origin of solute trapping in rapid solidification. In the limit  where the asymptotics has been performed, $\mu_s=\mu_l$, the magnitude of this jump can be controlled by choosing an appropriate value of $A$ as evident from Eqs.~(\ref{integral1}) and (\ref{integral2}).  We find this to be the case as well when $\mu_s\ne \mu_l$ and by choosing this form of the source term  and verifying numerically that trapping can indeed be controlled by the parameter $A$. While other choices of source terms are certainly possible, we choose the above form for the sake of simplicity. Moreover, such a simple expression for the sources considerably simplifies the subsequent asymptotic analysis. Thus the terms involving $\tilde{c}_s - \tilde{c}_l$ can be replaced by $(1-A)(c_s - c_l)$ in Eqs.~(\ref{model18}), (\ref{eq18.1}) and (\ref{eq18.2}).

 Once the choice of source terms is determined, it is easy to evaluate the first order correction to the Gibbs Thomson condition which is dependent on the form of the sources. The Gibbs Thomson condition up to first order in asymptotics is obtained as
 \begin{align}\label{math5}
     \mu^{\pm} = &\frac{f_s(c_{s,0})-f_{l}(c_{l,0})}{c_{s,0}-c_{l,0}} + \frac{\gamma \kappa}{c_{s,0}-c_{l,0}} \nonumber \\
     &+\frac{V}{(c_{s,0}-c_{l,0})} \frac{a_1}{M_{\phi}W}\Big[1-a_2^{\pm}\frac{W^2M_{\phi}}{D_l}\frac{(c_{s,0}-c_{l,0})^2}{\partial c_{l,0}/\partial \mu}\Big] \nonumber \\
     & + \frac{W}{D_l\partial c_{l,0}/\partial \mu}q_n^-\frac{\partial\mu_0}{\partial r}\Big|^- A\Delta F
 \end{align}
 where,
 \begin{equation}\label{math6}
     a_1 = I = \int_{-\infty}^{+\infty}\left(\frac{\partial\phi_0}{\partial\eta}\right)^2\mathrm{d}\eta,
 \end{equation}
 \begin{equation}\label{math7}
     a_2^{\pm} = \frac{F^{\pm} + M -(1-A)^2N + (1-A)P}{I},
 \end{equation}
 with the constants defined as
 \begin{equation}\label{math7.1}
     M = \int_{-\infty}^{+\infty}\left[\int_{0}^{\eta}\tilde{p}(\phi_0)\mathrm{d}\eta^{\prime}\right]g^{\prime}(\phi_0)\frac{\partial\phi_0}{\partial\eta}\mathrm{d}\eta,
 \end{equation}
 \begin{equation}\label{math8}
     N = M_{l,0}\int_{-\infty}^{+\infty}\frac{1}{M_s^n(\phi_0) + M_l^n(\phi_0)}\mathrm{d}\eta,
 \end{equation}
 \begin{equation}\label{math8.1}
     P = \int_{-\infty}^{+\infty}a(\phi_0)\tilde{p}(\phi_0)\mathrm{d}\eta,
 \end{equation}
 and
 \begin{equation}\label{math9}
     \Delta F = F^- - F^+.
 \end{equation}
 The value of the first order diffusion potential is different on either side of the interface due to difference in the values of the constants $F^+$ and $F^-$ in the expression of $a_2$. The correction of the mass conservation equation due to finite interface width can be obtained up to first order as
 \begin{align}\label{math10}
     q_n^+\frac{\partial\mu}{\partial r}\Big|^+ - q_n^-\frac{\partial\mu}{\partial r}\Big|^- = & -V(c_l - c_s) - VW\kappa\left(H^+ - H^-\right) \nonumber \\
     &-W\frac{\partial^2 \mu_0}{\partial s^2}D_l\frac{\partial c_{l,0}}{\partial\mu}\left(S^+ - S^-\right)
 \end{align}
 where,
 \begin{equation}\label{math11}
     H^{\pm} = \int_0^{\pm\infty} \left[g(\phi_0) - g(\phi_0|^{\pm})\right]\mathrm{d}\eta
 \end{equation}
 and
 \begin{equation}\label{math12}
     S^{\pm} = \frac{1}{M_{l,0}}\int_0^{\pm\infty}\left[q_t(\phi_0) - q_t(\phi_0|^{\pm})\right]\mathrm{d}\eta,
 \end{equation}
 with 
 \begin{equation}\label{math13}
     q_t(\phi_0) = M_s^t(\phi_0)g(\phi_0)^2 + M_l^t(\phi_0)\left\{1-g(\phi_0)\right\}^2
 \end{equation}
 The term $(c_l-c_s)$ on the right hand side of the above equation denotes jump in the concentration at the interface up to first order. The second and the third terms in Eq.~(\ref{math10}) represents the correction due to interface stretching and surface diffusion respectively. From Eqs.~(\ref{math12}) and (\ref{math13}) it is evident that anti-symmetric forms of $g(\phi_0)$ and $q_t(\phi_0)$ will satisy the constraints $H^+=H^-$ and $S^+ = S^-$. Therefore, $M_s^t(\phi_0)$ and $M_l^t(\phi_0)$ are chosen to give rise to anti-symmetric form of $q_t(\phi_0)$. The choice of the interpolation functions $g(\phi) = h(\phi)$, $M_s^t(\phi) = M_s/h(\phi)$ and $M_l^t/\left\{1-h(\phi)\right\}$ discussed in the Section \ref{interpolation_discussion} naturally satisfies the above constraints.

\subsection{Alternative choice of interpolation functions}\label{alternative_choice}
\begin{figure}
\includegraphics[scale = 0.75]{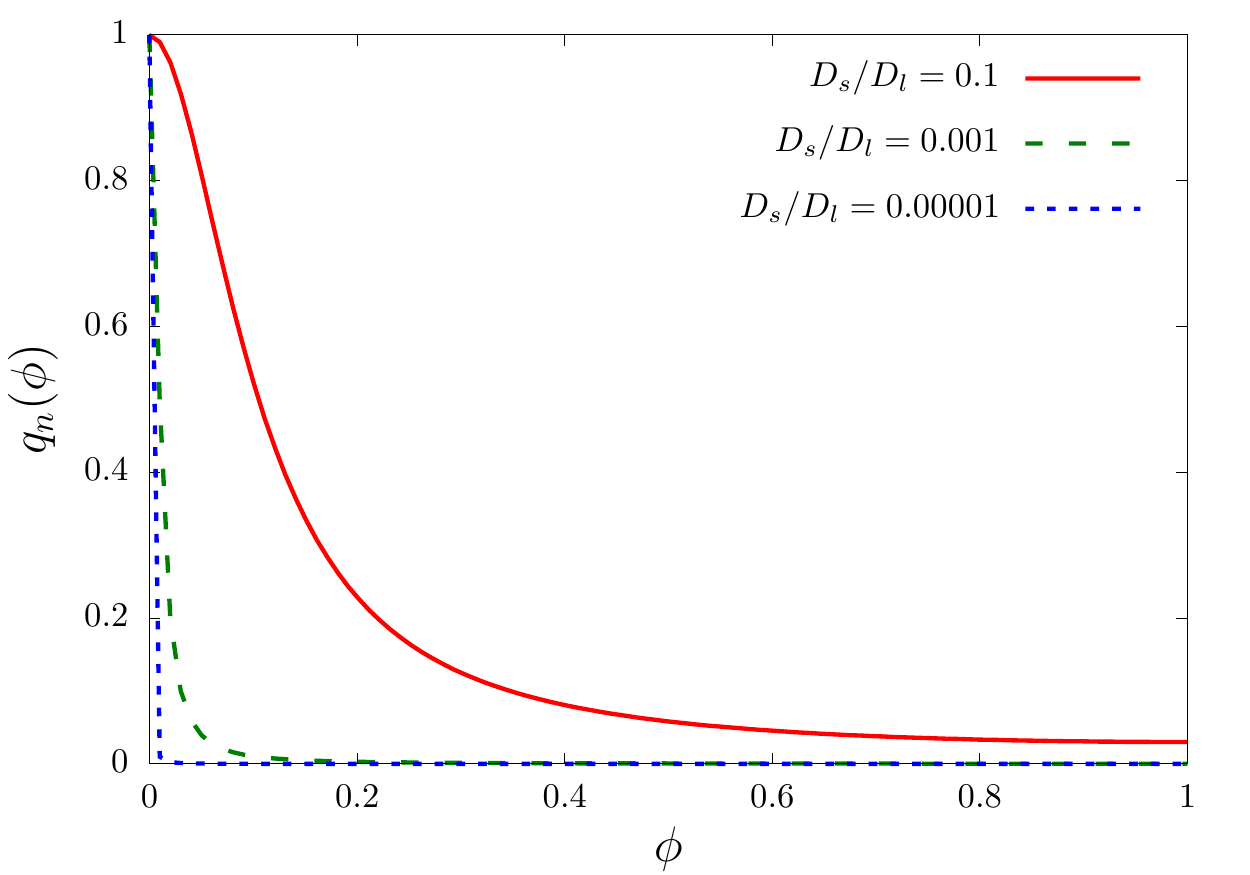}
\caption{\label{fig1} The variation of the diffusivity interpolation function $q_n(\phi)$ given by Eq.~(\ref{inverse}) for different ratios of solid to liquid diffusion coefficients. The profiles exhibit a sharp transition as $\phi$ approaches $0$ (i.e. liquid). As a result the interface diffusivity corresponding  to $\phi = 0.5$ is dictated by the diffusion coefficient in the solid. }
\end{figure}

All the thin-interface effects can be controlled or eliminated within the variational framework. However, the choice of single $h(\phi)$ function to interpolate mobilities to remove both the Kapitza jump and surface diffusion imposes a severe restriction on the numerical efficiency of the model. For instance, the inverse interpolation function Eq.~(\ref{inverse}) leads to an abrupt transition between the bulk values across the interface as shown in Fig.\ref{fig1}. The interface diffusivity is governed by the solid diffusivity. This restricts the choice of the interface width $W$ in the simulations which has to be progressively decreased with decreasing solid diffusivity to resolve the diffusion length in the solid. Therefore, we utilize additional degrees of freedom in the model by exploiting different interpolation functions for the mobilities in the normal and tangential direction to control the different thin-interface effects, while still retaining the variational character of the equations.

\begin{figure}
\includegraphics[scale = 0.45]{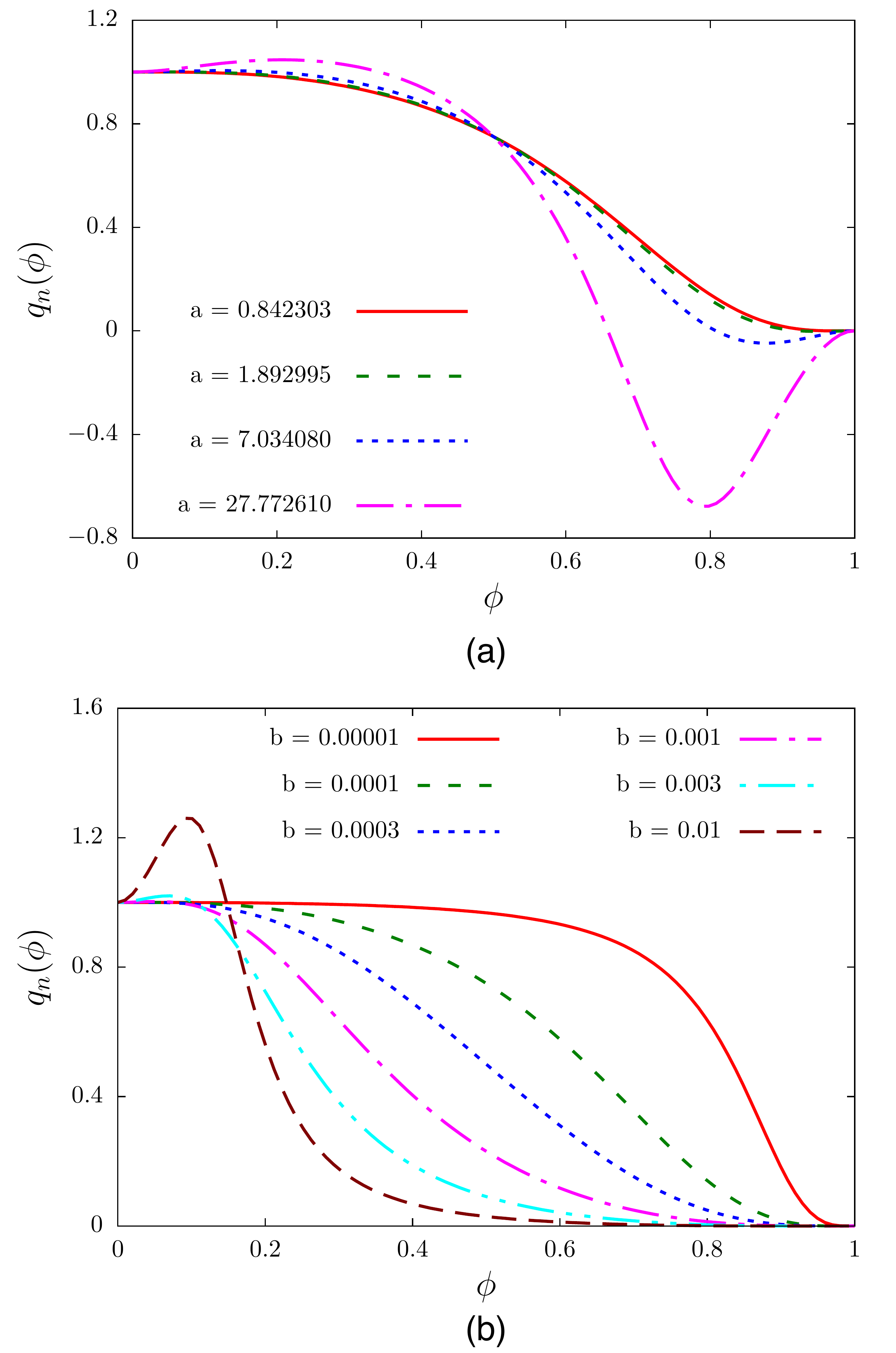}
\caption{\label{fig2}(a) The variation of the diffusivity interpolation function $q_n(\phi)$ given by Eq.~(\ref{normal2}) for different values of the roots of $a$ with $b = 0.001$ and $D_s/D_l = 0.001$ such that the constraint $G^+ = G^-$ is satisfied. Higher values of the root leads to non-monotonic variation across the interface. (b) The variation of the $q_n(\phi)$ across the interface for different values of $b$ and $D_s/D_l = 0.001$. The quintic interpolation function $h(\phi) = \phi^3(6\phi^2 - 15\phi + 10)$ has been employed in the plots.}
\end{figure}

The problem related to the abrupt variation of mobility in inverse interpolation function can be ameliorated by interpolating the mobilities in $q_n(\phi)$ in Eq.~(\ref{normal}) as $M_s^n(\phi) = M_s/h_n(\phi)$ and $M_l^n(\phi) = M_l/\{1-h_n(\phi)\}$ to allow for a smoother variation of $q_n(\phi)$ as
 \begin{equation}\label{normal2}
     \frac{1}{q_n(\phi)} = \frac{h_n(\phi)}{M_s} + \frac{1-h_n(\phi)}{M_l},
 \end{equation}
 where
 \begin{equation}\label{interpolation1}
     h_n(\phi) = \frac{b p(\phi)}{b p(\phi) + 1 - p(\phi)},
 \end{equation}
 where, $b$ is a constant which  determines the variation of the mobility across the interface and $p(\phi)$ is given by
 \begin{equation}\label{interpolation2}
     p(\phi) = h(\phi) + a(\phi-\frac{1}{2})\phi^2(1-\phi)^2.
 \end{equation}
 $p(\phi)$ comprises of a summation of two anti-symmetric functions with an adjustable parameter $a$. The constant $a$ is evaluated numerically such that the integrals $G^-$ and $G^+$ in Eq.~(\ref{integral3}) are equal. It can be verified that the function satisfies the constraints $p(\phi =0) = p(\phi = 1) = 0$ and $p^{\prime}(\phi = 0,1) = 0$. Usually multiple values of $a$ satisfy the constraint. The variation of $q_n(\phi)$ for different values of $a$ are shown in Fig.\ref{fig2} for $b = 0.001$ and $D_s/D_l = 0.001$. As $a$ increases the variation of $q_n(\phi)$ becomes non-monotonic. Therefore, we choose from the smaller roots of $a$ for the study. The effect of the parameter $b$ on variation of $q_n(\phi)$ is shown in Fig.\ref{fig2}(b). Values of $b$ not close to $D_s/D_l$ gives rise to asymmetric profiles, although a symmetric profile is not a strict requirement. However, the choice of $b$ and $a$ should lead to $q_n(\phi) \geq 0$ for the phase concentration equations to be stable.
 
To eliminate surface diffusion, we choose $M_s^t(\phi) = M_s/p(\phi)$ and $M_l^t = M_l/\{1-p(\phi)\}$ to interpolate the tangential component of the diffusion potentials since $p(\phi)$ is anti-symmetric. In addition we choose $g(\phi) = p(\phi)$ to interpolate the phase concentrations (Eq.~(\ref{model1})), free energies (Eq.~(\ref{model1.1})) and fluxes (Eq.~(\ref{model6})). The modified form of $a(\phi)$ (Eq.~(\ref{antitrapping})) is written as
\begin{equation}\label{shape_function}
    a(\phi) = \frac{(M_s - b M_l)p(\phi)\{1-p(\phi)\}}{M_s\{1-p(\phi)\} + M_l b p(\phi)}.
\end{equation}

The phase concentrations can thus be written as
\begin{align}\label{cs_nonvar}
    \frac{\partial c_s}{\partial t} =  \nabla \cdot \Bigg[& q_n(\phi)\nabla_n \Bigg(\frac{\partial f_s}{\partial c_s}p(\phi) + \frac{\partial f_l}{\partial c_l}\{1-p(\phi)\}\Bigg) \nonumber \\
    & + M_s \nabla_t \frac{\partial f_s}{\partial c_s} 
     + \frac{({c_s} - {c_l})(1-A)}{|\nabla\phi|}\frac{M_s \{1-p(\phi)\}}{\overline{M}(\phi)}\frac{\nabla\phi}{|\nabla\phi|}\frac{\partial \phi}{\partial t}\Bigg]\nonumber \\
     & - A(c_s - c_l) p^{\prime}(\phi)\frac{\partial\phi}{\partial t},
\end{align}

\begin{align}\label{cl_nonvar}
   \frac{\partial c_l}{\partial t} =  \nabla \cdot \Bigg[& q_n(\phi)\nabla_n \Bigg(\frac{\partial f_s}{\partial c_s}p(\phi) + \frac{\partial f_l}{\partial c_l}\{1-p(\phi)\}\Bigg) \nonumber \\
    & + M_l \nabla_t \frac{\partial f_l}{\partial c_l}  - \frac{({c_s} - {c_l})(1-A)}{|\nabla\phi|}\frac{b M_l p(\phi)}{\overline{M}(\phi)}\frac{\nabla\phi}{|\nabla\phi|}\frac{\partial \phi}{\partial t}\Bigg] \nonumber \\
    &- A(c_s - c_l)p^{\prime}(\phi)\frac{\partial\phi}{\partial t}
\end{align}
where, the terms in the square bracket in Eqs.~(\ref{cs_nonvar}) and (\ref{cl_nonvar}) are the solid ($J_s$) and liquid fluxes ($J_l$) respectively. The overall solute concentration is governed by the equation
\begin{align}\label{solute_nonvar}
    \frac{\partial c}{\partial t} = \nabla \cdot \Bigg[& q_n(\phi)\nabla_n\Bigg(\frac{\partial f_s}{\partial c_s}p(\phi) + \frac{\partial f_l}{\partial c_l}\{1-p(\phi)\}\Bigg) \nonumber \\
    &+M_s^t p(\phi)\nabla_t\frac{\partial f_s}{\partial c_s} + M_l^t\{1-p(\phi)\}\nabla_t\frac{\partial f_l}{\partial c_l} \nonumber \\
    &+\frac{({c_s} - {c_l})}{|\nabla\phi|}(1-A)a(\phi)\frac{\nabla\phi}{|\nabla\phi|}\frac{\partial \phi}{\partial t}\Bigg],
\end{align}

Similarly, the phase field equation is written as
\begin{align}\label{phasefield_nonvar}
     \Bigg[\frac{1}{M_{\phi}} + & \frac{(c_s-c_l)^2(1-A)^2 b p(\phi)\{1-p(\phi)\}}{\{\overline{M}(\phi)(bp(\phi)+1-p(\phi))\}}\frac{1}{|\nabla\phi|^2}\Bigg]\frac{\partial\phi}{\partial t} =  \sigma\nabla^2\phi - Hf_{\mathrm{dw}}^{\prime}(\phi)- \Delta G_{\mathrm{driv}}p^{\prime}(\phi) \nonumber \\
     & + \frac{(c_s-c_l)(1-A)p(\phi)\{1-p(\phi)\}}{\overline{M}(\phi)}\frac{1}{|\nabla\phi|}\Bigg( M_s \nabla\frac{\partial f_s}{\partial c_s} - b M_l\nabla\frac{\partial f_l}{\partial c_l}\Bigg)\cdot \frac{\nabla\phi}{|\nabla\phi|},
 \end{align}
 where,
 \begin{align}\label{drivingforce_nonvar}
    \Delta G_{\mathrm{driv}} =  f_s(c_s) - f_l(c_l) - \mu_l(c_s - c_l) - &(c_s-c_l)(\mu_s - \mu_l) \times \nonumber \\
    & \Big\{ \frac{M_s\{1-p(\phi)\}(A\{1-p(\phi)\}-1) - M_lp(\phi)Ap(\phi)}{\overline{M}(\phi)}\Big\}
 \end{align}
 with $\overline{M}(\phi) = M_s\{1-g(\phi)\}+b M_lg(\phi)$.

It can be verified that the asymptotic expressions given by Eqs.~(\ref{jump}), (\ref{integral1}), (\ref{integral2}) and (\ref{integral3}) remain unchanged but with new definitions of $q_n(\phi_0)$ and $a(\phi_0)$ given by Eqs.~(\ref{normal2}) and (\ref{shape_function}). Owing to the numerical efficiency, in the rest of the article we will employ Eqs.~(\ref{cs_nonvar}), (\ref{cl_nonvar}) and (\ref{phasefield_nonvar}) for numerical purposes.

The expression of the driving force Eq.~(\ref{drivingforce_nonvar}) warrants a discussion to make connection with the sharp-interface picture.The driving force in the classical expression for free energy dissipation from the CGM \cite{aziz1988continuous} can be written as
 \begin{equation}\label{free_energy1}
     \Delta G_{\mathrm{driv}} = \Delta G_c - \mathcal{D}\Delta G_D
 \end{equation}
 where, $\Delta G_c$ is the free energy of crystallization or the free energy change when solid of composition $c_s$ is formed from liquid of composition $c_l$. Thus,
 \begin{equation}\label{free_energy2}
     \Delta G_c = f_s - f_l - \mu_l(c_s - c_l).
 \end{equation}
 $\Delta G_D$ is the solute-drag free energy and is defined as the amount of energy dissipated in solute redistribution, given by,
 \begin{equation}\label{free_energy3}
     \Delta G_D = (c_s - c_l)(\mu_s-\mu_l).
 \end{equation}
 Substituting Eqs.~(\ref{free_energy2}) and (\ref{free_energy3}) in Eq.~(\ref{free_energy1}) yields
 \begin{equation}\label{free_energy4}
     \Delta G_{\mathrm{driv}} = f_s - f_l -\mu_l(c_s - c_l) - \mathcal{D}(c_s - c_l)(\mu_s - \mu_l).
 \end{equation}
 Comparing Eqs.~(\ref{drivingforce_nonvar}) and (\ref{free_energy4}) it can be seen that the drag coefficient in this phase field model can be written as
 \begin{equation}\label{drag_coefficient}
     \mathcal{D}(\phi) = \frac{M_s\{1-p(\phi)\}(A\{1-p(\phi)\}-1) - M_lp(\phi)Ap(\phi)}{M_s\{1-p(\phi)\}+M_lp(\phi)}.
 \end{equation}
It is evident that solute drag in the present model is dependent on the value of the model parameter $A$. Thus, $A$ not only controls solute trapping but also solute drag.

\section{Dilute alloy approximation}\label{dilute}
To perform numerical computation we have to specify the thermodynamics of the system. We specialize the equations by considering a dilute ideal solution alloy. The free energy of the bulk solid and liquid is written as
\begin{equation}\label{dil1}
    f_{i}(c_{i}) = f_{i}^{\mathrm{A}}(T) + \epsilon_{i}c_{i} + \frac{RT}{V_{\mathrm{m}}}(c_{i}\ln{} c_{i} - c_{i}) \hspace{2cm} i = s,l
\end{equation}
where $f_{i}^{\mathrm{A}}$ is the free energy of the pure A in the respective phase and the second and the third term represent the change of the internal energy density and the dilute form of entropy of mixing respectively. We linearize the free energy about $(T-T_{\mathrm{m}})$ to recover straight solidus and liquidus and retaining the first order terms to obtain
\begin{equation}\label{dil_linear}
    f_{i}(c_{i}) = f_{i}^A(T_{\mathrm{m}}) - s_{i}(T-T_{\mathrm{m}}) + \epsilon_{i}c_{i} + \frac{RT_{\mathrm{m}}}{V_{\mathrm{m}}}(c_{i}\ln c_{i} - c_{i})
\end{equation}
where $s_{i} = -\partial f_{i}^{\mathrm{A}}/\partial T$ are the entropy densities.

At equilibrium, from the condition of equality of diffusion potential it is easy to show
\begin{equation}\label{dil2}
    k_e = \frac{c_s^{\mathrm{eq}}}{c_l^{\mathrm{eq}}} = \mathrm{exp}\Bigg[ - \frac{\epsilon_s - \epsilon_l}{RT_{\mathrm{m}}/V_{\mathrm{m}}}\Bigg] .
\end{equation}
In addition, the equality of grand-potential i.e. 
\begin{equation}\label{dil4}
    \omega_{s} = f_s - \mu_s c_s = f_l - \mu_l c_l = \omega_l
\end{equation}
yields the relation
\begin{equation}\label{dil5}
    f_s^{\mathrm{A}}(T) - f_l^{\mathrm{A}}(T) = \frac{RT_{\mathrm{m}}}{V_{\mathrm{m}}}(c_s^{\mathrm{eq}} - c_l^{\mathrm{eq}}).
\end{equation}
The driving forces in the phase field equation Eq.~(\ref{model18}) are given by
\begin{equation}\label{dil5.1}
    f_s - f_l - (c_s - c_l)\frac{\partial f_s}{\partial c_s} = \frac{RT_{\mathrm{m}}}{V_{\mathrm{m}}}c_l \ln\Big(\frac{1}{k_e}\frac{c_s}{c_l}\Big) + \frac{RT_{\mathrm{m}}}{V_{\mathrm{m}}}\Big[c_l^{\mathrm{eq}}(k_e-1) - (c_s - c_l)\Big]
\end{equation}
and
\begin{equation}\label{dil5.2}
    f_s - f_l - (c_s - c_l)\frac{\partial f_l}{\partial c_l} = \frac{RT_{\mathrm{m}}}{V_{\mathrm{m}}}c_s \ln\Big(\frac{1}{k_e}\frac{c_s}{c_l}\Big) + \frac{RT_{\mathrm{m}}}{V_{\mathrm{m}}}\Big[c_l^{\mathrm{eq}}(k_e-1) - (c_s - c_l)\Big]
\end{equation}
The driving force can also be expressed in terms of undercooling. For a linearized phase diagram the straight solidus and liquidus lines have slopes $m/k_e$ and $m$ respectively. The equilibrium composition $c_l^e$ can be eliminated in favor of the temperature $T$ as
\begin{equation}\label{dil7}
    T = T_{\mathrm{m}} - |m|c_l^{\mathrm{eq}},
\end{equation}
where $T_{\mathrm{m}}$ is the melting temperature of pure $B$. Therefore,
\begin{equation}\label{dil8}
    f_s - f_l - (c_s - c_l)\frac{\partial f_s}{\partial c_s} =\frac{RT_{\mathrm{m}}}{V_{\mathrm{m}}}c_l \Big(\frac{1}{k_e}\frac{c_s}{c_l}\Big) + \frac{RT_{\mathrm{m}}}{V_{\mathrm{m}}}\Big[\frac{(T_{\mathrm{m}} - T)(k_e -1)}{|m|} - (c_s - c_l)\Big] .
\end{equation}
Similarly,
\begin{equation}\label{dil9}
    f_s - f_l - (c_s - c_l)\frac{\partial f_l}{\partial c_l} =\frac{RT_{\mathrm{m}}}{V_{\mathrm{m}}}c_s \Big(\frac{1}{k_e}\frac{c_s}{c_l}\Big) + \frac{RT_{\mathrm{m}}}{V_{\mathrm{m}}}\Big[\frac{(T_{\mathrm{m}} - T)(k_e -1)}{|m|} - (c_s - c_l)\Big] .
\end{equation}
\subsection{Directional solidification}
In experiments, it is difficult to maintain a fixed undercooling and far easier to control the interface speed through an arrangement known as the directional solidification (DS). In a DS set up, a temperature gradient is established across the sample, which is then pulled through the temperature field at constant velocity. The two variables that control the microstructure features, temperature gradient $G$ and pulling speed $v_p$ are controlled independently. In addition, a frozen temperature approximation is employed which assumes that the temperature field is unperturbed by the advancing interface such that
\begin{equation}\label{ds1}
    T(x,t) = T_{\mathrm{ref}} + G[(x-x_o) - v_pt].
\end{equation}
$T_{\mathrm{ref}}$ is the reference temperature at the initial position $x_o$ of the solid-liquid interface. It is customary to measure the local temperature and concentration with respect to a reference temperature and concentration. For a sample with composition $c_{\infty}$, the solidus temperature $T_s$ and the corresponding alloy composition $c_{\infty}$ is selected to be the reference temperature and concentration respectively. The driving forces Eqs.~(\ref{dil8}) and (\ref{dil9}) can be re-written with respect to ($c_{\infty},T_s$) are
\begin{equation}\label{ds2}
    f_s - f_l - (c_s - c_l)\frac{\partial f_s}{\partial c_s} =\frac{RT_{\mathrm{m}}}{V_{\mathrm{m}}}c_l \Big(\frac{1}{k_e}\frac{c_s}{c_l}\Big) + \frac{RT_{\mathrm{m}}}{V_{\mathrm{m}}}\Bigg[\frac{(T-T_s)(1-k_e)}{|m|} + \Big\{ (c_l - c_s) - \frac{c_{\infty}}{k_e}(1-k_e)\Big\}\Bigg]
\end{equation}
and 
\begin{equation}\label{ds3}
    f_s - f_l - (c_s - c_l)\frac{\partial f_l}{\partial c_l} =\frac{RT_{\mathrm{m}}}{V_{\mathrm{m}}}c_s \Big(\frac{1}{k_e}\frac{c_s}{c_l}\Big) + \frac{RT_{\mathrm{m}}}{V_{\mathrm{m}}}\Bigg[\frac{(T-T_s)(1-k_e)}{|m|} + \Big\{ (c_l - c_s) - \frac{c_{\infty}}{k_e}(1-k_e)\Big\}\Bigg]
\end{equation}

\section{Relation to other phase field models}\label{pfm}
In this section we show that under certain conditions a number of previously developed phase field models can be derived  as special cases of the present model.

\subsection{Wheeler-Boettinger-McFadden (WBM) model}
The WBM model \cite{wheeler1992phase} was the first phase field model developed to study isothermal binary alloy solidification. In the WBM model, the interface is defined as a mixture of solid and liquid of same composition i.e. $c_s = c_l = c$. The free energy is thus defined as
\begin{equation}\label{wbm1}
    f(c,\phi) = f_s(c)g(\phi) + f_l(c)\{1-g(\phi)\}.
\end{equation}
The constraint Eq.~(\ref{model2}) and decomposition to kinetic Eqs.~(\ref{model8}) and (\ref{model9}) is not required. The governing equations for $c$ and $\phi$ can be derived to guarantee monotonic decrease of the free energy functional Eq.~(\ref{model4}) as
\begin{equation}\label{wbm2}
    \frac{\partial c}{\partial t} = \nabla \cdot \Big[q_n(\phi)\nabla\frac{\delta \mathcal{F}}{\delta c}\Big],
\end{equation}
and 
\begin{equation}\label{wbm3}
    -\frac{1}{M_{\phi}}\frac{\partial\phi}{\partial t} = \frac{\delta \mathcal{F}}{\delta\phi}
\end{equation}
The diffusion potential $\mu = \frac{\delta \mathcal{F}}{\delta c}$ is defined as
\begin{equation}\label{wbm4}
    \mu = \frac{\partial f_s(c)}{\partial c} g(\phi) + \frac{\partial f_l(c)}{\partial c}\{1-g(\phi)\}
\end{equation}
Alternatively, we can set $c_s = c_l = c$ in Eq.~(\ref{model18}) to recover the phase field kinetic equation of the WBM model as
\begin{equation}\label{wbm5}
    \frac{1}{M_{\phi}}\frac{\partial\phi}{\partial t} = \sigma\nabla^2\phi - Hf_{\mathrm{dw}}^{\prime}(\phi) - \Big\{f_s(c) - f_l(c)\Big\}g^{\prime}(\phi).
\end{equation}
A direct reduction of the concentration equation to the WBM model is obtained by substituting $c_s = c_l = c$, $M_s^n(\phi) = M_s^t(\phi) = M_s^n/h_n(\phi)$, $M_l^n(\phi) = M_l^t(\phi) = M_s/\{1-h_n(\phi)\}$,  with $h_n(\phi) = M_s g(\phi)/\{M_s g(\phi) + M_l(1-g(\phi))\}$ in Eq.~(\ref{model19}) to yield
\begin{equation}
    \frac{\partial c}{\partial t} = \nabla \cdot \Big[ q(\phi) \nabla \Big\{ \frac{\partial f_s(c)}{\partial c} g(\phi) + \frac{\partial f_l(c)}{\partial c}\{1-g(\phi)\}\Big\}\Big],
\end{equation}
where $q(\phi) = M_s g(\phi)+M_l\{1-g(\phi)\}$. Thus with the choice of $g(\phi) = h(\phi)$ the WBM model can be recovered.

\subsection{Kim-Kim-Suzuki (KKS) model}
The KKS model \cite{kim1999phase} like the present model assumes the interface to be a mixture of solid and liquid with distinct compositions $c_s$ and $c_l$. However the phase concentrations are related via the equality of diffusion potentials i.e. $\frac{\partial f_s}{\partial c_s} = \frac{\partial f_l}{\partial c_l}$ at each spatial position in the interface. The kinetic equation for the KKS model can be recovered by substituting $\frac{\partial f_s}{\partial c_s} = \frac{\partial f_l}{\partial c_l}$ and setting the  source term $A=1$ in Eq.~(\ref{model18}) to obtain
\begin{equation}\label{KKS1}
    \frac{1}{M_{\phi}}\frac{\partial\phi}{\partial t} = \sigma\nabla^2\phi - Hf_{\mathrm{dw}}(\phi) - \Big\{f_s(c_s) - f_l(c_l) - \mu_l(c_s-c_l)\Big\}g^{\prime}(\phi).
\end{equation}
The concentration equation for the KKS model can be recovered by additionally assuming $M_s^n(\phi) = M_s^t(\phi) = M_s^n/h_n(\phi)$, $M_l^n(\phi) = M_l^t(\phi) = M_s/\{1-h_n(\phi)\}$,  with $h_n(\phi) = M_s g(\phi)/\{M_s g(\phi) + M_l(1-g(\phi))\}$ in Eq.~(\ref{model19}) and letting $g(\phi) = h(\phi)$ to obtain
\begin{equation}\label{KKS2}
    \frac{\partial c}{\partial t} = \nabla \cdot \Big[ q(\phi)\nabla\mu_l \Big]
\end{equation}
where $q(\phi) = M_s h(\phi) + M_l\{1-h(\phi)\}$.
It is to be noted that the equality of diffusion potential is solved at each spatial point and time iteration to relate the phase concentrations. This additional step in KKS can become a computation bottle-neck depending on the alloy model in multi-phase, multi-component systems. In contrast in the present model, phase concentrations are governed by separate kinetic equations and can handle arbitrary phase diagrams with equal ease. One can thus expect to gain some  computational expedience especially for complex free energies. However, a comparative study on the numerical performance of both the models is outside the scope of the present work.

\subsection{Echebarria-Folch-Karma-Plapp (EFKP) model}
EFKP \cite{echebarria2004quantitative} is a specialized form of KKS with dilute alloy approximation as shown in Ref.\cite{ohno2009quantitative}. To make connection with the EFKP model we define the solute field in terms of local supersaturation $U = (c_l - c_l^{\mathrm{eq}})/(c_l^{\mathrm{eq}} - c_s^{\mathrm{eq}})$. Furthermore, the equality of diffusion potential for the dilute alloy approximation relates the phase concentrations as $c_s/c_l = k_e$, where, $k_e$ is the equilibrium partition coefficient. Assuming the source terms to be $A = 1$ Eqs.~(\ref{KKS1}) and (\ref{KKS2}) can be re-written employing Eq.~(\ref{ds2}) as
\begin{equation}\label{EFKP1}
    \frac{1}{M_{\phi}}\frac{\partial\phi}{\partial t} = \sigma\nabla^2\phi - Hf_{\mathrm{dw}}^{\prime}(\phi) - \tilde{\lambda}\Big[U + \frac{T-T_s}{|m|c_l^{\mathrm{eq}}(1-k_e)}\Big]h^{\prime}(\phi)
\end{equation}
where $\tilde{\lambda} = \frac{RT_{\mathrm{m}}}{V_{\mathrm{m}}}c_l^{\mathrm{eq}}(1-k_e)^2$ and
\begin{equation}\label{EFKP2}
    \Big[1-h(\phi)(1-k_e)\Big]\frac{\partial U}{\partial t} = \nabla \cdot \Big[ q(\phi)\nabla U \Big] + [1+(1-k_e)U]h^{\prime}(\phi)\frac{\partial\phi}{\partial t}.
\end{equation}
which are the variational form of the governing equations employed in the EFKP model. The non-variational form of the model employs an anti-trapping current that is added to Eq.~(\ref{EFKP2}) to maintain local equilibrium at the interface such that the above Eq. is modified as
\begin{equation}
   \Big[1-h(\phi)(1-k_e)\Big]\frac{\partial U}{\partial t} = \nabla \cdot \Big[ q(\phi)\nabla U + \{1+(1-k_e)U\}a(\phi)\frac{\nabla\phi}{|\nabla\phi|}\frac{\partial\phi}{\partial t}\Big] + [1+(1-k_e)U]h^{\prime}(\phi)\frac{\partial\phi}{\partial t}. 
\end{equation}
The form of $a(\phi)$ is deduced based on the thin-interface analysis of the model. The model by Pinomaa and Provatas \cite{pinomaa2019quantitative} tune the form of $a(\phi)$ to control trapping.

\subsection{Non-diagonal model of Ohno et al.}
One can derive the non-diagonal model of Ohno et al. \cite{ohno2017variational} which was employed to study equilibrium solidification processes by setting $\mu_s = \mu_l$ and choosing the source terms to be zero i.e. $S_s = S_l = 0$ (i.e. $A = 0$) with the choice of $M_s^n(\phi) = M_s^t(\phi) = M_s/h(\phi)$, $M_l^n(\phi) = M_l^t(\phi) = M_l/\{1-h(\phi)\}$ and $g(\phi) = h(\phi)$ in Eqs.~(\ref{model18}) and (\ref{model19}) to obtain
\begin{align}\label{ohno1}
     \Bigg[\frac{1}{M_{\phi}} + & \frac{(c_s-c_l)^2 h(\phi)\{1-h(\phi)\}}{M_s\{1-h(\phi)\}+M_l h(\phi)}\frac{1}{|\nabla\phi|^2}\Bigg]\frac{\partial\phi}{\partial t} =  \sigma\nabla^2\phi - Hf_{\mathrm{dw}}^{\prime}(\phi)- \Big[f_s(c_s) - f_l(c_l)  \nonumber \\
     & - \mu_l(c_s - c_l)\Big]h^{\prime}(\phi) + \frac{(c_s-c_l)}{|\nabla\phi|}a(\phi)\nabla\mu_l\cdot \frac{\nabla\phi}{|\nabla\phi|},
 \end{align}
 and,
\begin{align}\label{ohno2}
    \frac{\partial c}{\partial t} = &\nabla \cdot \Bigg[ q_n(\phi)\nabla_n \mu_l + q_t(\phi)\nabla_t \mu_l + \frac{({c_s} - {c_l})}{|\nabla\phi|}a(\phi)\frac{\nabla\phi}{|\nabla\phi|}\frac{\partial \phi}{\partial t}\Bigg],
\end{align}
where,
\begin{equation}
    q_n(\phi) = \frac{M_s M_l}{M_s\{1-h(\phi)\}+M_l h(\phi)},
\end{equation}
\begin{equation}
    q_t(\phi) = M_s h(\phi) + M_l \{1-h(\phi)\}
\end{equation}
and
\begin{equation}
    a(\phi) = \frac{(M_s - M_l)h(\phi)\{1-h(\phi)\}}{M_s\{1-h(\phi)\}+M_l h(\phi)}
\end{equation}
Trapping is present in the model at high velocities but cannot be controlled. The present model tunes solute trapping by the judicious choice of the interfacial source terms.

\subsection{Finite Interface Dissipation (FID) Model}
The finite interface dissipation model by Steinbach et al. \cite{steinbach2012phase} similar to the present work treat the phase concentrations $c_s$ and $c_l$ as independent variables. While the FID model cannot be derived as a special case from the present model, we seek to highlight the difference and similarities in the structure of both the models. The starting point of both the present and the FID models are Eqs.~(\ref{model1})-(\ref{model4}). The subsequent derivation of FID follows under three assumptions: (i) the spatial domain is divided into distinct reference volumes (RV), (ii) each RV is insulated from outside and (iii) processes within a RV and exchange of atoms between adjacent RVs are independent and can be superimposed. The evolution of the phase concentrations inside an RV are chosen according to Eq.~(\ref{model4}) as
\begin{equation}\label{fid1}
    \frac{\partial c_s}{\partial t} = -P_s(\phi)\left[\frac{\delta F}{\delta c_s} - \lambda g(\phi)\right]
\end{equation}
\begin{equation}\label{fid2}
    \frac{\partial c_l}{\partial t} = -P_l(\phi)\left[\frac{\delta F}{\delta c_l} - \lambda\{1-g(\phi)\}\right],
\end{equation}
where, $P_s(\phi)$ and $P_l(\phi)$ are the kinetic coefficients termed as interface permeability for redistribution fluxes and chosen to be of the form $P_s(\phi) = P/g(\phi)$ and $P_l(\phi) = P/\{1-g(\phi)\}$. Based on assumption (ii), $\lambda$ is evaluated from the condition 
\begin{equation}\label{fid3}
    \frac{\partial c}{\partial t} = 0.
\end{equation}
Substituting Eqs.~(\ref{model1}), (\ref{fid1}) and (\ref{fid2}) in Eq.~(\ref{fid3}) we obtain
\begin{equation}\label{fid4}
    \lambda = \frac{\partial f_s}{\partial c_s}g(\phi) + \frac{\partial f_l}{\partial c_l}\{1-g(\phi)\} - \frac{(c_s-c_l)}{P}g^{\prime}(\phi)\frac{\partial\phi}{\partial t}
\end{equation}
Substituting Eq.~(\ref{fid4}) in Eqs.~(\ref{fid1}) and (\ref{fid2}) and adding the diffusion fluxes between adjacent RVs the kinetic equations for phase concentrations can be obtained as
\begin{align}\label{fid5}
    g(\phi)\frac{\partial c_s}{\partial t} = & \nabla \cdot \left(g(\phi)M_s\nabla \frac{\partial f_s}{\partial c_s}\right) + Pg(\phi)\{1-g(\phi)\}\left(\frac{\partial f_l}{\partial c_l}-\frac{\partial f_s}{\partial c_s}\right) \nonumber \\
    & + (c_l-c_s)g(\phi)g^{\prime}(\phi)\frac{\partial\phi}{\partial t}
\end{align}
\begin{align}\label{fid6}
    \{1-g(\phi)\}\frac{\partial c_l}{\partial t} = &\nabla \cdot \left(\{1-g(\phi)\}M_l\nabla \frac{\partial f_l}{\partial c_l}\right) + Pg(\phi)\{1-g(\phi)\}\left(\frac{\partial f_s}{\partial c_s}-\frac{\partial f_l}{\partial c_l}\right) \nonumber \\
    & + (c_l-c_s)\{1-g(\phi)\}g^{\prime}(\phi)\frac{\partial\phi}{\partial t}.
\end{align}
It is worth noting that diffusion fluxes from only the respective phases (first term in the above equations) are added during superposition. Adding Eqs.~(\ref{fid5}) and (\ref{fid6}) the overall solute conservation writes as
\begin{equation}\label{fid7}
    \frac{\partial c}{\partial t} = \nabla \cdot \left(g(\phi)M_s\nabla\frac{\partial f_s}{\partial c_s}\right) + \nabla \cdot \left(\{1-g(\phi)\}M_l\nabla\frac{\partial f_l}{\partial c_l}\right).
\end{equation}
The evolution equation for the phase-field follows from Eq.~(\ref{model4}) as
\begin{equation}\label{fid8}
    \frac{\partial\phi}{\partial t} = -M_{\phi}\left[\frac{\delta F}{\delta \phi} - \lambda(c_s-c_l)g^{\prime}(\phi)\right].
\end{equation}
Substituting Eq.~(\ref{fid4}) in the above equation yields the equation of motion as
\begin{align}\label{fid9}
    &\left[\frac{1}{M_{\phi}}  +  \frac{(c_s-c_l)^2}{P}g^{\prime}(\phi)^2\right]\frac{\partial\phi}{\partial t} = \frac{\sigma}{2}\nabla^2\phi - Hf_{\mathrm{dw}}^{\prime}(\phi) \nonumber \\
    & - \left[(f_s - f_l) - \left\{\frac{\partial f_s}{\partial c_s}g(\phi)+\frac{\partial f_l}{\partial c_l}\left(1-g(\phi)\right)\right\}(c_s-c_l)\right]g^{\prime}(\phi).
\end{align}
Let us first compare the structure of the phase concentration equations.
In the FID model diffusion flux in the kinetic equations of the phase concentrations are governed by gradient of the diffusion potential of the respective phases (first term in Eqs.~(\ref{fid5}) and (\ref{fid6})). The source expressions consist of terms driven by the difference between the diffusion potentials of the two phases (second term) and the solute redistribution term due to the moving interface or phase transformation (third term). This result is borne out of the assumptions (i) and (iii). On the other hand, Eqs.~(\ref{cs_nonvar}) and (\ref{cl_nonvar}) in the present model suggest that the phase concentrations are driven by the gradient of the diffusion potential of both the phases. This difference arises because of the specific decomposition of the flux in the interfacial region according to Eq.~(\ref{model6}) and the subsequent derivation of the form of the fluxes that guarantee free energy decrease. The sources in the present work only account for solute redistribution due to phase change and controls partitioning. The magnitude of the diffusion potential jump in FID model is controlled by the combination of parameters $P$ and $W$. While the spirit of gaining flexibility to tune solute trapping in both the models is similar i.e. decomposition of independent kinetic equations for phase concentrations leading to additional parameters namely source terms and interface permeability, the connection of the FID model to the sharp interface description is not as apparent as the present work. The authors presented an expression to relate the diffusion potential jump to $P$ and $W$ but limited the treatment to a stationary interface. This is also evident by careful consideration of Eqs.~(\ref{fid5}) and (\ref{fid6}) revealing that the magnitude of diffusion potential jump at a stationary interface ($\frac{\partial\phi}{\partial t} = 0$) is governed by $P$. It is similar to the second term in Eq.~(\ref{jump}) of the present analysis and arises due to mass flux across the interface. Solute trapping during rapid solidification is a result of suppressed solute partitioning due to high interface velocities. The source terms in the present work have a direct connection to this phenomenon as evident from the first term in Eq.~(\ref{jump}). In the FID model, although solute partitioning during rapid solidification is controlled by $P$ for enlarged interface widths, it is difficult to separately control the two contributions of the diffusion potential jump. A rigorous mathematical analysis of the governing equations may help in this regard but is outside the scope of the present work.

We next compare the equations of motion for the phase-field in both the models. The derivation of both the models lead to a new mobility correction term as evident on the left hand sides of Eqs.~(\ref{phasefield_nonvar}) and (\ref{fid9}). This is not entirely fortuitous as the starting form of the phase-field equation arises from the free energy functional Eq.~(\ref{model4}). The differences in the rest of the equations are due to the different assumptions made while evaluating $\lambda$. The driving force in FID model (the terms in the square bracket on the right hand side of Eq.~(\ref{fid9})) can be rearranged to obtain
\begin{equation}\label{fid10}
    \Delta G_{\mathrm{driv}} = f_s - f_l -\frac{\partial f_l}{\partial c_l}(c_s - c_l) - \left(\frac{\partial f_s}{\partial c_s} - \frac{\partial f_l}{\partial c_l}\right)(c_s-c_l)g(\phi)
\end{equation}
which has a similar structure to the $\Delta G_{\mathrm{driv}}$ of the present model as seen from Eq.~(\ref{drivingforce_nonvar}). A comparison with the sharp interface expression Eq.~(\ref{free_energy4}) suggests that the drag coefficient in the FID model is given by $\mathcal{D}(\phi) = g(\phi)$. While the magnitude of the jump $\left(\frac{\partial f_s}{\partial c_s}-\frac{\partial f_l}{\partial c_l}\right)$ is controlled by $P$, its effect on solute drag is not apparent from the above expression. In contrast in the present work, it is evident from Eq.~(\ref{drag_coefficient}) that $A$ controls solute drag in addition to solute trapping.

\begin{figure}
\includegraphics[scale = 0.45]{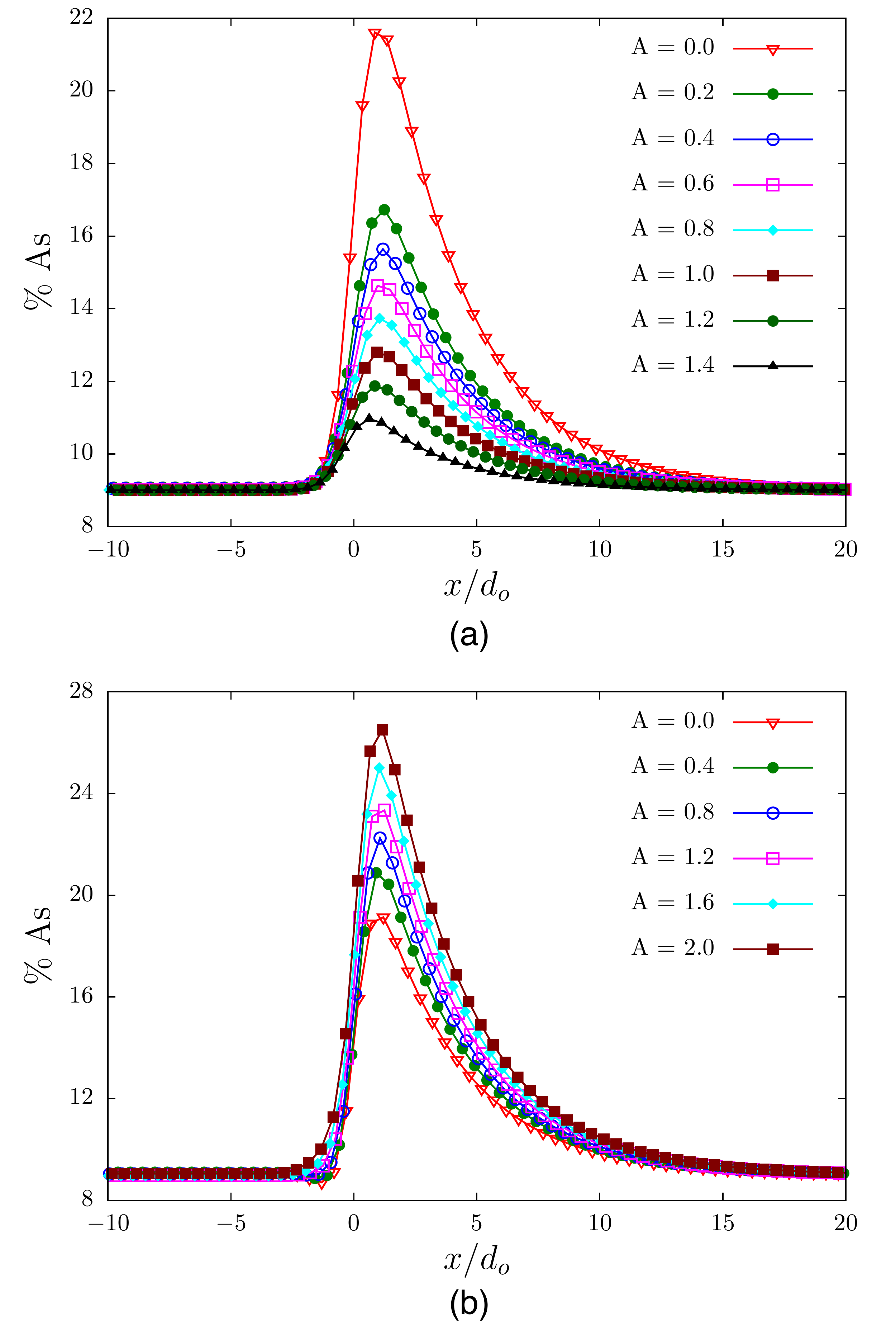}
\caption{\label{fig3} Effect of solute trapping parameter $A$ on the concentration profiles with $v_p = 0.32$ \SI{}{\meter\per\second} and $W = 13.5$ nm for (a) $D_s/D_l = 0.1$ and $b = 1.0$ and (b) $D_s/D_l = 0.001$ and $b = 0.00001$. In the parameter space where $M_s < bM_l$ is satisfied, increasing the value of $A$ increases solute segregation, while the trend is reversed when $M_s > bM_l$, due to the sign change in the expression for $a(\phi)$ given by Eq.~(\ref{shape_function}).}
\end{figure}

\section{Results and Discussion}\label{results}
\subsection{Numerical technique}
\begin{table}
 \caption{\label{table1} Thermophysical properties of \ce{Si}--\SI{9}{\percent}\ce{As} and \ce{Ni}--\SI{5}{\percent}\ce{Cu} and directional solidification parameters used in the present study.}
 \begin{ruledtabular}
 \begin{tabular}{l c c}
   Parameter & \ce{Si}--\SI{9}{\percent}\ce{As} \cite{pinomaa2019quantitative} &  \ce{Ni}--\SI{5}{\percent}\ce{Cu} \cite{algoso2003solidification} \\
   \hline
   Equilibrium partition coefficient, $k_e$ & 0.3 & 0.5\\
   Melting temperature, $T_{\mathrm{m}}$[\SI{}{\kelvin}] & 1685 & 1728 \\
   Equilibrium liquidus slope $m$[\SI{}{\kelvin\per\percent}] & -4.0 & -2.0 \\
   Alloy concentration, $c_{\mathrm{\infty}}$[\SI{}{\percent}] & 9 & 5\\
   Gibbs Thomson coefficient, $\Gamma$[\SI{}{\kelvin\meter}] & 3.4 $\times$ $10^{-7}$ & 2.6 $\times$ $10^{-7}$\\
   Capillary length, $d_o$[\SI{}{m}] & 13.5 $\times$ $10^{-9}$ & 26 $\times$ $10^{-9}$\\
   Liquid diffusion coefficient, $D_l$[\SI{}{\meter\squared\per\second}] & 15 $\times$ $10^{-9}$ & 3 $\times$ $10^{-9}$\\
   Kinetic coefficient, $\beta$[\SI{}{\second\per\meter}] & 0.595  & 0.2\\
   \hline
   Solid diffusion coefficient, $D_s$[\SI{}{\meter\squared\per\second}] & $10^{-3} {D_l}$ to $0.1 {D_l}$ \\
   
   Thermal gradient, $G$[\SI{}{\kelvin\per\meter}] & 1.4 $\times$ $10^{7}$ \\
   Pulling speed, $v_p$[\SI{}{\meter\per\second}] & 0.02 to 1 \\
 \end{tabular}
 \end{ruledtabular}
 \end{table}
To perform numerical tests we choose the binary alloys \ce{Si}--\SI{9}{\percent}\ce{As} and \ce{Ni}--\SI{5}{\percent}\ce{Cu} as model systems with compositions expressed as atomic percentages. The thermodynamic and kinetic data along with the solidification parameters are presented in Table \ref{table1}. The governing Eqs.~(\ref{cs_nonvar}), (\ref{cl_nonvar}) and (\ref{phasefield_nonvar}) are written in non-dimensional forms for numerical implementation. The non-dimensionalization procedure is presented in the supplementary material. The governing equations have been discretized using an explicit finite-difference technique on a staggered-grid. All the spatial derivatives have been resolved to second-order accuracy. The gradients and the divergence operators in the diffusion equation are discretized using a combination of forward and backward differences, with the mobility and the anti-trapping function calculated at the the center of the two grid points. For temporal discretization, we have employed the  order Euler scheme. The grid spacing in most simulations is selected to be $\Delta x = 0.5 W$, however, occasionally we have also employed a coarser mesh with $\Delta x = 0.8 W$ which is the lowest resolution for which the results are well converged. The simulation is initiated with an equilibrium phase field profile and a homogeneous equilibrium phase concentrations of $c_l = c_{\infty}$ and $c_s = k_ec_{\infty}$ at each point. A no-flux boundary condition is employed at the cold end (however, see the next paragraph, as this is mooted by the shifting algorithm), while the concentration and phase field are kept fixed to their respective values at the hot end. The temperature profile is pulled with a velocity $v_p$ resulting in solute partitioning.

In addition, we have employed a shifting algorithm to reduce computation time. This is implemented by shifting the solute and phase field profiles by an amount equal to $\Delta x$, whenever the product of pulling speed and time ($v_p t$) corresponds to a multiple of this distance. This essentially implies cutting off a layer from the cold end and adding a fresh layer at the hot end with a value of $\phi = 0$, and $c_l = c_{\infty}$.

\begin{figure}
\includegraphics[scale = 0.45]{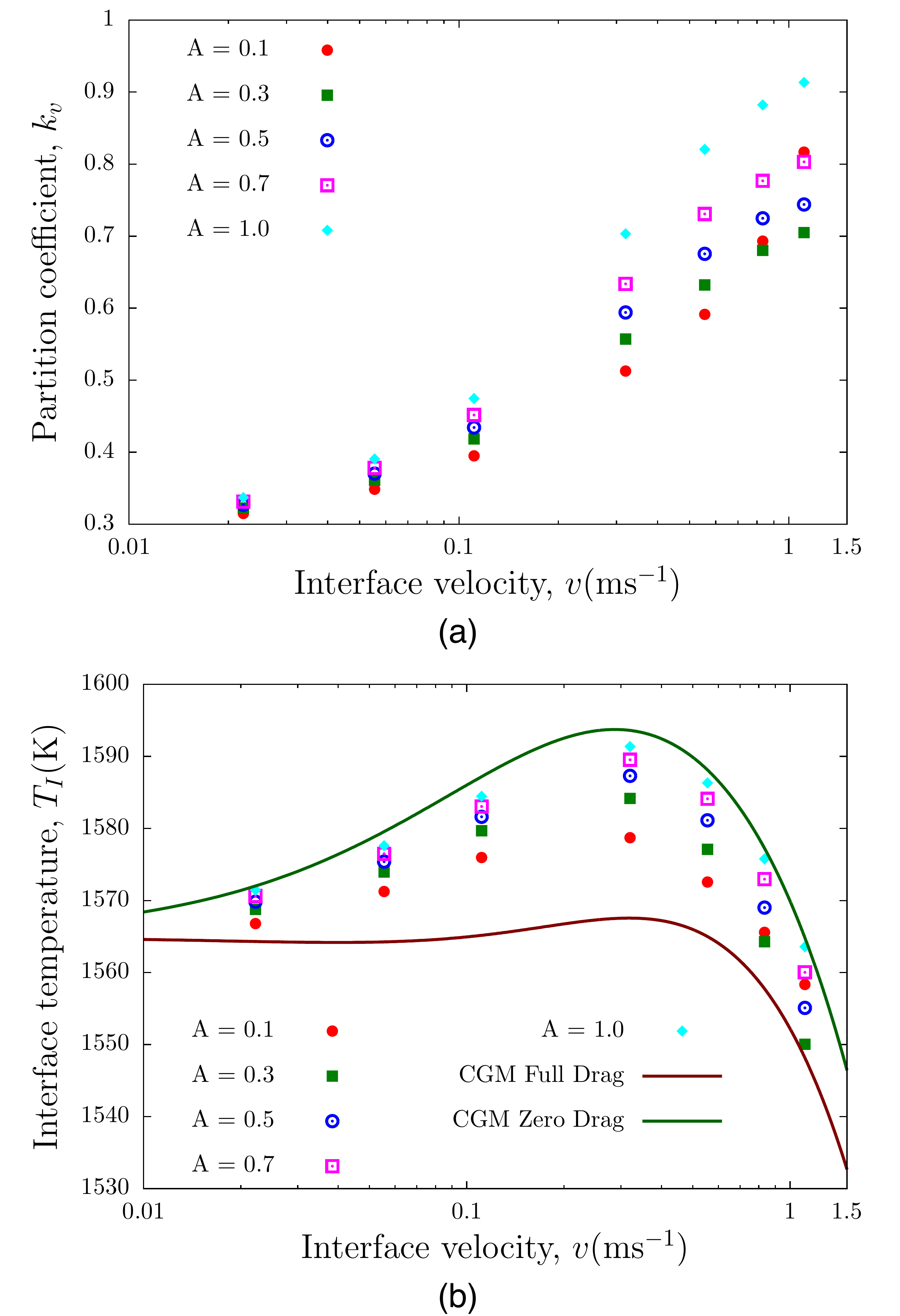}
\caption{\label{fig4}(a) Comparison of the partition coefficient as a function of velocity obtained from phase field simulations for different values of parameter $A$. (b) Comparison of the interface temperature as function of velocity computed for different values of $A$. The solid lines represent the prediction from the CGM which are provided for the sake of comparison. $W$ is fixed in the simulations and chosen to be $13.5$ nm and the value of $b$ is selected as $1$ . In addition to trapping, $A$ controls the solute drag in the system.}
\end{figure}

\subsection{Effect of parameter $A$ on solute profiles}
The effect of solute trapping parameter $A$ on the solute profiles corresponding to $D_s/D_l = 0.1$, $b=1.0$ and $v_p = 0.32$ \SI{}{\meter\per\second} for the Si-9 \% As system is shown in Fig.\ref{fig3}(a). The interface width has been fixed for all the simulations and is chosen as $W = 13.5$ nm. The maximum of the solute profiles decreases with increase in the value of $A$ indicating a reduction of interfacial solute segregation. The effect of $A$ on solute segregation also depends on the sign of the expression $a(\phi)$ given by Eq.~(\ref{shape_function}). If the condition $M_s < b M_l$ is satisfied (which is true for the above set of parameters), increasing the value of $A$ increases the level of trapping. For the solute profiles shown in Fig.~\ref{fig3}(b), the parameters correspond to $D_s/D_l = 0.001$ and $b = 0.00001$ for which the condition $M_s > b M_l$ is satisfied. In such a case increasing the value of $A$ decreases the level of solute segregation. Thus, the parameter $A$ (or the source terms), in addition to the interfacial width $W$, provides a degree of freedom to tailor solute partitioning. 

\begin{figure}
\includegraphics[scale = 0.45]{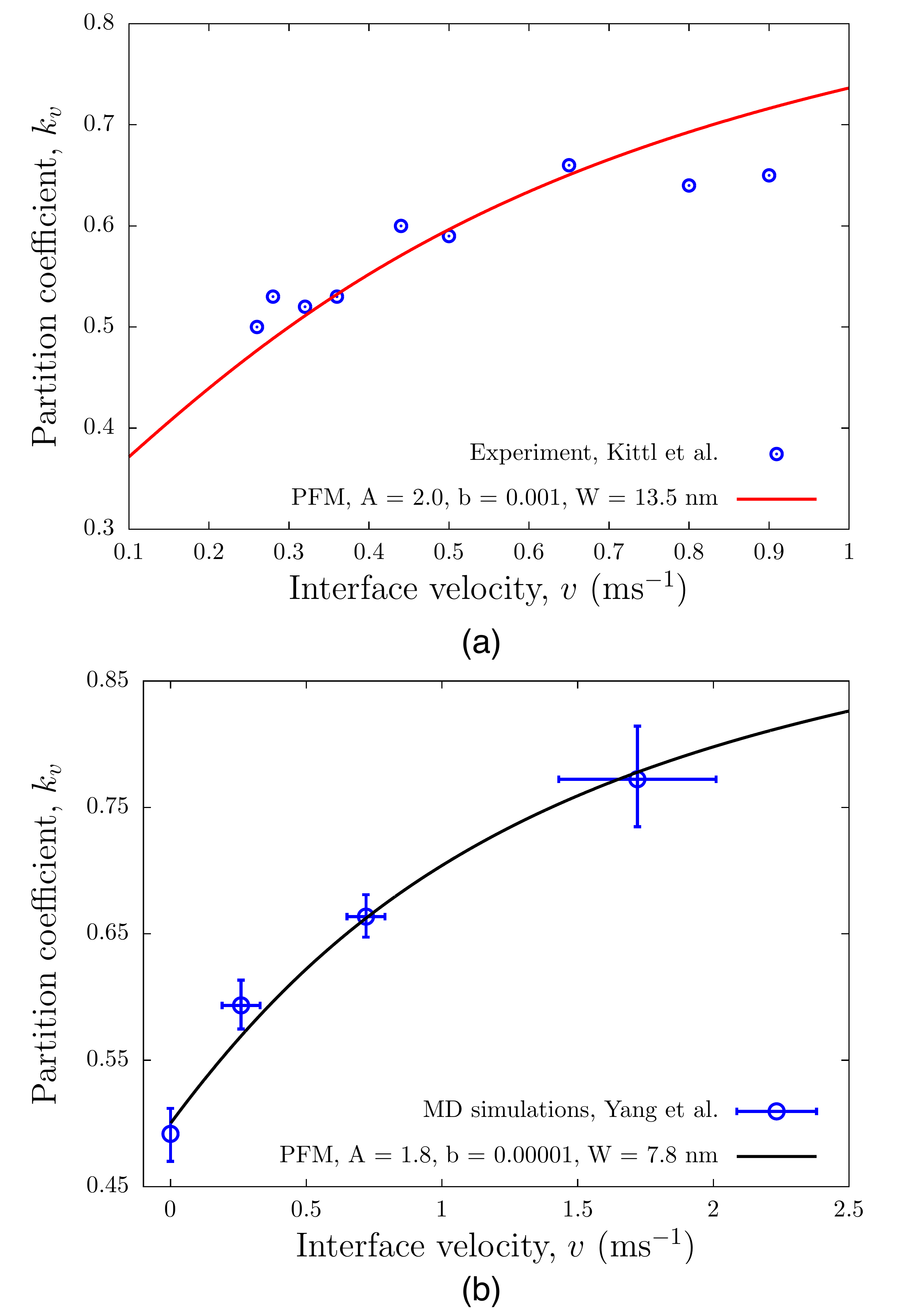}
\caption{\label{fig5}  Comparison of the partition coefficient from the phase field asymptotics obtained from the numerical solution of Eq.~(\ref{partition1}) with $A$, $b$ and $W$ as a fitting parameter and (a) the experimental results from Kittl et al. for \ce{Si}--\SI{9}{\%}\ce{As} \cite{kittl2000complete} and (b) Molecular Dynamics data for \ce{Ni}--\SI{5}{\%}\ce{Cu} \cite{yang2011atomistic}.}
\end{figure}

\begin{figure}
\includegraphics[scale = 0.48]{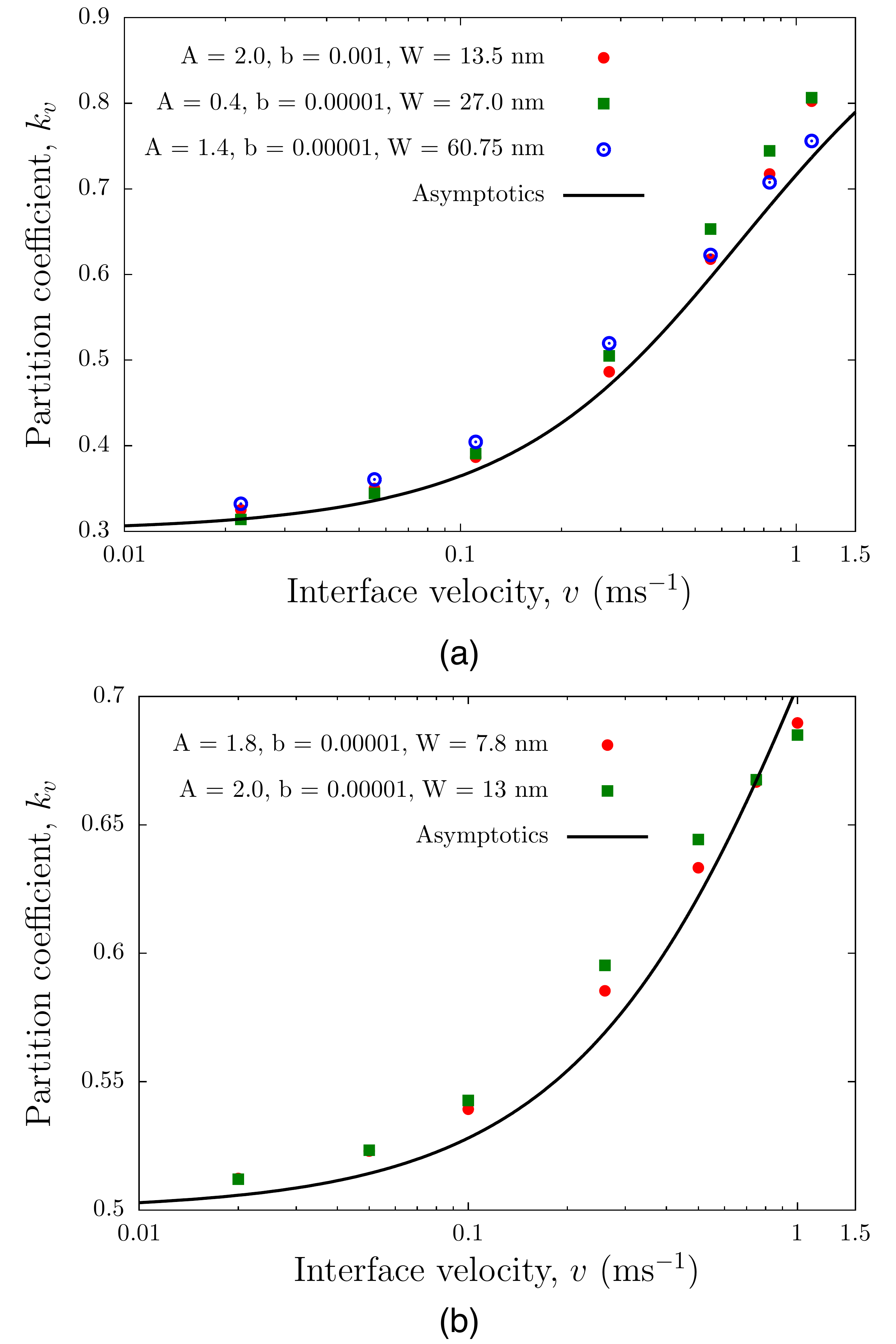}
\caption{\label{fig6} Comparison of the partition coefficient from the phase field asymptotics obtained from the numerical solution of Eq.~(\ref{partition1}) and numerical solution of the governing equations for different combinations of $A$, $b$ and $W$ for (a) \ce{Si}--\SI{9}{\%}\ce{As} and (b) \ce{Ni}--\SI{5}{\%}\ce{Cu}. Same level of trapping can be maintained for larger interface widths by appropriately tuning $A$ and $b$. }
\end{figure} 

\subsection{Effect of parameter $A$ on velocity-dependent partition coefficient and interface temperature}
The effect of model parameter $A$ on the velocity-dependent partition coefficient $k_v$ is shown in Fig.\ref{fig4}(a). The simulations have been performed with $W = 13.5$ nm, $D_s/D_l = 0.1$ and $b = 1.0$. The partition coefficient is defined from the overall solute profile according to Ahmad et al. as \cite{ahmad1998solute}
\begin{equation}
    k_v = \frac{c_s}{c_l} = \frac{c_{\infty}}{c_{\mathrm{max}}} .
\end{equation}
where $c_{\mathrm{max}}$ denotes the maximum value in the solute concentration profile. Different values of $A$ yield distinct $k_v$ curves. At low velocities ($v \sim 0.02$ \SI{}{\meter\per\second}), $A$ has little effect on $k_v$ and all the curves tend to the equilibrium partition coefficient ($k_v=k_e = 0.3$). As the velocities increase, the $k_v$ curve corresponding to higher values of $A$, in general, exhibits higher levels of trapping. Thus, the parameter $A$ can be tuned to achieve a desired level of velocity-dependent partitioning. 

We next focus on the variation of interface temperature with the velocity. The interface temperature $T_I$ corresponding to different values of $A$ are shown in Fig.\ref{fig4}(b). The interface temperature in phase field simulations is defined as the temperature corresponding to $\phi = 0.5$. The interface temperature is velocity-dependent and different choice of $A$ leads to a distinct interface temperature-velocity relationship. The interface temperature corresponds to the solidus temperature at low velocities. With increase in velocity, the interface temperature initially rises and then drops at larger velocities when the effect of interface kinetics become significant. The result is consistent with most sharp-interface models, for instance CGM which predict a similar velocity-dependent interface temperature. The CGM model predicts the interfacial temperature of the form \cite{aziz1994transition},
\begin{equation}\label{interfacetemp1}
    T_I = T_{\mathrm{m}} + \frac{m c_{\infty}}{k_v}\Big[ \frac{1 - k_v + \{k_v + \mathcal{D}(1-k_v)\}\ln(\frac{k_v}{k_e})}{1-k_e} \Big] + \frac{V}{V_o}\frac{m}{1-k_e} .
\end{equation}

The quantity in the square bracket represents the velocity- dependent liquidus slope and the third term denotes the correction due to interface kinetics. Another prediction of the CGM is that the interface temperature depends on solute drag through the drag parameter $\mathcal{D}$ in Eq.~(\ref{interfacetemp1}). The value of $\mathcal{D}$ lies between $0$ and $1$ and the case $\mathcal{D}=0$ and $\mathcal{D} =1$ corresponds to zero and full drag respectively. From Fig.\ref{fig4}(b), it is evident that the model parameter $A$ controls the solute drag as well. This can also be realized from Eq.~(\ref{drag_coefficient}). Thus, a combination of value of $A$ and $W$ uniquely determines the velocity-dependent partition coefficient and interface temperature. We contrast the prediction from the phase field model with CGM which predicts a drag independent partition coefficient. The result from the phase field model is consistent with the recent sharp-interface non-equilibrium model by Hareland et al. \cite{hareland2022thermodynamics} which shows that the solute drag affects the interfacial composition of the phases. A comparison of the interface temperature between the phase field model and the CGM indicates that as the value of $A$ increases, the amount of solute drag decreases in the system.

\begin{figure}
   \includegraphics[scale = 0.45]{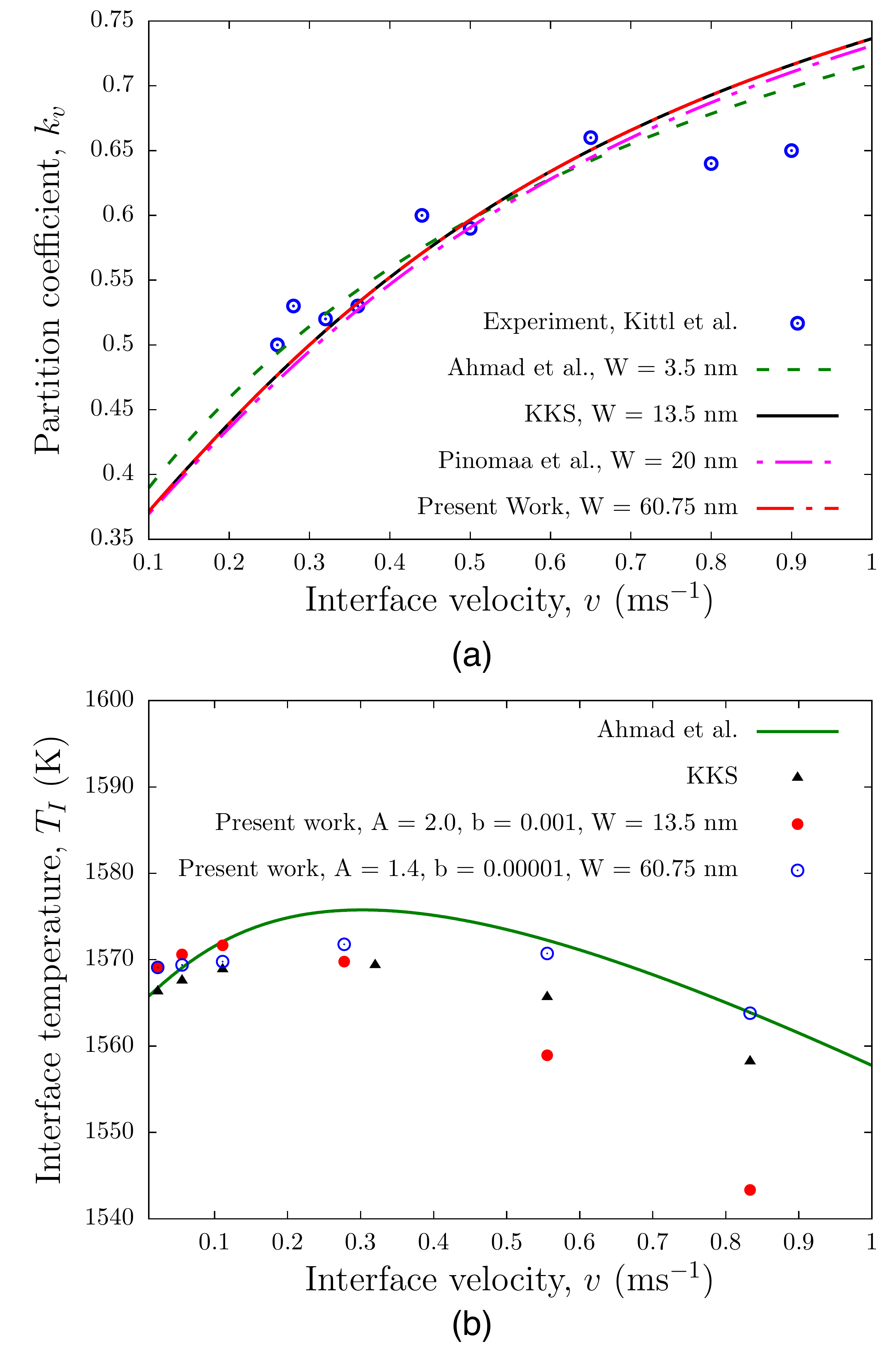}
   \caption{\label{fig7} (a) Comparison of the present with previously developed phase field models and the allowable interface widths that emulates the experimental data from Kittl et al. \cite{kittl2000complete} in Si-9 \% As and (b) Comparison of the interface temperature from different phase field models.}
\end{figure}
 
\subsection{Tailoring the partition coefficient: Selection of $A$}
By construction,  the model parameter $A$ controls the level of solute segregation. The next question that naturally arises is how to choose a value of $A$ that captures the ``correct" amount of trapping. While no closed form expression for the partition coefficient can be deduced for the model $\mu_s \neq \mu_l$, we again proceed by employing the expressions derived for the model with the assumption $\mu_s = \mu_l$. The diffusion potential jump expression Eq.~(\ref{jump}) can be re-written for a dilute alloy in dimensional form as,
\begin{equation}\label{partition1}
    \ln\left(\frac{k_v}{k_e}\right) = \frac{VW}{D_l}(1-k_v)\Delta F
\end{equation}
 where $\Delta F = F^- - F^+$, giving a transcendental equation which can be solved numerically to give a velocity-dependent $k_v$ curve for a given combination of $A$, $b$ and $W$. This implies that one can increase $W$ for computational convenience and still obtain an ``experimentally relevant trapping" by appropriately choosing the constants $A$ and $b$. This observation is also true for the general model where $\mu_s \neq \mu_l$. There are several ways in which an optimal combination of the model parameters can be chosen. The first obvious choice is to fit the $k_v$ from Eq.~(\ref{partition1}) to an experimental dataset for a range of velocities. For the sake of comparison we choose the alloy system \ce{Si}--\SI{9}{\%}\ce{As}, for which the experimental data is readily available in the literature \cite{kittl2000complete}. With $A$, $b$ and $W$ as a fitting parameter, the numerical solution of Eq.~(\ref{partition1}) is fit to the experimental data within the velocity range of \SI{0.1}{\meter\per\second} to \SI{1}{\meter\per\second}. A value of $A = 2.0$, $b = 0.001$ and $W = 13.5$ nm with $D_s/D_l = 10^{-3}$ fits the data with reasonable agreement as presented in Fig.\ref{fig5}(a). Other combinations of the model parameters can also be selected to obtain the same level of trapping. We emphasize that the above procedure gives an approximate or a first estimate of the model parameters, since Eq.~(\ref{partition1}) is derived for the model $\mu_s = \mu_l$. In our experience, we have observed that the $k_v$ curve for the model $\mu_s = \mu_l$ closely resembles the prediction of the general model upto velocities of about $0.3$ \SI{}{\meter\per\second} in most cases for the materials parameters given in Table 1. 

In absence of experimental data for an alloy, one can rely on atomistic simulation such as molecular dynamics (MD) or sharp-interface theories to choose an appropriate value of the model parameters. Fig.\ref{fig5}(b) represents the fit from solution of Eq.~(\ref{partition1}) to the
 MD simulation data on \ce{Ni}--\SI{5}{\%}\ce{Cu} from Yang et al. \cite{yang2011atomistic} in the velocity range of $v = \SI{0}{\meter\per\second}$ to $\SI{2}{\meter\per\second}$. In Fig.\ref{fig5}(b) $A = 1.8$, $b = 0.00001$ and $W = 7.8$ nm has been chosen to best fit the data from the MD simulation.

Once an estimate of the model parameters are derived, numerical simulations are performed to check the validity of the asymptotic expressions. A comparison of the prediction of Eq.~(\ref{partition1}) and numerical results from the phase field model for different combinations of $A$, $b$ and $W$ are shown in Fig.\ref{fig6}(a) for the system \ce{Si}--\SI{9}{\%}\ce{As} and for the \ce{Ni}--\SI{5}{\%}\ce{Cu} system in Fig.\ref{fig6}(b). The results for the phase field model are within $15\%$ of the prediction of the analytical expression. At higher velocities, the choice of $W$ is restricted by the requirement to resolve the diffusion length which is given by the ratio of interface diffusivity and pulling speed. It is evident from Fig.\ref{fig6} that the interface width $W$ can be increased without changing the partition coefficient by simultaneously tuning the values of parameters $A$ and $b$. It is to be noted that the choice of $A$, $b$ and $W$ should also be chosen to capture the appropriate (measured experimentally or through atomistic simulations) velocity-dependent interface temperature.

\begin{figure}
\centering
  \includegraphics[scale = 0.5]{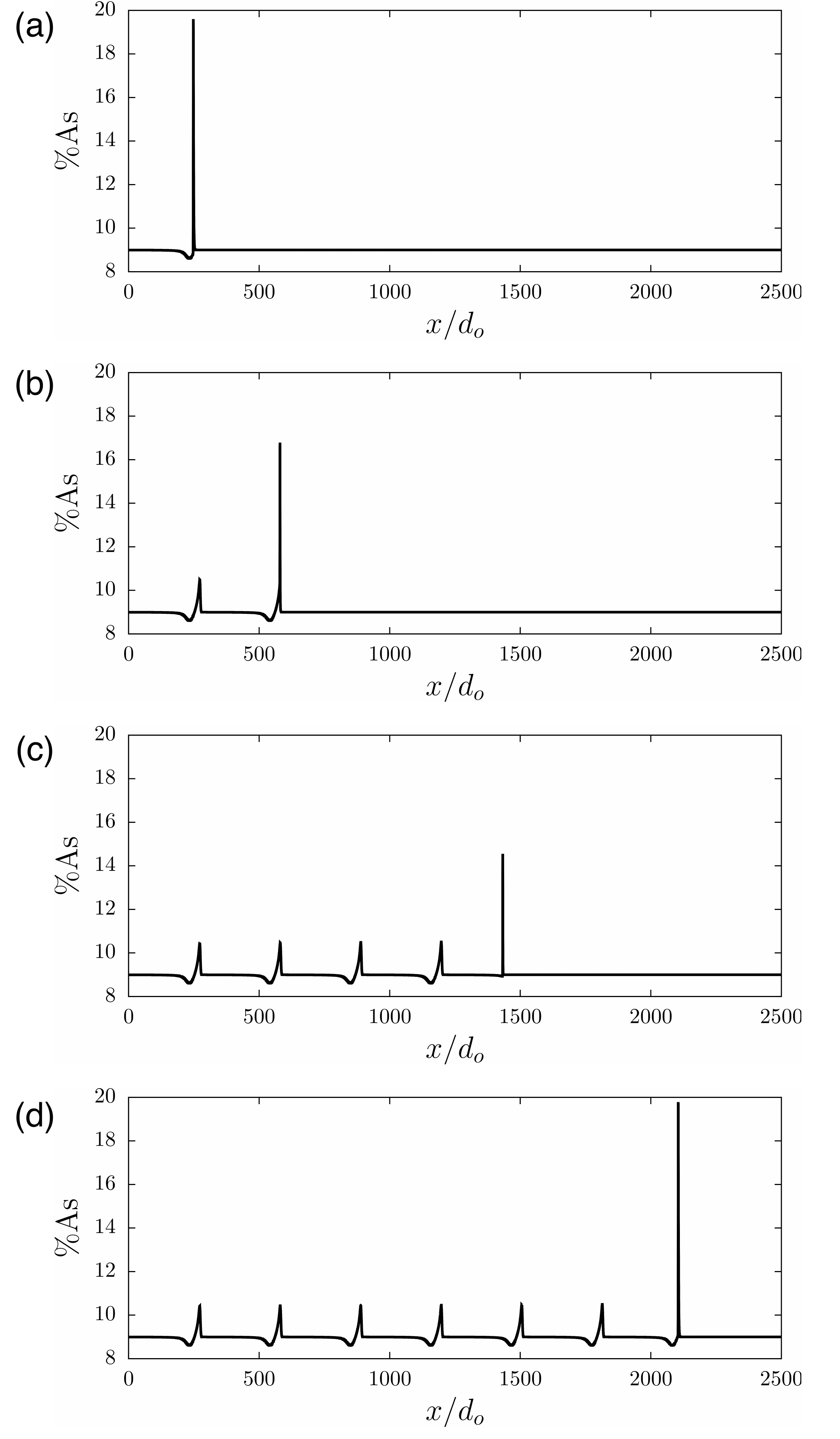}
  \caption{\label{fig8} Solute profiles exhibiting solute bands corresponding to (a) $t = 1.92 \times 10^{-4}$ sec, (b) $t = 2.56 \times 10^{-4}$ sec, (c) $t = 3.84 \times 10^{-4}$ sec and (d) $t = 5.12 \times 10^{-4}$ sec. Time periodic oscillations of the (e) interface velocity and (f) interface temperature. (h) Variation of oscillation wavelength with the temperature gradient.}
\end{figure}

\subsection{Comparison with other phase field models}
We next compare the present model to a number of previously developed phase field models. The comparison is restricted to models which arise as a special case of the present model. A direct numerical comparison with these models allows us to quantify the progress made due to the addition of interfacial source terms. The intention is to compare the order of interface widths $W$ that can be employed in different phase field models. We first examine the \ce{Si}--\ce{As} system in Fig.\ref{fig7}(a). The curve denoted by Ahmad et al. is obtained from the high velocity asymptotics of the WBM model~\cite{ahmad1998solute}. $k_v$ has a similar form to the Aziz trapping function \cite{aziz1988continuous} given by
\begin{equation}
    k_v = \frac{k_v + V/V_D}{1 + V/V_D},
\end{equation}
where $V_D$ is given by
\begin{equation}\label{WBMkv}
    V_D = \frac{3}{8}\frac{D_I}{W}\frac{\ln(1/k_e)}{1-k_e}
\end{equation}
Since $W$ is the only free parameter in the model, to emulate $V_D = \SI{0.68}{\meter\per\second}$, one can use a value of $W$ no larger than  $3.5$ nm which was used in the numerical study of Danilov and Nestler \cite{danilov2006phase} employing the WBM model. For the KKS model, which is a special case of the present model (with $\mu_s = \mu_l$ and $A=1$), the asymptotic analysis reveals that the maximum $W$ that can be chosen is about $13.5$ nm. Since the present model has an additional degrees of freedom (in $A$ and $b$), one can employ roughly four and half times the interface thickness ($60.75$ nm) and flush out the extra solute from the interface by appropriately choosing $A$ and $b$. While the asymptotic analysis of the Pinomaa model \cite{pinomaa2019quantitative} predicts the choice of $W = 20$ nm, numerical studies revealed a good convergence only until velocity of about \SI{0.45}{\meter\per\second} \cite{pinomaa2019quantitative}. For the present model convergence up to \SI{0.75}{\meter\per\second} can be seen in Fig.\ref{fig6}. 

The choice of $W$ which is treated as a fitting parameter in phase field models strictly depends on the solute trapping behaviour of the material system of interest. The solute trapping tendency of the system can be characterized by $V_D$. A lower value of $V_D$ implies a higher trapping tendency of the system or a steeper increase of the partition coefficient with velocity and vice versa. Indeed, if a similar analysis is repeated for the \ce{Ni}--\ce{Cu} system which has $V_D = \SI{1.5}{\meter\per\second}$, Eq.~(\ref{WBMkv}) predicts $W = 0.5$ nm which brings it closer to the values that were employed in numerical studies in Ref.\cite{ahmad1998solute}. An interface width of about $1$ nm can be employed for the KKS model as evident from the asymptotic analysis (with $A = 1$ and assuming $\mu_s =\mu_l$) and numerical studies for the same system in Ref.\cite{kim1999phase}. The present model allows one to choose $W$ in the range of $7.85$ nm to $13$ nm as evident from the convergence studies presented in Fig.\ref{fig6}(b). Since the smallest grid spacing needed for numeric computation scales with interface thickness, increasing the interface thickness permits use of coarser grid spacing and also the allowable timestep width. For an explicit finite difference algorithm, with uniform grid spacing, the number of floating point operations increases proportionally with the number of grid points which, in turn, scales as $1/W^d$, where, $d$ denotes the number of dimensions. The number of floating point operations also depends on the number of simulation timesteps which scales inversely with the timestep width $\Delta t$ which in turn scales as $W^2$. Therefore, the number of floating point operations scales with the interface width as $1/W^{d+2}$. Therefore, increasing the interface width by a factor of two speeds up the computation time by a factor of $16$ in two dimensions and $32$ in three dimensions. The above analysis implies a tremendous gain in computation time when the model is applied to simulate microstructures in two and three dimensions.

A comparison of the interface temperature for the \ce{Si}--\ce{As} system is presented in Fig.\ref{fig7}(b). The interface thickness for the different models are listed in Fig.\ref{fig7}(a). Not all phase field models have the same degree of freedom in tuning the interface temperature. The  interface temperature can be controlled in the present model by the parameters $A$, $b$ and $W$. Different choice of the model parameters gives rise to distinct curves as shown in Fig.\ref{fig7}(b).  The high velocity asymptotics of the WBM model predicts the interface temperature of the form given by Eq.~(\ref{interfacetemp1}) with partial drag ($\mathcal{D} = 24/35$) embedded in the model. The interface temperature was only presented until velocity of \SI{0.2}{\meter\per\second} in the study of Pinomaa et al.~ \cite{pinomaa2019quantitative} and hence not included here. Different interface temperatures for different models indicate that the solute drag is sensitive to the specifics of the model construction. We remark that the interface temperature can be made to fall on the high-velocity asymptotic solution by appropriately choosing the model parameters $A = 1.4$, $b = 0.00001$ and $W$ to $60.75$ nm without appreciably changing the $k_v$ curve as shown earlier in Fig.(\ref{fig6}). However, interface temperature alone cannot be changed without simultaneously changing $k_v$ if $W$ is altered in the WBM and KKS model. Thus, while the model by Ahmad et al. is restricted by the choice of $W = 3.5$ nm to reproduce the experimental $k_v$ curve, the present model allows to increase the interface width by $60.75$ nm to emulate the same $k_v$ and interface temperature curve.

\begin{figure}
\centering
  \includegraphics[scale = 0.55]{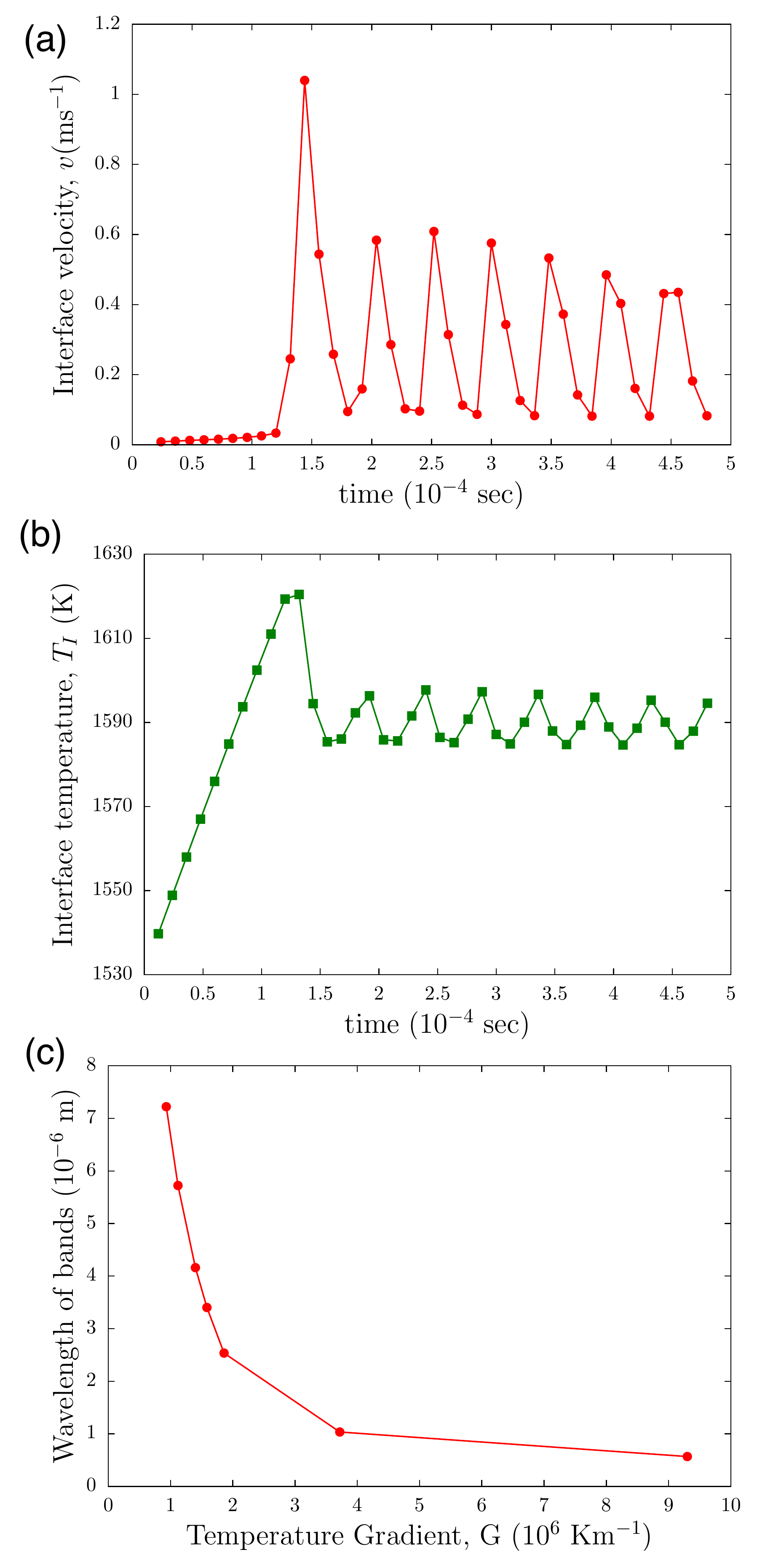}
  \caption{\label{fig9} Time periodic oscillations of the (a) interface velocity and (b) interface temperature. (c) Variation of oscillation wavelength with the temperature gradient.}
\end{figure}

\subsection{Oscillations and solute bands}
Hitherto, we have focused on the steady state interface motion. In rapid solidification experiments on metallic alloys such as Ag-Cu, Al-Cu and Al-Fe systems \cite{boettinger1984effect, kurz1996banded}, found microstructures that are periodic along the growth direction. Such structures have been termed as "banded structure" . The bands consists of alternate dark bands of either cellular-dendritic or eutectic structure, depending on the alloy composition and light bands free from any microsegregation. The origin of banding has been rationalized in terms of an oscillatory instability of the planar  solid-liquid interface with zero wavenumber which is characterized by large variations of the interface velocity. Incorporating non-equilibrium effects into the Mullins-Sekerka theory \cite{mullins1964stability} i.e. velocity dependent partition coefficient and interface temperature, Coriell and Sekerka \cite{coriell1983oscillatory}, Merchant and Davis \cite{merchant1990morphological} and Karma and Sarkissian \cite{karma1993interface} have independently identified an oscillatory instability with a zero wavenumber along the solid-liquid front.

Our phase field model, which naturally incorporates the non-equilibrium effects, is a perfect tool to investigate such an instability. We present solute profiles exhibiting solute bands in Fig.\ref{fig8} corresponding to $G=  1.4 \times 10^6$ \SI{}{\kelvin\per\meter} and $v_p =$ \SI{0.3}{\meter\per\second}. The velocity is selected such that it corresponds to $\frac{\mathrm{d}T}{\mathrm{d}V} > 0$ regime of the interface temperature-velocity curve, where such instabilities are expected. As the solid-liquid front advances, it leaves behind a trail of solute rich and solute poor zones. The region between the bands is free of solute segregation. The amplitude of the peaks and troughs of the bands are almost constant owing to the negligible solute diffusion in solid ($D_s/D_l = 0.001$).  Furthermore, the solid-liquid front undergoes time-periodic oscillation as characterized by the interface velocity and temperature shown in Fig.\ref{fig9}(a) and (b). The liquid composition of the interface characterized by the maximum of the front composition also undergoes time-periodic variation as evident from the solute profiles(Fig.\ref{fig8}(a)-(d)).The bands are evenly spaced with a wavelength of about \SI{4}{\micro\meter}.  While we do not aim to make a quantitative comparison with the aforementioned analytical theories here, the qualitative dependence of the oscillation frequency on $G$ can be observed in Fig.\ref{fig9}(c). As predicted by Merchant and Davis, the oscillation wavelength decreases with increase in the value of $G$.

\section{Conclusions}\label{conclusion}
We have presented a variational phase field model in which non-equilibrium effects such as solute trapping, drag and interface kinetics can be introduced in a controlled manner. The key feature of the model is the derivation of separate kinetic equations for the phase concentrations with interfacial source terms which can be selected to control solute trapping and drag. Separate governing equations for the phase concentrations allows for a straightforward incorporation of arbitrary alloy model without having to solve the condition of pointwise equality of diffusion potentials which is often the computation bottle-neck in the previous works. To perform simulations with the present model following steps needs to be performed:
\begin{itemize}
    \item For a given alloy system, the free energies of the solid and the liquid phases have to be specified first. While the free energies in the present work have been selected as an ideal dilute alloy with straight solidus and liquidus, the model is capable to take input from thermodynamic databases such as CALPHAD to accommodate more realistic free energies. Since $\mu_s$ is not constrained to be equal to $\mu_l$, there is no need to solve a coupled set of  nonlinear equations to determine effective phase concentrations as in the KKS model. 
    
    \item Next, we select the mobility interpolation functions in the normal and tangential directions to counter Kapitza jump and surface diffusion respectively based on the discussion in section \ref{alternative_choice}. Depending on the ratio of the chosen diffusivity of the phases $D_s/D_l$, the parameters $b$ and $a$ in $q_n(\phi)$ (Eq.~(\ref{normal2})) are evaluated such that $G^+ = G^-$ in Eq.~(\ref{integral3}) is satisfied. $b$ is usually chosen such that the diffusivity at the interfacial region is high which is true if $b < D_s/D_l$. The choice of $b$ and $a$ should also satisfy $q_n(\phi) \geq 0$ for the phase concentration equations to be stable. 
    
    \item To capture a specific partition coefficient and interface temperature as a function of velocity, the source term $A$, the diffusivity shape parameter $b$ and interface width $W$ can be tuned by comparing the numerical solution of asymptotics Eq.~(\ref{partition1}) for partition coefficient to an experimental dataset, atomistic simulations or sharp-interface theories. While this method provides various combinations of the model parameters, the choice of $A$, $b$ and $W$ should additionally capture the desired interface temperature (or solute drag).
\end{itemize}

 Numerical performance of the model was demonstrated through study of the partition coefficient and interface temperature at different interface widths. A comparison with the previous phase field models reveal that ``experimentally-relevant" trapping can be achieved with interface widths of about $25$ times larger than the WBM model \cite{ahmad1998solute}, about four times that can be employed in the KKS \cite{kim1999phase} and  about $3$ times the Pinomaa \cite{pinomaa2019quantitative} model depending upon the material system at hand. Our phase field model was able to capture the oscillatory instability during rapid solidification leading to solute bands which are in qualitative agreement with the analytical and experimental predictions. Future two-dimensional studies is currently underway to quantitatively identify the regimes of cellular, dendrite and bands depending on the rapid solidification parameters (such as temperature gradient and pulling speeds). 

 \begin{acknowledgments}
 This work was performed under financial assistance award 70NANB14H012 from the U.S. Department of Commerce, National Institute of Standards and Technology as part of the Center for Hierarchical Materials Design (CHiMaD).
\end{acknowledgments}

\bibliography{apssamp}

\begin{thebibliography}{67}%
\makeatletter
\providecommand \@ifxundefined [1]{%
 \@ifx{#1\undefined}
}%
\providecommand \@ifnum [1]{%
 \ifnum #1\expandafter \@firstoftwo
 \else \expandafter \@secondoftwo
 \fi
}%
\providecommand \@ifx [1]{%
 \ifx #1\expandafter \@firstoftwo
 \else \expandafter \@secondoftwo
 \fi
}%
\providecommand \natexlab [1]{#1}%
\providecommand \enquote  [1]{``#1''}%
\providecommand \bibnamefont  [1]{#1}%
\providecommand \bibfnamefont [1]{#1}%
\providecommand \citenamefont [1]{#1}%
\providecommand \href@noop [0]{\@secondoftwo}%
\providecommand \href [0]{\begingroup \@sanitize@url \@href}%
\providecommand \@href[1]{\@@startlink{#1}\@@href}%
\providecommand \@@href[1]{\endgroup#1\@@endlink}%
\providecommand \@sanitize@url [0]{\catcode `\\12\catcode `\$12\catcode
  `\&12\catcode `\#12\catcode `\^12\catcode `\_12\catcode `\%12\relax}%
\providecommand \@@startlink[1]{}%
\providecommand \@@endlink[0]{}%
\providecommand \url  [0]{\begingroup\@sanitize@url \@url }%
\providecommand \@url [1]{\endgroup\@href {#1}{\urlprefix }}%
\providecommand \urlprefix  [0]{URL }%
\providecommand \Eprint [0]{\href }%
\providecommand \doibase [0]{https://doi.org/}%
\providecommand \selectlanguage [0]{\@gobble}%
\providecommand \bibinfo  [0]{\@secondoftwo}%
\providecommand \bibfield  [0]{\@secondoftwo}%
\providecommand \translation [1]{[#1]}%
\providecommand \BibitemOpen [0]{}%
\providecommand \bibitemStop [0]{}%
\providecommand \bibitemNoStop [0]{.\EOS\space}%
\providecommand \EOS [0]{\spacefactor3000\relax}%
\providecommand \BibitemShut  [1]{\csname bibitem#1\endcsname}%
\let\auto@bib@innerbib\@empty
%</preamble>
\bibitem [{\citenamefont {Cui}\ \emph {et~al.}(2021)\citenamefont {Cui},
  \citenamefont {Zhang}, \citenamefont {Zhang}, \citenamefont {Chen},
  \citenamefont {Zhang},\ and\ \citenamefont {Dong}}]{cui2021additive}%
  \BibitemOpen
  \bibfield  {author} {\bibinfo {author} {\bibfnamefont {X.}~\bibnamefont
  {Cui}}, \bibinfo {author} {\bibfnamefont {S.}~\bibnamefont {Zhang}}, \bibinfo
  {author} {\bibfnamefont {C.}~\bibnamefont {Zhang}}, \bibinfo {author}
  {\bibfnamefont {J.}~\bibnamefont {Chen}}, \bibinfo {author} {\bibfnamefont
  {J.}~\bibnamefont {Zhang}},\ and\ \bibinfo {author} {\bibfnamefont
  {S.}~\bibnamefont {Dong}},\ }\bibfield  {title} {\bibinfo {title} {Additive
  manufacturing of 24{C}r{N}i{M}o low alloy steel by selective laser melting:
  Influence of volumetric energy density on densification, microstructure and
  hardness},\ }\href@noop {} {\bibfield  {journal} {\bibinfo  {journal}
  {Materials Science and Engineering: A}\ }\textbf {\bibinfo {volume} {809}},\
  \bibinfo {pages} {140957} (\bibinfo {year} {2021})}\BibitemShut {NoStop}%
\bibitem [{\citenamefont {Wolf}\ \emph {et~al.}(2020)\citenamefont {Wolf},
  \citenamefont {E.~Silva}, \citenamefont {Zepon}, \citenamefont {Kiminami},
  \citenamefont {Bolfarini},\ and\ \citenamefont {Botta}}]{wolf2020single}%
  \BibitemOpen
  \bibfield  {author} {\bibinfo {author} {\bibfnamefont {W.}~\bibnamefont
  {Wolf}}, \bibinfo {author} {\bibfnamefont {L.~P.~M.}\ \bibnamefont
  {E.~Silva}}, \bibinfo {author} {\bibfnamefont {G.}~\bibnamefont {Zepon}},
  \bibinfo {author} {\bibfnamefont {C.~S.}\ \bibnamefont {Kiminami}}, \bibinfo
  {author} {\bibfnamefont {C.}~\bibnamefont {Bolfarini}},\ and\ \bibinfo
  {author} {\bibfnamefont {W.~J.}\ \bibnamefont {Botta}},\ }\bibfield  {title}
  {\bibinfo {title} {Single step fabrication by spray forming of large volume
  {A}l-based composites reinforced with quasicrystals},\ }\href@noop {}
  {\bibfield  {journal} {\bibinfo  {journal} {Scripta Materialia}\ }\textbf
  {\bibinfo {volume} {181}},\ \bibinfo {pages} {86} (\bibinfo {year}
  {2020})}\BibitemShut {NoStop}%
\bibitem [{\citenamefont {Oliveira}\ \emph {et~al.}(2020)\citenamefont
  {Oliveira}, \citenamefont {Santos},\ and\ \citenamefont
  {Miranda}}]{oliveira2020revisiting}%
  \BibitemOpen
  \bibfield  {author} {\bibinfo {author} {\bibfnamefont {J.}~\bibnamefont
  {Oliveira}}, \bibinfo {author} {\bibfnamefont {T.}~\bibnamefont {Santos}},\
  and\ \bibinfo {author} {\bibfnamefont {R.}~\bibnamefont {Miranda}},\
  }\bibfield  {title} {\bibinfo {title} {Revisiting fundamental welding
  concepts to improve additive manufacturing: from theory to practice},\
  }\href@noop {} {\bibfield  {journal} {\bibinfo  {journal} {Progress in
  Materials Science}\ }\textbf {\bibinfo {volume} {107}},\ \bibinfo {pages}
  {100590} (\bibinfo {year} {2020})}\BibitemShut {NoStop}%
\bibitem [{\citenamefont {Kobold}\ \emph {et~al.}(2017)\citenamefont {Kobold},
  \citenamefont {Kuang}, \citenamefont {Wang}, \citenamefont {Hornfeck},
  \citenamefont {Kolbe},\ and\ \citenamefont {Herlach}}]{kobold2017dendrite}%
  \BibitemOpen
  \bibfield  {author} {\bibinfo {author} {\bibfnamefont {R.}~\bibnamefont
  {Kobold}}, \bibinfo {author} {\bibfnamefont {W.}~\bibnamefont {Kuang}},
  \bibinfo {author} {\bibfnamefont {H.}~\bibnamefont {Wang}}, \bibinfo {author}
  {\bibfnamefont {W.}~\bibnamefont {Hornfeck}}, \bibinfo {author}
  {\bibfnamefont {M.}~\bibnamefont {Kolbe}},\ and\ \bibinfo {author}
  {\bibfnamefont {D.~M.}\ \bibnamefont {Herlach}},\ }\bibfield  {title}
  {\bibinfo {title} {Dendrite growth velocity in the undercooled melt of glass
  forming {N}i50{Z}r50 compound},\ }\href@noop {} {\bibfield  {journal}
  {\bibinfo  {journal} {Philosophical Magazine Letters}\ }\textbf {\bibinfo
  {volume} {97}},\ \bibinfo {pages} {249} (\bibinfo {year} {2017})}\BibitemShut
  {NoStop}%
\bibitem [{\citenamefont {Ghosh}\ \emph {et~al.}(2018)\citenamefont {Ghosh},
  \citenamefont {Ma}, \citenamefont {Levine}, \citenamefont {Ricker},
  \citenamefont {Stoudt}, \citenamefont {Heigel},\ and\ \citenamefont
  {Guyer}}]{ghosh2018single}%
  \BibitemOpen
  \bibfield  {author} {\bibinfo {author} {\bibfnamefont {S.}~\bibnamefont
  {Ghosh}}, \bibinfo {author} {\bibfnamefont {L.}~\bibnamefont {Ma}}, \bibinfo
  {author} {\bibfnamefont {L.~E.}\ \bibnamefont {Levine}}, \bibinfo {author}
  {\bibfnamefont {R.~E.}\ \bibnamefont {Ricker}}, \bibinfo {author}
  {\bibfnamefont {M.~R.}\ \bibnamefont {Stoudt}}, \bibinfo {author}
  {\bibfnamefont {J.~C.}\ \bibnamefont {Heigel}},\ and\ \bibinfo {author}
  {\bibfnamefont {J.~E.}\ \bibnamefont {Guyer}},\ }\bibfield  {title} {\bibinfo
  {title} {Single-track melt-pool measurements and microstructures in {I}nconel
  625},\ }\href@noop {} {\bibfield  {journal} {\bibinfo  {journal} {JOM}\
  }\textbf {\bibinfo {volume} {70}},\ \bibinfo {pages} {1011} (\bibinfo {year}
  {2018})}\BibitemShut {NoStop}%
\bibitem [{\citenamefont {Chen}\ \emph {et~al.}(2021)\citenamefont {Chen},
  \citenamefont {Du}, \citenamefont {Fu}, \citenamefont {He}, \citenamefont
  {Huang}, \citenamefont {Shi},\ and\ \citenamefont {Xie}}]{chen2021formation}%
  \BibitemOpen
  \bibfield  {author} {\bibinfo {author} {\bibfnamefont {M.}~\bibnamefont
  {Chen}}, \bibinfo {author} {\bibfnamefont {Q.}~\bibnamefont {Du}}, \bibinfo
  {author} {\bibfnamefont {H.}~\bibnamefont {Fu}}, \bibinfo {author}
  {\bibfnamefont {X.}~\bibnamefont {He}}, \bibinfo {author} {\bibfnamefont
  {Z.}~\bibnamefont {Huang}}, \bibinfo {author} {\bibfnamefont
  {R.}~\bibnamefont {Shi}},\ and\ \bibinfo {author} {\bibfnamefont
  {J.}~\bibnamefont {Xie}},\ }\bibfield  {title} {\bibinfo {title} {The
  formation of secondary phase in sub-rapid solidification process of
  {A}l--{M}g--{S}i alloys},\ }\href@noop {} {\bibfield  {journal} {\bibinfo
  {journal} {Materialia}\ }\textbf {\bibinfo {volume} {15}},\ \bibinfo {pages}
  {101022} (\bibinfo {year} {2021})}\BibitemShut {NoStop}%
\bibitem [{\citenamefont {Enrique}\ \emph {et~al.}(2021)\citenamefont
  {Enrique}, \citenamefont {Keshavarzkermani},\ and\ \citenamefont
  {Zhou}}]{enrique2021effect}%
  \BibitemOpen
  \bibfield  {author} {\bibinfo {author} {\bibfnamefont {P.~D.}\ \bibnamefont
  {Enrique}}, \bibinfo {author} {\bibfnamefont {A.}~\bibnamefont
  {Keshavarzkermani}},\ and\ \bibinfo {author} {\bibfnamefont {N.~Y.}\
  \bibnamefont {Zhou}},\ }\bibfield  {title} {\bibinfo {title} {Effect of
  microsegregation on high-temperature microstructure evolution in rapid
  solidification processed {N}b-rich {N}i superalloys},\ }\href@noop {}
  {\bibfield  {journal} {\bibinfo  {journal} {Advanced Engineering Materials}\
  }\textbf {\bibinfo {volume} {23}},\ \bibinfo {pages} {2001396} (\bibinfo
  {year} {2021})}\BibitemShut {NoStop}%
\bibitem [{\citenamefont {Keller}\ \emph {et~al.}(2017)\citenamefont {Keller},
  \citenamefont {Lindwall}, \citenamefont {Ghosh}, \citenamefont {Ma},
  \citenamefont {Lane}, \citenamefont {Zhang}, \citenamefont {Kattner},
  \citenamefont {Lass}, \citenamefont {Heigel}, \citenamefont {Idell} \emph
  {et~al.}}]{keller2017application}%
  \BibitemOpen
  \bibfield  {author} {\bibinfo {author} {\bibfnamefont {T.}~\bibnamefont
  {Keller}}, \bibinfo {author} {\bibfnamefont {G.}~\bibnamefont {Lindwall}},
  \bibinfo {author} {\bibfnamefont {S.}~\bibnamefont {Ghosh}}, \bibinfo
  {author} {\bibfnamefont {L.}~\bibnamefont {Ma}}, \bibinfo {author}
  {\bibfnamefont {B.~M.}\ \bibnamefont {Lane}}, \bibinfo {author}
  {\bibfnamefont {F.}~\bibnamefont {Zhang}}, \bibinfo {author} {\bibfnamefont
  {U.~R.}\ \bibnamefont {Kattner}}, \bibinfo {author} {\bibfnamefont {E.~A.}\
  \bibnamefont {Lass}}, \bibinfo {author} {\bibfnamefont {J.~C.}\ \bibnamefont
  {Heigel}}, \bibinfo {author} {\bibfnamefont {Y.}~\bibnamefont {Idell}}, \emph
  {et~al.},\ }\bibfield  {title} {\bibinfo {title} {Application of finite
  element, phase-field, and {CALPHAD}-based methods to additive manufacturing
  of {N}i-based superalloys},\ }\href@noop {} {\bibfield  {journal} {\bibinfo
  {journal} {Acta Materialia}\ }\textbf {\bibinfo {volume} {139}},\ \bibinfo
  {pages} {244} (\bibinfo {year} {2017})}\BibitemShut {NoStop}%
\bibitem [{\citenamefont {Ghosh}\ \emph {et~al.}(2017)\citenamefont {Ghosh},
  \citenamefont {Ma}, \citenamefont {Ofori-Opoku},\ and\ \citenamefont
  {Guyer}}]{ghosh2017primary}%
  \BibitemOpen
  \bibfield  {author} {\bibinfo {author} {\bibfnamefont {S.}~\bibnamefont
  {Ghosh}}, \bibinfo {author} {\bibfnamefont {L.}~\bibnamefont {Ma}}, \bibinfo
  {author} {\bibfnamefont {N.}~\bibnamefont {Ofori-Opoku}},\ and\ \bibinfo
  {author} {\bibfnamefont {J.~E.}\ \bibnamefont {Guyer}},\ }\bibfield  {title}
  {\bibinfo {title} {On the primary spacing and microsegregation of cellular
  dendrites in laser deposited {N}i--{N}b alloys},\ }\href@noop {} {\bibfield
  {journal} {\bibinfo  {journal} {Modelling and Simulation in Materials Science
  and Engineering}\ }\textbf {\bibinfo {volume} {25}},\ \bibinfo {pages}
  {065002} (\bibinfo {year} {2017})}\BibitemShut {NoStop}%
\bibitem [{\citenamefont {Karayagiz}\ \emph {et~al.}(2020)\citenamefont
  {Karayagiz}, \citenamefont {Johnson}, \citenamefont {Seede}, \citenamefont
  {Attari}, \citenamefont {Zhang}, \citenamefont {Huang}, \citenamefont
  {Ghosh}, \citenamefont {Duong}, \citenamefont {Karaman}, \citenamefont
  {Elwany} \emph {et~al.}}]{karayagiz2020finite}%
  \BibitemOpen
  \bibfield  {author} {\bibinfo {author} {\bibfnamefont {K.}~\bibnamefont
  {Karayagiz}}, \bibinfo {author} {\bibfnamefont {L.}~\bibnamefont {Johnson}},
  \bibinfo {author} {\bibfnamefont {R.}~\bibnamefont {Seede}}, \bibinfo
  {author} {\bibfnamefont {V.}~\bibnamefont {Attari}}, \bibinfo {author}
  {\bibfnamefont {B.}~\bibnamefont {Zhang}}, \bibinfo {author} {\bibfnamefont
  {X.}~\bibnamefont {Huang}}, \bibinfo {author} {\bibfnamefont
  {S.}~\bibnamefont {Ghosh}}, \bibinfo {author} {\bibfnamefont
  {T.}~\bibnamefont {Duong}}, \bibinfo {author} {\bibfnamefont
  {I.}~\bibnamefont {Karaman}}, \bibinfo {author} {\bibfnamefont
  {A.}~\bibnamefont {Elwany}}, \emph {et~al.},\ }\bibfield  {title} {\bibinfo
  {title} {Finite interface dissipation phase field modeling of {N}i--{N}b
  under additive manufacturing conditions},\ }\href@noop {} {\bibfield
  {journal} {\bibinfo  {journal} {Acta Materialia}\ }\textbf {\bibinfo {volume}
  {185}},\ \bibinfo {pages} {320} (\bibinfo {year} {2020})}\BibitemShut
  {NoStop}%
\bibitem [{\citenamefont {Chadwick}\ and\ \citenamefont
  {Voorhees}(2021)}]{chadwick2021development}%
  \BibitemOpen
  \bibfield  {author} {\bibinfo {author} {\bibfnamefont {A.~F.}\ \bibnamefont
  {Chadwick}}\ and\ \bibinfo {author} {\bibfnamefont {P.~W.}\ \bibnamefont
  {Voorhees}},\ }\bibfield  {title} {\bibinfo {title} {The development of grain
  structure during additive manufacturing},\ }\href@noop {} {\bibfield
  {journal} {\bibinfo  {journal} {Acta Materialia}\ }\textbf {\bibinfo {volume}
  {211}},\ \bibinfo {pages} {116862} (\bibinfo {year} {2021})}\BibitemShut
  {NoStop}%
\bibitem [{\citenamefont {Wang}\ \emph {et~al.}(2019)\citenamefont {Wang},
  \citenamefont {Liu}, \citenamefont {Ji}, \citenamefont {Liu}, \citenamefont
  {Horstemeyer},\ and\ \citenamefont {Chen}}]{wang2019investigation}%
  \BibitemOpen
  \bibfield  {author} {\bibinfo {author} {\bibfnamefont {X.}~\bibnamefont
  {Wang}}, \bibinfo {author} {\bibfnamefont {P.}~\bibnamefont {Liu}}, \bibinfo
  {author} {\bibfnamefont {Y.}~\bibnamefont {Ji}}, \bibinfo {author}
  {\bibfnamefont {Y.}~\bibnamefont {Liu}}, \bibinfo {author} {\bibfnamefont
  {M.}~\bibnamefont {Horstemeyer}},\ and\ \bibinfo {author} {\bibfnamefont
  {L.}~\bibnamefont {Chen}},\ }\bibfield  {title} {\bibinfo {title}
  {Investigation on microsegregation of {IN}718 alloy during additive
  manufacturing via integrated phase-field and finite-element modeling},\
  }\href@noop {} {\bibfield  {journal} {\bibinfo  {journal} {Journal of
  Materials Engineering and Performance}\ }\textbf {\bibinfo {volume} {28}},\
  \bibinfo {pages} {657} (\bibinfo {year} {2019})}\BibitemShut {NoStop}%
\bibitem [{\citenamefont {Kurz}\ and\ \citenamefont
  {Trivedi}(1994)}]{kurz1994rapid}%
  \BibitemOpen
  \bibfield  {author} {\bibinfo {author} {\bibfnamefont {W.}~\bibnamefont
  {Kurz}}\ and\ \bibinfo {author} {\bibfnamefont {R.}~\bibnamefont {Trivedi}},\
  }\bibfield  {title} {\bibinfo {title} {Rapid solidification processing and
  microstructure formation},\ }\href@noop {} {\bibfield  {journal} {\bibinfo
  {journal} {Materials Science and Engineering: A}\ }\textbf {\bibinfo {volume}
  {179}},\ \bibinfo {pages} {46} (\bibinfo {year} {1994})}\BibitemShut
  {NoStop}%
\bibitem [{\citenamefont {Merchant}\ and\ \citenamefont
  {Davis}(1990)}]{merchant1990morphological}%
  \BibitemOpen
  \bibfield  {author} {\bibinfo {author} {\bibfnamefont {G.}~\bibnamefont
  {Merchant}}\ and\ \bibinfo {author} {\bibfnamefont {S.}~\bibnamefont
  {Davis}},\ }\bibfield  {title} {\bibinfo {title} {Morphological instability
  in rapid directional solidification},\ }\href@noop {} {\bibfield  {journal}
  {\bibinfo  {journal} {Acta Metallurgica et Materialia}\ }\textbf {\bibinfo
  {volume} {38}},\ \bibinfo {pages} {2683} (\bibinfo {year}
  {1990})}\BibitemShut {NoStop}%
\bibitem [{\citenamefont {Aziz}(1982)}]{aziz1982model}%
  \BibitemOpen
  \bibfield  {author} {\bibinfo {author} {\bibfnamefont {M.~J.}\ \bibnamefont
  {Aziz}},\ }\bibfield  {title} {\bibinfo {title} {Model for solute
  redistribution during rapid solidification},\ }\href@noop {} {\bibfield
  {journal} {\bibinfo  {journal} {Journal of Applied Physics}\ }\textbf
  {\bibinfo {volume} {53}},\ \bibinfo {pages} {1158} (\bibinfo {year}
  {1982})}\BibitemShut {NoStop}%
\bibitem [{\citenamefont {Aziz}\ and\ \citenamefont
  {Kaplan}(1988)}]{aziz1988continuous}%
  \BibitemOpen
  \bibfield  {author} {\bibinfo {author} {\bibfnamefont {M.~J.}\ \bibnamefont
  {Aziz}}\ and\ \bibinfo {author} {\bibfnamefont {T.}~\bibnamefont {Kaplan}},\
  }\bibfield  {title} {\bibinfo {title} {Continuous growth model for interface
  motion during alloy solidification},\ }\href@noop {} {\bibfield  {journal}
  {\bibinfo  {journal} {Acta Metallurgica}\ }\textbf {\bibinfo {volume} {36}},\
  \bibinfo {pages} {2335} (\bibinfo {year} {1988})}\BibitemShut {NoStop}%
\bibitem [{\citenamefont {Aziz}\ and\ \citenamefont
  {Boettinger}(1994)}]{aziz1994transition}%
  \BibitemOpen
  \bibfield  {author} {\bibinfo {author} {\bibfnamefont {M.~J.}\ \bibnamefont
  {Aziz}}\ and\ \bibinfo {author} {\bibfnamefont {W.}~\bibnamefont
  {Boettinger}},\ }\bibfield  {title} {\bibinfo {title} {On the transition from
  short-range diffusion-limited to collision-limited growth in alloy
  solidification},\ }\href@noop {} {\bibfield  {journal} {\bibinfo  {journal}
  {Acta Metallurgica et Materialia}\ }\textbf {\bibinfo {volume} {42}},\
  \bibinfo {pages} {527} (\bibinfo {year} {1994})}\BibitemShut {NoStop}%
\bibitem [{\citenamefont {{\AA}gren}(1989)}]{aagren1989simplified}%
  \BibitemOpen
  \bibfield  {author} {\bibinfo {author} {\bibfnamefont {J.}~\bibnamefont
  {{\AA}gren}},\ }\bibfield  {title} {\bibinfo {title} {A simplified treatment
  of the transition from diffusion controlled to diffusion-less growth},\
  }\href@noop {} {\bibfield  {journal} {\bibinfo  {journal} {Acta
  Metallurgica}\ }\textbf {\bibinfo {volume} {37}},\ \bibinfo {pages} {181}
  (\bibinfo {year} {1989})}\BibitemShut {NoStop}%
\bibitem [{\citenamefont {Jackson}\ \emph {et~al.}(2004)\citenamefont
  {Jackson}, \citenamefont {Beatty},\ and\ \citenamefont
  {Gudgel}}]{jackson2004analytical}%
  \BibitemOpen
  \bibfield  {author} {\bibinfo {author} {\bibfnamefont {K.~A.}\ \bibnamefont
  {Jackson}}, \bibinfo {author} {\bibfnamefont {K.~M.}\ \bibnamefont
  {Beatty}},\ and\ \bibinfo {author} {\bibfnamefont {K.~A.}\ \bibnamefont
  {Gudgel}},\ }\bibfield  {title} {\bibinfo {title} {An analytical model for
  non-equilibrium segregation during crystallization},\ }\href@noop {}
  {\bibfield  {journal} {\bibinfo  {journal} {Journal of Crystal Growth}\
  }\textbf {\bibinfo {volume} {271}},\ \bibinfo {pages} {481} (\bibinfo {year}
  {2004})}\BibitemShut {NoStop}%
\bibitem [{\citenamefont {Baker}\ and\ \citenamefont
  {Cahn}(1971)}]{baker1971solidification}%
  \BibitemOpen
  \bibfield  {author} {\bibinfo {author} {\bibfnamefont {J.}~\bibnamefont
  {Baker}}\ and\ \bibinfo {author} {\bibfnamefont {J.}~\bibnamefont {Cahn}},\
  }\bibfield  {title} {\bibinfo {title} {Solidification, {A}merican {S}ociety
  for {M}etals},\ }\href@noop {} {\bibfield  {journal} {\bibinfo  {journal}
  {Metals Park, OH}\ }\textbf {\bibinfo {volume} {23}} (\bibinfo {year}
  {1971})}\BibitemShut {NoStop}%
\bibitem [{\citenamefont {Hillert}\ and\ \citenamefont
  {Sundman}(1977)}]{hillert1977solute}%
  \BibitemOpen
  \bibfield  {author} {\bibinfo {author} {\bibfnamefont {M.}~\bibnamefont
  {Hillert}}\ and\ \bibinfo {author} {\bibfnamefont {B.}~\bibnamefont
  {Sundman}},\ }\bibfield  {title} {\bibinfo {title} {A solute-drag treatment
  of the transition from diffusion-controlled to diffusionless
  solidification},\ }\href@noop {} {\bibfield  {journal} {\bibinfo  {journal}
  {Acta Metallurgica}\ }\textbf {\bibinfo {volume} {25}},\ \bibinfo {pages}
  {11} (\bibinfo {year} {1977})}\BibitemShut {NoStop}%
\bibitem [{\citenamefont {Baker}(1970)}]{baker1970interfacial}%
  \BibitemOpen
  \bibfield  {author} {\bibinfo {author} {\bibfnamefont {J.~C.}\ \bibnamefont
  {Baker}},\ }\emph {\bibinfo {title} {Interfacial partitioning during
  solidification}},\ \href@noop {} {Ph.D. thesis},\ \bibinfo  {school}
  {Massachusetts Institute of Technology} (\bibinfo {year} {1970})\BibitemShut
  {NoStop}%
\bibitem [{\citenamefont {Boettinger}\ \emph {et~al.}(2002)\citenamefont
  {Boettinger}, \citenamefont {Warren}, \citenamefont {Beckermann},\ and\
  \citenamefont {Karma}}]{boettinger2002phase}%
  \BibitemOpen
  \bibfield  {author} {\bibinfo {author} {\bibfnamefont {W.~J.}\ \bibnamefont
  {Boettinger}}, \bibinfo {author} {\bibfnamefont {J.~A.}\ \bibnamefont
  {Warren}}, \bibinfo {author} {\bibfnamefont {C.}~\bibnamefont {Beckermann}},\
  and\ \bibinfo {author} {\bibfnamefont {A.}~\bibnamefont {Karma}},\ }\bibfield
   {title} {\bibinfo {title} {Phase-field simulation of solidification},\
  }\href@noop {} {\bibfield  {journal} {\bibinfo  {journal} {Annual Review of
  Materials Research}\ }\textbf {\bibinfo {volume} {32}},\ \bibinfo {pages}
  {163} (\bibinfo {year} {2002})}\BibitemShut {NoStop}%
\bibitem [{\citenamefont {Tourret}\ \emph {et~al.}(2022)\citenamefont
  {Tourret}, \citenamefont {Liu},\ and\ \citenamefont
  {Llorca}}]{tourret2022phase}%
  \BibitemOpen
  \bibfield  {author} {\bibinfo {author} {\bibfnamefont {D.}~\bibnamefont
  {Tourret}}, \bibinfo {author} {\bibfnamefont {H.}~\bibnamefont {Liu}},\ and\
  \bibinfo {author} {\bibfnamefont {J.}~\bibnamefont {Llorca}},\ }\bibfield
  {title} {\bibinfo {title} {Phase-field modeling of microstructure evolution:
  Recent applications, perspectives and challenges},\ }\href@noop {} {\bibfield
   {journal} {\bibinfo  {journal} {Progress in Materials Science}\ }\textbf
  {\bibinfo {volume} {123}},\ \bibinfo {pages} {100810} (\bibinfo {year}
  {2022})}\BibitemShut {NoStop}%
\bibitem [{\citenamefont {Chen}(2002)}]{chen2002phase}%
  \BibitemOpen
  \bibfield  {author} {\bibinfo {author} {\bibfnamefont {L.-Q.}\ \bibnamefont
  {Chen}},\ }\bibfield  {title} {\bibinfo {title} {Phase-field models for
  microstructure evolution},\ }\href@noop {} {\bibfield  {journal} {\bibinfo
  {journal} {Annual Review of Materials Research}\ }\textbf {\bibinfo {volume}
  {32}},\ \bibinfo {pages} {113} (\bibinfo {year} {2002})}\BibitemShut
  {NoStop}%
\bibitem [{\citenamefont {Moelans}\ \emph {et~al.}(2008)\citenamefont
  {Moelans}, \citenamefont {Blanpain},\ and\ \citenamefont
  {Wollants}}]{moelans2008introduction}%
  \BibitemOpen
  \bibfield  {author} {\bibinfo {author} {\bibfnamefont {N.}~\bibnamefont
  {Moelans}}, \bibinfo {author} {\bibfnamefont {B.}~\bibnamefont {Blanpain}},\
  and\ \bibinfo {author} {\bibfnamefont {P.}~\bibnamefont {Wollants}},\
  }\bibfield  {title} {\bibinfo {title} {An introduction to phase-field
  modeling of microstructure evolution},\ }\href@noop {} {\bibfield  {journal}
  {\bibinfo  {journal} {Calphad}\ }\textbf {\bibinfo {volume} {32}},\ \bibinfo
  {pages} {268} (\bibinfo {year} {2008})}\BibitemShut {NoStop}%
\bibitem [{\citenamefont {Steinbach}(2013)}]{steinbach2013phase}%
  \BibitemOpen
  \bibfield  {author} {\bibinfo {author} {\bibfnamefont {I.}~\bibnamefont
  {Steinbach}},\ }\bibfield  {title} {\bibinfo {title} {Phase-field model for
  microstructure evolution at the mesoscopic scale},\ }\href@noop {} {\bibfield
   {journal} {\bibinfo  {journal} {Annual Review of Materials Research}\
  }\textbf {\bibinfo {volume} {43}},\ \bibinfo {pages} {89} (\bibinfo {year}
  {2013})}\BibitemShut {NoStop}%
\bibitem [{\citenamefont {Karma}(2001)}]{karma2001phase}%
  \BibitemOpen
  \bibfield  {author} {\bibinfo {author} {\bibfnamefont {A.}~\bibnamefont
  {Karma}},\ }\bibfield  {title} {\bibinfo {title} {Phase-field formulation for
  quantitative modeling of alloy solidification},\ }\href@noop {} {\bibfield
  {journal} {\bibinfo  {journal} {Physical Review Letters}\ }\textbf {\bibinfo
  {volume} {87}},\ \bibinfo {pages} {115701} (\bibinfo {year}
  {2001})}\BibitemShut {NoStop}%
\bibitem [{\citenamefont {Almgren}(1999)}]{almgren1999second}%
  \BibitemOpen
  \bibfield  {author} {\bibinfo {author} {\bibfnamefont {R.~F.}\ \bibnamefont
  {Almgren}},\ }\bibfield  {title} {\bibinfo {title} {Second-order phase field
  asymptotics for unequal conductivities},\ }\href@noop {} {\bibfield
  {journal} {\bibinfo  {journal} {SIAM Journal on Applied Mathematics}\
  }\textbf {\bibinfo {volume} {59}},\ \bibinfo {pages} {2086} (\bibinfo {year}
  {1999})}\BibitemShut {NoStop}%
\bibitem [{\citenamefont {Karma}\ and\ \citenamefont
  {Rappel}(1998)}]{karma1998quantitative}%
  \BibitemOpen
  \bibfield  {author} {\bibinfo {author} {\bibfnamefont {A.}~\bibnamefont
  {Karma}}\ and\ \bibinfo {author} {\bibfnamefont {W.-J.}\ \bibnamefont
  {Rappel}},\ }\bibfield  {title} {\bibinfo {title} {Quantitative phase-field
  modeling of dendritic growth in two and three dimensions},\ }\href@noop {}
  {\bibfield  {journal} {\bibinfo  {journal} {Physical Review E}\ }\textbf
  {\bibinfo {volume} {57}},\ \bibinfo {pages} {4323} (\bibinfo {year}
  {1998})}\BibitemShut {NoStop}%
\bibitem [{\citenamefont {Ohno}\ and\ \citenamefont
  {Matsuura}(2009)}]{ohno2009quantitative}%
  \BibitemOpen
  \bibfield  {author} {\bibinfo {author} {\bibfnamefont {M.}~\bibnamefont
  {Ohno}}\ and\ \bibinfo {author} {\bibfnamefont {K.}~\bibnamefont
  {Matsuura}},\ }\bibfield  {title} {\bibinfo {title} {Quantitative phase-field
  modeling for dilute alloy solidification involving diffusion in the solid},\
  }\href@noop {} {\bibfield  {journal} {\bibinfo  {journal} {Physical Review
  E}\ }\textbf {\bibinfo {volume} {79}},\ \bibinfo {pages} {031603} (\bibinfo
  {year} {2009})}\BibitemShut {NoStop}%
\bibitem [{\citenamefont {Zhang}\ \emph {et~al.}(2013)\citenamefont {Zhang},
  \citenamefont {Danilova}, \citenamefont {Steinbach}, \citenamefont
  {Medvedev},\ and\ \citenamefont {Galenko}}]{zhang2013diffuse}%
  \BibitemOpen
  \bibfield  {author} {\bibinfo {author} {\bibfnamefont {L.}~\bibnamefont
  {Zhang}}, \bibinfo {author} {\bibfnamefont {E.~V.}\ \bibnamefont {Danilova}},
  \bibinfo {author} {\bibfnamefont {I.}~\bibnamefont {Steinbach}}, \bibinfo
  {author} {\bibfnamefont {D.}~\bibnamefont {Medvedev}},\ and\ \bibinfo
  {author} {\bibfnamefont {P.~K.}\ \bibnamefont {Galenko}},\ }\bibfield
  {title} {\bibinfo {title} {Diffuse-interface modeling of solute trapping in
  rapid solidification: Predictions of the hyperbolic phase-field model and
  parabolic model with finite interface dissipation},\ }\href@noop {}
  {\bibfield  {journal} {\bibinfo  {journal} {Acta Materialia}\ }\textbf
  {\bibinfo {volume} {61}},\ \bibinfo {pages} {4155} (\bibinfo {year}
  {2013})}\BibitemShut {NoStop}%
\bibitem [{\citenamefont {Mullis}\ \emph {et~al.}(2010)\citenamefont {Mullis},
  \citenamefont {Rosam},\ and\ \citenamefont {Jimack}}]{mullis2010solute}%
  \BibitemOpen
  \bibfield  {author} {\bibinfo {author} {\bibfnamefont {A.}~\bibnamefont
  {Mullis}}, \bibinfo {author} {\bibfnamefont {J.}~\bibnamefont {Rosam}},\ and\
  \bibinfo {author} {\bibfnamefont {P.}~\bibnamefont {Jimack}},\ }\bibfield
  {title} {\bibinfo {title} {Solute trapping and the effects of anti-trapping
  currents on phase-field models of coupled thermo-solutal solidification},\
  }\href@noop {} {\bibfield  {journal} {\bibinfo  {journal} {Journal of Crystal
  Growth}\ }\textbf {\bibinfo {volume} {312}},\ \bibinfo {pages} {1891}
  (\bibinfo {year} {2010})}\BibitemShut {NoStop}%
\bibitem [{\citenamefont {Ahmad}\ \emph {et~al.}(1998)\citenamefont {Ahmad},
  \citenamefont {Wheeler}, \citenamefont {Boettinger},\ and\ \citenamefont
  {McFadden}}]{ahmad1998solute}%
  \BibitemOpen
  \bibfield  {author} {\bibinfo {author} {\bibfnamefont {N.}~\bibnamefont
  {Ahmad}}, \bibinfo {author} {\bibfnamefont {A.}~\bibnamefont {Wheeler}},
  \bibinfo {author} {\bibfnamefont {W.~J.}\ \bibnamefont {Boettinger}},\ and\
  \bibinfo {author} {\bibfnamefont {G.~B.}\ \bibnamefont {McFadden}},\
  }\bibfield  {title} {\bibinfo {title} {Solute trapping and solute drag in a
  phase-field model of rapid solidification},\ }\href@noop {} {\bibfield
  {journal} {\bibinfo  {journal} {Physical Review E}\ }\textbf {\bibinfo
  {volume} {58}},\ \bibinfo {pages} {3436} (\bibinfo {year}
  {1998})}\BibitemShut {NoStop}%
\bibitem [{\citenamefont {Wheeler}\ \emph {et~al.}(1992)\citenamefont
  {Wheeler}, \citenamefont {Boettinger},\ and\ \citenamefont
  {McFadden}}]{wheeler1992phase}%
  \BibitemOpen
  \bibfield  {author} {\bibinfo {author} {\bibfnamefont {A.~A.}\ \bibnamefont
  {Wheeler}}, \bibinfo {author} {\bibfnamefont {W.~J.}\ \bibnamefont
  {Boettinger}},\ and\ \bibinfo {author} {\bibfnamefont {G.~B.}\ \bibnamefont
  {McFadden}},\ }\bibfield  {title} {\bibinfo {title} {Phase-field model for
  isothermal phase transitions in binary alloys},\ }\href@noop {} {\bibfield
  {journal} {\bibinfo  {journal} {Physical Review A}\ }\textbf {\bibinfo
  {volume} {45}},\ \bibinfo {pages} {7424} (\bibinfo {year}
  {1992})}\BibitemShut {NoStop}%
\bibitem [{\citenamefont {Kim}\ \emph {et~al.}(1999)\citenamefont {Kim},
  \citenamefont {Kim},\ and\ \citenamefont {Suzuki}}]{kim1999phase}%
  \BibitemOpen
  \bibfield  {author} {\bibinfo {author} {\bibfnamefont {S.~G.}\ \bibnamefont
  {Kim}}, \bibinfo {author} {\bibfnamefont {W.~T.}\ \bibnamefont {Kim}},\ and\
  \bibinfo {author} {\bibfnamefont {T.}~\bibnamefont {Suzuki}},\ }\bibfield
  {title} {\bibinfo {title} {Phase-field model for binary alloys},\ }\href@noop
  {} {\bibfield  {journal} {\bibinfo  {journal} {Physical Review E}\ }\textbf
  {\bibinfo {volume} {60}},\ \bibinfo {pages} {7186} (\bibinfo {year}
  {1999})}\BibitemShut {NoStop}%
\bibitem [{\citenamefont {Danilov}\ and\ \citenamefont
  {Nestler}(2006)}]{danilov2006phase}%
  \BibitemOpen
  \bibfield  {author} {\bibinfo {author} {\bibfnamefont {D.}~\bibnamefont
  {Danilov}}\ and\ \bibinfo {author} {\bibfnamefont {B.}~\bibnamefont
  {Nestler}},\ }\bibfield  {title} {\bibinfo {title} {Phase-field modelling of
  solute trapping during rapid solidification of a {S}i--{A}s alloy},\
  }\href@noop {} {\bibfield  {journal} {\bibinfo  {journal} {Acta Materialia}\
  }\textbf {\bibinfo {volume} {54}},\ \bibinfo {pages} {4659} (\bibinfo {year}
  {2006})}\BibitemShut {NoStop}%
\bibitem [{\citenamefont {Conti}(1997)}]{conti1997solute}%
  \BibitemOpen
  \bibfield  {author} {\bibinfo {author} {\bibfnamefont {M.}~\bibnamefont
  {Conti}},\ }\bibfield  {title} {\bibinfo {title} {Solute trapping in
  directional solidification at high speed: A one-dimensional study with the
  phase-field model},\ }\href@noop {} {\bibfield  {journal} {\bibinfo
  {journal} {Physical Review E}\ }\textbf {\bibinfo {volume} {56}},\ \bibinfo
  {pages} {3717} (\bibinfo {year} {1997})}\BibitemShut {NoStop}%
\bibitem [{\citenamefont {Kim}\ and\ \citenamefont {Kim}(2001)}]{kim2001phase}%
  \BibitemOpen
  \bibfield  {author} {\bibinfo {author} {\bibfnamefont {S.~G.}\ \bibnamefont
  {Kim}}\ and\ \bibinfo {author} {\bibfnamefont {W.~T.}\ \bibnamefont {Kim}},\
  }\bibfield  {title} {\bibinfo {title} {Phase-field modeling of rapid
  solidification},\ }\href@noop {} {\bibfield  {journal} {\bibinfo  {journal}
  {Materials Science and Engineering: A}\ }\textbf {\bibinfo {volume} {304}},\
  \bibinfo {pages} {281} (\bibinfo {year} {2001})}\BibitemShut {NoStop}%
\bibitem [{\citenamefont {Glasner}(2001)}]{glasner2001solute}%
  \BibitemOpen
  \bibfield  {author} {\bibinfo {author} {\bibfnamefont {K.}~\bibnamefont
  {Glasner}},\ }\bibfield  {title} {\bibinfo {title} {Solute trapping and the
  non-equilibrium phase diagram for solidification of binary alloys},\
  }\href@noop {} {\bibfield  {journal} {\bibinfo  {journal} {Physica D:
  Nonlinear Phenomena}\ }\textbf {\bibinfo {volume} {151}},\ \bibinfo {pages}
  {253} (\bibinfo {year} {2001})}\BibitemShut {NoStop}%
\bibitem [{\citenamefont {Wheeler}\ \emph {et~al.}(1993)\citenamefont
  {Wheeler}, \citenamefont {Boettinger},\ and\ \citenamefont
  {McFadden}}]{wheeler1993phase}%
  \BibitemOpen
  \bibfield  {author} {\bibinfo {author} {\bibfnamefont {A.~A.}\ \bibnamefont
  {Wheeler}}, \bibinfo {author} {\bibfnamefont {W.~J.}\ \bibnamefont
  {Boettinger}},\ and\ \bibinfo {author} {\bibfnamefont {G.~B.}\ \bibnamefont
  {McFadden}},\ }\bibfield  {title} {\bibinfo {title} {Phase-field model of
  solute trapping during solidification},\ }\href@noop {} {\bibfield  {journal}
  {\bibinfo  {journal} {Physical Review E}\ }\textbf {\bibinfo {volume} {47}},\
  \bibinfo {pages} {1893} (\bibinfo {year} {1993})}\BibitemShut {NoStop}%
\bibitem [{\citenamefont {Boettinger}\ and\ \citenamefont
  {Warren}(1999)}]{boettinger1999simulation}%
  \BibitemOpen
  \bibfield  {author} {\bibinfo {author} {\bibfnamefont {W.~J.}\ \bibnamefont
  {Boettinger}}\ and\ \bibinfo {author} {\bibfnamefont {J.~A.}\ \bibnamefont
  {Warren}},\ }\bibfield  {title} {\bibinfo {title} {Simulation of the cell to
  plane front transition during directional solidification at high velocity},\
  }\href@noop {} {\bibfield  {journal} {\bibinfo  {journal} {Journal of Crystal
  Growth}\ }\textbf {\bibinfo {volume} {200}},\ \bibinfo {pages} {583}
  (\bibinfo {year} {1999})}\BibitemShut {NoStop}%
\bibitem [{\citenamefont {Steinbach}\ \emph {et~al.}(2012)\citenamefont
  {Steinbach}, \citenamefont {Zhang},\ and\ \citenamefont
  {Plapp}}]{steinbach2012phase}%
  \BibitemOpen
  \bibfield  {author} {\bibinfo {author} {\bibfnamefont {I.}~\bibnamefont
  {Steinbach}}, \bibinfo {author} {\bibfnamefont {L.}~\bibnamefont {Zhang}},\
  and\ \bibinfo {author} {\bibfnamefont {M.}~\bibnamefont {Plapp}},\ }\bibfield
   {title} {\bibinfo {title} {Phase-field model with finite interface
  dissipation},\ }\href@noop {} {\bibfield  {journal} {\bibinfo  {journal}
  {Acta Materialia}\ }\textbf {\bibinfo {volume} {60}},\ \bibinfo {pages}
  {2689} (\bibinfo {year} {2012})}\BibitemShut {NoStop}%
\bibitem [{\citenamefont {Pinomaa}\ and\ \citenamefont
  {Provatas}(2019)}]{pinomaa2019quantitative}%
  \BibitemOpen
  \bibfield  {author} {\bibinfo {author} {\bibfnamefont {T.}~\bibnamefont
  {Pinomaa}}\ and\ \bibinfo {author} {\bibfnamefont {N.}~\bibnamefont
  {Provatas}},\ }\bibfield  {title} {\bibinfo {title} {Quantitative phase field
  modeling of solute trapping and continuous growth kinetics in quasi-rapid
  solidification},\ }\href@noop {} {\bibfield  {journal} {\bibinfo  {journal}
  {Acta Materialia}\ }\textbf {\bibinfo {volume} {168}},\ \bibinfo {pages}
  {167} (\bibinfo {year} {2019})}\BibitemShut {NoStop}%
\bibitem [{\citenamefont {Echebarria}\ \emph {et~al.}(2004)\citenamefont
  {Echebarria}, \citenamefont {Folch}, \citenamefont {Karma},\ and\
  \citenamefont {Plapp}}]{echebarria2004quantitative}%
  \BibitemOpen
  \bibfield  {author} {\bibinfo {author} {\bibfnamefont {B.}~\bibnamefont
  {Echebarria}}, \bibinfo {author} {\bibfnamefont {R.}~\bibnamefont {Folch}},
  \bibinfo {author} {\bibfnamefont {A.}~\bibnamefont {Karma}},\ and\ \bibinfo
  {author} {\bibfnamefont {M.}~\bibnamefont {Plapp}},\ }\bibfield  {title}
  {\bibinfo {title} {Quantitative phase-field model of alloy solidification},\
  }\href@noop {} {\bibfield  {journal} {\bibinfo  {journal} {Physical Review
  E}\ }\textbf {\bibinfo {volume} {70}},\ \bibinfo {pages} {061604} (\bibinfo
  {year} {2004})}\BibitemShut {NoStop}%
\bibitem [{\citenamefont {Ji}\ \emph {et~al.}(2022)\citenamefont {Ji},
  \citenamefont {Dorari}, \citenamefont {Clarke},\ and\ \citenamefont
  {Karma}}]{ji2022microstructural}%
  \BibitemOpen
  \bibfield  {author} {\bibinfo {author} {\bibfnamefont {K.}~\bibnamefont
  {Ji}}, \bibinfo {author} {\bibfnamefont {E.}~\bibnamefont {Dorari}}, \bibinfo
  {author} {\bibfnamefont {A.~J.}\ \bibnamefont {Clarke}},\ and\ \bibinfo
  {author} {\bibfnamefont {A.}~\bibnamefont {Karma}},\ }\bibfield  {title}
  {\bibinfo {title} {Microstructural pattern formation during
  far-from-equilibrium alloy solidification},\ }\href@noop {} {\bibfield
  {journal} {\bibinfo  {journal} {arXiv preprint arXiv:2209.11352}\ } (\bibinfo
  {year} {2022})}\BibitemShut {NoStop}%
\bibitem [{\citenamefont {Brener}\ and\ \citenamefont
  {Boussinot}(2012)}]{brener2012kinetic}%
  \BibitemOpen
  \bibfield  {author} {\bibinfo {author} {\bibfnamefont {E.~A.}\ \bibnamefont
  {Brener}}\ and\ \bibinfo {author} {\bibfnamefont {G.}~\bibnamefont
  {Boussinot}},\ }\bibfield  {title} {\bibinfo {title} {Kinetic cross coupling
  between nonconserved and conserved fields in phase field models},\
  }\href@noop {} {\bibfield  {journal} {\bibinfo  {journal} {Physical Review
  E}\ }\textbf {\bibinfo {volume} {86}},\ \bibinfo {pages} {060601} (\bibinfo
  {year} {2012})}\BibitemShut {NoStop}%
\bibitem [{\citenamefont {Boussinot}\ and\ \citenamefont
  {Brener}(2014)}]{boussinot2014achieving}%
  \BibitemOpen
  \bibfield  {author} {\bibinfo {author} {\bibfnamefont {G.}~\bibnamefont
  {Boussinot}}\ and\ \bibinfo {author} {\bibfnamefont {E.~A.}\ \bibnamefont
  {Brener}},\ }\bibfield  {title} {\bibinfo {title} {Achieving realistic
  interface kinetics in phase-field models with a diffusional contrast},\
  }\href@noop {} {\bibfield  {journal} {\bibinfo  {journal} {Physical Review
  E}\ }\textbf {\bibinfo {volume} {89}},\ \bibinfo {pages} {060402} (\bibinfo
  {year} {2014})}\BibitemShut {NoStop}%
\bibitem [{\citenamefont {Wang}\ \emph {et~al.}(2013)\citenamefont {Wang},
  \citenamefont {Liu}, \citenamefont {Ehlen},\ and\ \citenamefont
  {Herlach}}]{wang2013application}%
  \BibitemOpen
  \bibfield  {author} {\bibinfo {author} {\bibfnamefont {H.}~\bibnamefont
  {Wang}}, \bibinfo {author} {\bibfnamefont {F.}~\bibnamefont {Liu}}, \bibinfo
  {author} {\bibfnamefont {G.}~\bibnamefont {Ehlen}},\ and\ \bibinfo {author}
  {\bibfnamefont {D.}~\bibnamefont {Herlach}},\ }\bibfield  {title} {\bibinfo
  {title} {Application of the maximal entropy production principle to rapid
  solidification: a multi-phase-field model},\ }\href@noop {} {\bibfield
  {journal} {\bibinfo  {journal} {Acta Materialia}\ }\textbf {\bibinfo {volume}
  {61}},\ \bibinfo {pages} {2617} (\bibinfo {year} {2013})}\BibitemShut
  {NoStop}%
\bibitem [{\citenamefont {Fang}\ and\ \citenamefont
  {Mi}(2013)}]{fang2013recovering}%
  \BibitemOpen
  \bibfield  {author} {\bibinfo {author} {\bibfnamefont {A.}~\bibnamefont
  {Fang}}\ and\ \bibinfo {author} {\bibfnamefont {Y.}~\bibnamefont {Mi}},\
  }\bibfield  {title} {\bibinfo {title} {Recovering thermodynamic consistency
  of the antitrapping model: A variational phase-field formulation for alloy
  solidification},\ }\href@noop {} {\bibfield  {journal} {\bibinfo  {journal}
  {Physical Review E}\ }\textbf {\bibinfo {volume} {87}},\ \bibinfo {pages}
  {012402} (\bibinfo {year} {2013})}\BibitemShut {NoStop}%
\bibitem [{\citenamefont {Ohno}\ \emph {et~al.}(2016)\citenamefont {Ohno},
  \citenamefont {Takaki},\ and\ \citenamefont {Shibuta}}]{ohno2016variational}%
  \BibitemOpen
  \bibfield  {author} {\bibinfo {author} {\bibfnamefont {M.}~\bibnamefont
  {Ohno}}, \bibinfo {author} {\bibfnamefont {T.}~\bibnamefont {Takaki}},\ and\
  \bibinfo {author} {\bibfnamefont {Y.}~\bibnamefont {Shibuta}},\ }\bibfield
  {title} {\bibinfo {title} {Variational formulation and numerical accuracy of
  a quantitative phase-field model for binary alloy solidification with
  two-sided diffusion},\ }\href@noop {} {\bibfield  {journal} {\bibinfo
  {journal} {Physical Review E}\ }\textbf {\bibinfo {volume} {93}},\ \bibinfo
  {pages} {012802} (\bibinfo {year} {2016})}\BibitemShut {NoStop}%
\bibitem [{\citenamefont {Ohno}\ \emph {et~al.}(2017)\citenamefont {Ohno},
  \citenamefont {Takaki},\ and\ \citenamefont {Shibuta}}]{ohno2017variational}%
  \BibitemOpen
  \bibfield  {author} {\bibinfo {author} {\bibfnamefont {M.}~\bibnamefont
  {Ohno}}, \bibinfo {author} {\bibfnamefont {T.}~\bibnamefont {Takaki}},\ and\
  \bibinfo {author} {\bibfnamefont {Y.}~\bibnamefont {Shibuta}},\ }\bibfield
  {title} {\bibinfo {title} {Variational formulation of a quantitative
  phase-field model for nonisothermal solidification in a multicomponent
  alloy},\ }\href@noop {} {\bibfield  {journal} {\bibinfo  {journal} {Physical
  Review E}\ }\textbf {\bibinfo {volume} {96}},\ \bibinfo {pages} {033311}
  (\bibinfo {year} {2017})}\BibitemShut {NoStop}%
\bibitem [{\citenamefont {Wang}\ \emph {et~al.}(2020)\citenamefont {Wang},
  \citenamefont {Boussinot}, \citenamefont {H{\"u}ter}, \citenamefont
  {Brener},\ and\ \citenamefont {Spatschek}}]{wang2020modeling}%
  \BibitemOpen
  \bibfield  {author} {\bibinfo {author} {\bibfnamefont {K.}~\bibnamefont
  {Wang}}, \bibinfo {author} {\bibfnamefont {G.}~\bibnamefont {Boussinot}},
  \bibinfo {author} {\bibfnamefont {C.}~\bibnamefont {H{\"u}ter}}, \bibinfo
  {author} {\bibfnamefont {E.~A.}\ \bibnamefont {Brener}},\ and\ \bibinfo
  {author} {\bibfnamefont {R.}~\bibnamefont {Spatschek}},\ }\bibfield  {title}
  {\bibinfo {title} {Modeling of dendritic growth using a quantitative
  nondiagonal phase field model},\ }\href@noop {} {\bibfield  {journal}
  {\bibinfo  {journal} {Physical Review Materials}\ }\textbf {\bibinfo {volume}
  {4}},\ \bibinfo {pages} {033802} (\bibinfo {year} {2020})}\BibitemShut
  {NoStop}%
\bibitem [{\citenamefont {Wang}\ \emph {et~al.}(2021)\citenamefont {Wang},
  \citenamefont {Boussinot}, \citenamefont {Brener},\ and\ \citenamefont
  {Spatschek}}]{wang2021quantitative}%
  \BibitemOpen
  \bibfield  {author} {\bibinfo {author} {\bibfnamefont {K.}~\bibnamefont
  {Wang}}, \bibinfo {author} {\bibfnamefont {G.}~\bibnamefont {Boussinot}},
  \bibinfo {author} {\bibfnamefont {E.~A.}\ \bibnamefont {Brener}},\ and\
  \bibinfo {author} {\bibfnamefont {R.}~\bibnamefont {Spatschek}},\ }\bibfield
  {title} {\bibinfo {title} {Quantitative nondiagonal phase field modeling of
  eutectic and eutectoid transformations},\ }\href@noop {} {\bibfield
  {journal} {\bibinfo  {journal} {Physical Review B}\ }\textbf {\bibinfo
  {volume} {103}},\ \bibinfo {pages} {184111} (\bibinfo {year}
  {2021})}\BibitemShut {NoStop}%
\bibitem [{\citenamefont {Nicoli}\ \emph {et~al.}(2011)\citenamefont {Nicoli},
  \citenamefont {Plapp},\ and\ \citenamefont {Henry}}]{nicoli2011tensorial}%
  \BibitemOpen
  \bibfield  {author} {\bibinfo {author} {\bibfnamefont {M.}~\bibnamefont
  {Nicoli}}, \bibinfo {author} {\bibfnamefont {M.}~\bibnamefont {Plapp}},\ and\
  \bibinfo {author} {\bibfnamefont {H.}~\bibnamefont {Henry}},\ }\bibfield
  {title} {\bibinfo {title} {Tensorial mobilities for accurate solution of
  transport problems in models with diffuse interfaces},\ }\href@noop {}
  {\bibfield  {journal} {\bibinfo  {journal} {Physical Review E}\ }\textbf
  {\bibinfo {volume} {84}},\ \bibinfo {pages} {046707} (\bibinfo {year}
  {2011})}\BibitemShut {NoStop}%
\bibitem [{\citenamefont {Gugenberger}\ \emph {et~al.}(2008)\citenamefont
  {Gugenberger}, \citenamefont {Spatschek},\ and\ \citenamefont
  {Kassner}}]{gugenberger2008comparison}%
  \BibitemOpen
  \bibfield  {author} {\bibinfo {author} {\bibfnamefont {C.}~\bibnamefont
  {Gugenberger}}, \bibinfo {author} {\bibfnamefont {R.}~\bibnamefont
  {Spatschek}},\ and\ \bibinfo {author} {\bibfnamefont {K.}~\bibnamefont
  {Kassner}},\ }\bibfield  {title} {\bibinfo {title} {Comparison of phase-field
  models for surface diffusion},\ }\href@noop {} {\bibfield  {journal}
  {\bibinfo  {journal} {Physical Review E}\ }\textbf {\bibinfo {volume} {78}},\
  \bibinfo {pages} {016703} (\bibinfo {year} {2008})}\BibitemShut {NoStop}%
\bibitem [{\citenamefont {Ahmed}\ \emph {et~al.}(2013)\citenamefont {Ahmed},
  \citenamefont {Yablinsky}, \citenamefont {Schulte}, \citenamefont {Allen},\
  and\ \citenamefont {El-Azab}}]{ahmed2013phase}%
  \BibitemOpen
  \bibfield  {author} {\bibinfo {author} {\bibfnamefont {K.}~\bibnamefont
  {Ahmed}}, \bibinfo {author} {\bibfnamefont {C.}~\bibnamefont {Yablinsky}},
  \bibinfo {author} {\bibfnamefont {A.}~\bibnamefont {Schulte}}, \bibinfo
  {author} {\bibfnamefont {T.}~\bibnamefont {Allen}},\ and\ \bibinfo {author}
  {\bibfnamefont {A.}~\bibnamefont {El-Azab}},\ }\bibfield  {title} {\bibinfo
  {title} {Phase field modeling of the effect of porosity on grain growth
  kinetics in polycrystalline ceramics},\ }\href@noop {} {\bibfield  {journal}
  {\bibinfo  {journal} {Modelling and Simulation in Materials Science and
  Engineering}\ }\textbf {\bibinfo {volume} {21}},\ \bibinfo {pages} {065005}
  (\bibinfo {year} {2013})}\BibitemShut {NoStop}%
\bibitem [{\citenamefont {Provatas}\ and\ \citenamefont
  {Elder}(2011)}]{provatas2011phase}%
  \BibitemOpen
  \bibfield  {author} {\bibinfo {author} {\bibfnamefont {N.}~\bibnamefont
  {Provatas}}\ and\ \bibinfo {author} {\bibfnamefont {K.}~\bibnamefont
  {Elder}},\ }\href@noop {} {\emph {\bibinfo {title} {Phase-field methods in
  materials science and engineering}}}\ (\bibinfo  {publisher} {John Wiley \&
  Sons},\ \bibinfo {year} {2011})\BibitemShut {NoStop}%
\bibitem [{\citenamefont {Algoso}\ \emph {et~al.}(2003)\citenamefont {Algoso},
  \citenamefont {Hofmeister},\ and\ \citenamefont
  {Bayuzick}}]{algoso2003solidification}%
  \BibitemOpen
  \bibfield  {author} {\bibinfo {author} {\bibfnamefont {P.}~\bibnamefont
  {Algoso}}, \bibinfo {author} {\bibfnamefont {W.}~\bibnamefont {Hofmeister}},\
  and\ \bibinfo {author} {\bibfnamefont {R.}~\bibnamefont {Bayuzick}},\
  }\bibfield  {title} {\bibinfo {title} {Solidification velocity of undercooled
  {N}i--{C}u alloys},\ }\href@noop {} {\bibfield  {journal} {\bibinfo
  {journal} {Acta Materialia}\ }\textbf {\bibinfo {volume} {51}},\ \bibinfo
  {pages} {4307} (\bibinfo {year} {2003})}\BibitemShut {NoStop}%
\bibitem [{\citenamefont {Kittl}\ \emph {et~al.}(2000)\citenamefont {Kittl},
  \citenamefont {Sanders}, \citenamefont {Aziz}, \citenamefont {Brunco},\ and\
  \citenamefont {Thompson}}]{kittl2000complete}%
  \BibitemOpen
  \bibfield  {author} {\bibinfo {author} {\bibfnamefont {J.}~\bibnamefont
  {Kittl}}, \bibinfo {author} {\bibfnamefont {P.}~\bibnamefont {Sanders}},
  \bibinfo {author} {\bibfnamefont {M.}~\bibnamefont {Aziz}}, \bibinfo {author}
  {\bibfnamefont {D.}~\bibnamefont {Brunco}},\ and\ \bibinfo {author}
  {\bibfnamefont {M.}~\bibnamefont {Thompson}},\ }\bibfield  {title} {\bibinfo
  {title} {Complete experimental test of kinetic models for rapid alloy
  solidification},\ }\href@noop {} {\bibfield  {journal} {\bibinfo  {journal}
  {Acta Materialia}\ }\textbf {\bibinfo {volume} {48}},\ \bibinfo {pages}
  {4797} (\bibinfo {year} {2000})}\BibitemShut {NoStop}%
\bibitem [{\citenamefont {Yang}\ \emph {et~al.}(2011)\citenamefont {Yang},
  \citenamefont {Humadi}, \citenamefont {Buta}, \citenamefont {Laird},
  \citenamefont {Sun}, \citenamefont {Hoyt},\ and\ \citenamefont
  {Asta}}]{yang2011atomistic}%
  \BibitemOpen
  \bibfield  {author} {\bibinfo {author} {\bibfnamefont {Y.}~\bibnamefont
  {Yang}}, \bibinfo {author} {\bibfnamefont {H.}~\bibnamefont {Humadi}},
  \bibinfo {author} {\bibfnamefont {D.}~\bibnamefont {Buta}}, \bibinfo {author}
  {\bibfnamefont {B.~B.}\ \bibnamefont {Laird}}, \bibinfo {author}
  {\bibfnamefont {D.}~\bibnamefont {Sun}}, \bibinfo {author} {\bibfnamefont
  {J.~J.}\ \bibnamefont {Hoyt}},\ and\ \bibinfo {author} {\bibfnamefont
  {M.}~\bibnamefont {Asta}},\ }\bibfield  {title} {\bibinfo {title} {Atomistic
  simulations of nonequilibrium crystal-growth kinetics from alloy melts},\
  }\href@noop {} {\bibfield  {journal} {\bibinfo  {journal} {Physical Review
  Letters}\ }\textbf {\bibinfo {volume} {107}},\ \bibinfo {pages} {025505}
  (\bibinfo {year} {2011})}\BibitemShut {NoStop}%
\bibitem [{\citenamefont {Hareland}\ \emph {et~al.}(2022)\citenamefont
  {Hareland}, \citenamefont {Guillemot}, \citenamefont {Gandin},\ and\
  \citenamefont {Voorhees}}]{hareland2022thermodynamics}%
  \BibitemOpen
  \bibfield  {author} {\bibinfo {author} {\bibfnamefont {C.~A.}\ \bibnamefont
  {Hareland}}, \bibinfo {author} {\bibfnamefont {G.}~\bibnamefont {Guillemot}},
  \bibinfo {author} {\bibfnamefont {C.-A.}\ \bibnamefont {Gandin}},\ and\
  \bibinfo {author} {\bibfnamefont {P.~W.}\ \bibnamefont {Voorhees}},\
  }\bibfield  {title} {\bibinfo {title} {The thermodynamics of non-equilibrium
  interfaces during phase transformations in concentrated multicomponent
  alloys},\ }\href@noop {} {\bibfield  {journal} {\bibinfo  {journal} {Acta
  Materialia}\ ,\ \bibinfo {pages} {118407}} (\bibinfo {year}
  {2022})}\BibitemShut {NoStop}%
\bibitem [{\citenamefont {Boettinger}\ \emph {et~al.}(1984)\citenamefont
  {Boettinger}, \citenamefont {Shechtman}, \citenamefont {Schaefer},\ and\
  \citenamefont {Biancaniello}}]{boettinger1984effect}%
  \BibitemOpen
  \bibfield  {author} {\bibinfo {author} {\bibfnamefont {W.}~\bibnamefont
  {Boettinger}}, \bibinfo {author} {\bibfnamefont {D.}~\bibnamefont
  {Shechtman}}, \bibinfo {author} {\bibfnamefont {R.}~\bibnamefont
  {Schaefer}},\ and\ \bibinfo {author} {\bibfnamefont {F.}~\bibnamefont
  {Biancaniello}},\ }\bibfield  {title} {\bibinfo {title} {The effect of rapid
  solidification velocity on the microstructure of {A}g-{C}u alloys},\
  }\href@noop {} {\bibfield  {journal} {\bibinfo  {journal} {Metallurgical
  Transactions A}\ }\textbf {\bibinfo {volume} {15}},\ \bibinfo {pages} {55}
  (\bibinfo {year} {1984})}\BibitemShut {NoStop}%
\bibitem [{\citenamefont {Kurz}\ and\ \citenamefont
  {Trivedi}(1996)}]{kurz1996banded}%
  \BibitemOpen
  \bibfield  {author} {\bibinfo {author} {\bibfnamefont {W.}~\bibnamefont
  {Kurz}}\ and\ \bibinfo {author} {\bibfnamefont {R.}~\bibnamefont {Trivedi}},\
  }\bibfield  {title} {\bibinfo {title} {Banded solidification
  microstructures},\ }\href@noop {} {\bibfield  {journal} {\bibinfo  {journal}
  {Metallurgical and Materials Transactions A}\ }\textbf {\bibinfo {volume}
  {27}},\ \bibinfo {pages} {625} (\bibinfo {year} {1996})}\BibitemShut
  {NoStop}%
\bibitem [{\citenamefont {Mullins}\ and\ \citenamefont
  {Sekerka}(1964)}]{mullins1964stability}%
  \BibitemOpen
  \bibfield  {author} {\bibinfo {author} {\bibfnamefont {W.~W.}\ \bibnamefont
  {Mullins}}\ and\ \bibinfo {author} {\bibfnamefont {R.}~\bibnamefont
  {Sekerka}},\ }\bibfield  {title} {\bibinfo {title} {Stability of a planar
  interface during solidification of a dilute binary alloy},\ }\href@noop {}
  {\bibfield  {journal} {\bibinfo  {journal} {Journal of Applied Physics}\
  }\textbf {\bibinfo {volume} {35}},\ \bibinfo {pages} {444} (\bibinfo {year}
  {1964})}\BibitemShut {NoStop}%
\bibitem [{\citenamefont {Coriell}\ and\ \citenamefont
  {Sekerka}(1983)}]{coriell1983oscillatory}%
  \BibitemOpen
  \bibfield  {author} {\bibinfo {author} {\bibfnamefont {S.}~\bibnamefont
  {Coriell}}\ and\ \bibinfo {author} {\bibfnamefont {R.}~\bibnamefont
  {Sekerka}},\ }\bibfield  {title} {\bibinfo {title} {Oscillatory morphological
  instabilities due to non-equilibrium segregation},\ }\href@noop {} {\bibfield
   {journal} {\bibinfo  {journal} {Journal of Crystal Growth}\ }\textbf
  {\bibinfo {volume} {61}},\ \bibinfo {pages} {499} (\bibinfo {year}
  {1983})}\BibitemShut {NoStop}%
\bibitem [{\citenamefont {Karma}\ and\ \citenamefont
  {Sarkissian}(1993)}]{karma1993interface}%
  \BibitemOpen
  \bibfield  {author} {\bibinfo {author} {\bibfnamefont {A.}~\bibnamefont
  {Karma}}\ and\ \bibinfo {author} {\bibfnamefont {A.}~\bibnamefont
  {Sarkissian}},\ }\bibfield  {title} {\bibinfo {title} {Interface dynamics and
  banding in rapid solidification},\ }\href@noop {} {\bibfield  {journal}
  {\bibinfo  {journal} {Physical Review E}\ }\textbf {\bibinfo {volume} {47}},\
  \bibinfo {pages} {513} (\bibinfo {year} {1993})}\BibitemShut {NoStop}%
\end{thebibliography}%

\end{document}

%\preprint{}

\title{Supplementary Material to the article ``A quantitative variational phase field framework"}

\author{Arnab Mukherjee$^{1,2}$} \email[]{arnab.mukherjee@northwestern.edu}
\author{James A. Warren$^2$}
\author{Peter W. Voorhees$^{1,3,4}$}

%\thanks{}
\affiliation{$^1$Center for Hierarchical Materials Design, Northwestern University, Evanston, Illinois 60208, USA}
\affiliation{$^2$Material Measurement Laboratory, National Institute of Standards and Technology, Gaithersburg, Maryland 20899, USA}
\affiliation{$^3$Department of Materials Science and Engineering, Northwestern University, Evanston, Illinois 60208, USA}
\affiliation{$^4$Department of Engineering Sciences and Applied Mathematics, Northwestern University, Evanston, Illinois 60208, USA}


%\thanks{}


\date{\today}

\begin{abstract}
In the supplementary material, we study the equilibrium properties of the phase-field model and derive expression for the interfacial energy. The non-dimesionalization procedure and the complete thin-interface asymptotic analysis (until second order) for the model with the assumption $\mu_s = \mu_l$ are derived. The asymptotic analysis for the model $\mu_s \neq \mu_l$ until first order is also presented.
\end{abstract}

%\keywords{}

\maketitle

\section{Equilibrium properties}
\subsection{Interfacial profile}
The free energy of the system is given by
\begin{eqnarray}\label{eq1}
    \mathcal{\overline{F}} = \int \Big[Hf_{\mathrm{dw}}(\phi) + \frac{\sigma}{2}|\nabla\phi|^2 + f_s(c_s)h(\phi) + f_l(c_l)\{1-h(\phi)\}\Big] \mathrm{d}V  \nonumber \\
    + \int \lambda \Big[c - \{ c_s h(\phi) + c_l(1-h(\phi))\}\Big] \mathrm{d}V 
    + \int \lambda^{\prime}(c-c_o) \mathrm{d} V
\end{eqnarray}
where, the last integral takes into account the mass conservation with $c_o$ being the alloy composition. At equilibrium, the individual variation with respect to each variable will be equal to zero which implies
\begin{equation}\label{eq2}
    \frac{\delta \mathcal{\overline{F}}}{\delta c}  =  \lambda - \lambda^{\prime} = 0 ,
\end{equation}
\begin{equation}\label{eq3}
   \Rightarrow \lambda  =  \lambda^{\prime} .
\end{equation}
\begin{equation}\label{eq4}
    \frac{\delta \mathcal{\overline{F}}}{\delta c_s}  =  \frac{\partial f_s(c_s)}{\partial c_s}h(\phi_o) - \lambda h(\phi_o) = 0 ,
\end{equation}
\begin{equation}\label{eq5}
   \Rightarrow \frac{\partial f_s(c_s)}{\partial c_s}  =  \lambda .
\end{equation}
\begin{equation}\label{eq6}
    \frac{\delta \mathcal{\overline{F}}}{\delta c_l} = \frac{\partial f_l(c_l)}{\partial c_l}\{1-h(\phi_o)\} - \lambda \{1-h(\phi_o)\} = 0 ,
\end{equation}
\begin{equation}\label{eq7}
    \Rightarrow \frac{\partial f_l(c_l)}{\partial c_l} = \lambda .
\end{equation}
where $\phi_o$ denotes the equilibrium phase field profile and Eqs.~~(\ref{eq5}) and (\ref{eq7}) imply 
\begin{equation}\label{eq8}
    c_s(x) = c_s^e
\end{equation}
\begin{equation}\label{eq9}
    c_l(x) = c_l^e
\end{equation}
where $c_s^e$ and $c_l^e$ are the equilibrium solid and liquid concentrations respectively. Moving on to the variation with respect to $\phi$ we have
\begin{equation}\label{eq10}
    \frac{\delta \mathcal{\overline{F}}}{\delta \phi} = Hf_{\mathrm{dw}}^{\prime}(\phi_o) - \sigma \frac{\mathrm{d}^2\phi_o}{\mathrm{d}x^2} + \{f_s(c_s^e) - f_l(c_l^e) - \lambda(c_s^e - c_l^e)\}h^{\prime}(\phi_o) = 0 .
\end{equation}
Multiplying both sides by $\frac{\mathrm{d}\phi_o}{\mathrm{d}x}$ and integrating within the limits $x = -\infty$ and $x = +\infty$,
\begin{eqnarray}\label{eq11}
    \int_{-\infty}^{+\infty}Hf_{\mathrm{dw}}^{\prime}(\phi_o)\frac{\mathrm{d}\phi_o}{\mathrm{d}x}\mathrm{d}x + \int_{-\infty}^{+\infty}\sigma \frac{\mathrm{d}^2\phi_o}{\mathrm{d}x^2}\frac{\mathrm{d}\phi_o}{\mathrm{d}x}\mathrm{d}x \nonumber \\
    +\{f_s(c_s^e) - f_l(c_l^e) - \lambda(c_s^e-c_l^e)\}\int_{-\infty}^{+\infty} h^{\prime}(\phi_o)\frac{\mathrm{d}\phi_o}{\mathrm{d}x}\mathrm{d}x = 0
\end{eqnarray}
Simplifying and expressing in terms of perfect integrals we obtain
\begin{eqnarray}\label{eq12}
    \int_{-\infty}^{+\infty} H \frac{\mathrm{d}f_{\mathrm{dw}}(\phi_o)}{\mathrm{d}x}\mathrm{d}x + \int_{-\infty}^{+\infty}\frac{\sigma}{2}\frac{\mathrm{d}}{\mathrm{d}x}\Big(\frac{\mathrm{d}\phi_o}{\mathrm{d}x}\Big)^2\mathrm{d}x \nonumber \\
    +\{f_s(c_s^e) - f_l(c_l^e) - \lambda(c_s^e-c_l^e)\}\int_{-\infty}^{+\infty}\frac{\mathrm{d}h(\phi_o)}{\mathrm{d}x}\mathrm{d}x = 0.
\end{eqnarray}
Integrating we obtain
\begin{eqnarray}\label{eq13}
    H f_{\mathrm{dw}}(\phi_o)\Big|_{-\infty}^{+\infty} + \frac{\sigma}{2}\Big(\frac{\mathrm{d}\phi_o}{\mathrm{d}x}\Big)^2\Big|_{-\infty}^{+\infty}  +\{f_s(c_s^e) - f_l(c_l^e) - \lambda(c_s^e-c_l^e)\} h(\phi_o)\Big|_{-\infty}^{+\infty} = 0 .
\end{eqnarray}
Using the boundary condition $\phi = 1$ and $\frac{\mathrm{d}\phi_o}{\mathrm{d}x} = 0$  at $x = -\infty$ and $\phi = 0$ and $\frac{\mathrm{d}\phi_o}{\mathrm{d}x} = 0$ at $x = +\infty$ we have
\begin{equation}\label{eq14}
    \lambda = \frac{f_s(c_s^e) - f_l(c_l^e)}{c_s^e - c_l^e}.
\end{equation}
Therefore, $c_s^e$ and $c_l^e$ can be found from the common tangent construction. Substituting the value of $\lambda$ in the equilibrium condition of phase-field we have
\begin{equation}\label{eq15}
    H f_{\mathrm{dw}}^{\prime}(\phi_o) - \sigma \frac{\mathrm{d}^2\phi_o}{\mathrm{d}x^2} = 0
\end{equation}
Introducing the notation $W^2 = \sigma /H$ which is the measure of the interfacial width we have
\begin{equation}\label{eq16}
    f_{\mathrm{dw}}^{\prime}(\phi_o) - W^2\frac{\mathrm{d}^2\phi_o}{\mathrm{d}x^2} = 0.
\end{equation}
Multiplying both sides by $\frac{\mathrm{d}\phi_o}{\mathrm{d}x}$ and integrating we have
\begin{equation}\label{eq17}
    \frac{\mathrm{d}\phi_o}{\mathrm{d}x} = \frac{\sqrt{2}}{W}\sqrt{f_{\mathrm{dw}}(\phi_o)},
\end{equation}
where the integration constant is deduced to be zero. Integrating again we obtain the equilibrium phase-field profile as
\begin{equation}\label{eq18}
    \phi_o(x) = \frac{1}{2}\Bigg[ 1 - \tanh \left(\frac{x}{\sqrt{2}W}\right)\Bigg],
\end{equation}
where we have employed the relation $\phi_o = 1/2$ at $x = 0$ to remove the translation degree of freedom of the profile. The equilibrium concentration profile writes as
\begin{equation}\label{eq19}
    c_o(x) = c_s^e h(\phi_o(x)) + c_l^e\{1- h(\phi_o(x)) \}
\end{equation}

\begin{figure}
    \centering
    \includegraphics[scale = 0.28]{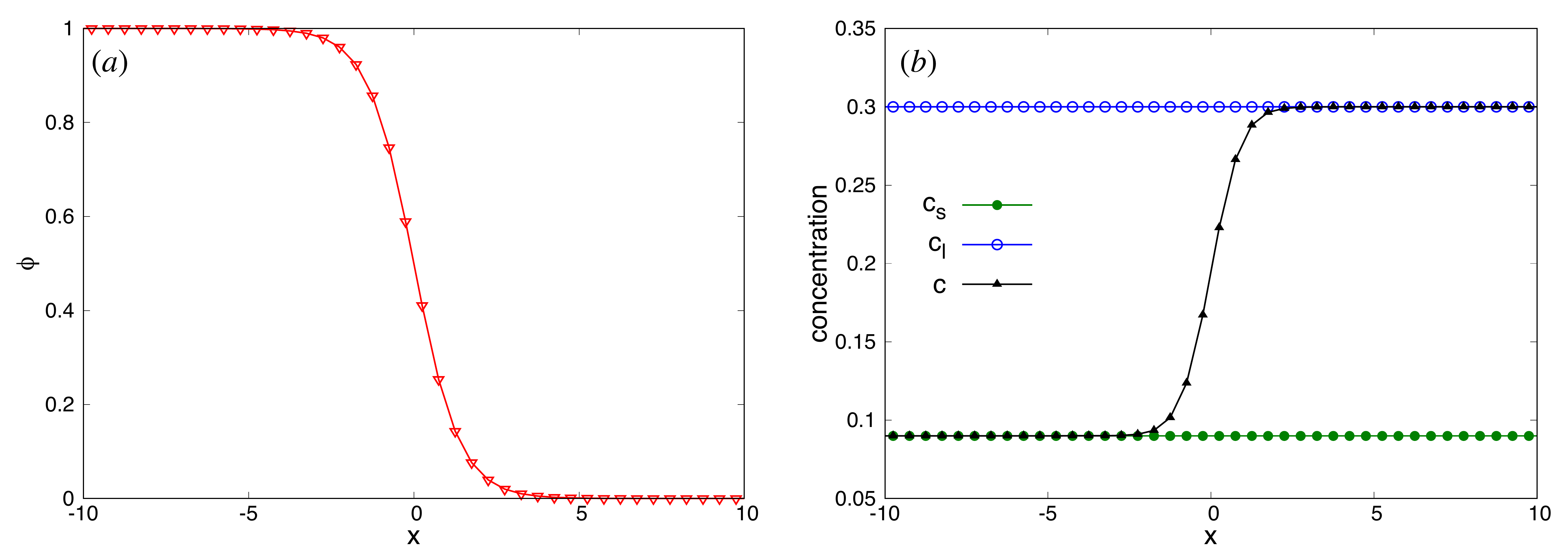}
    \caption{Numerically evaluated (a) equilibrium phase-field profile and (b) equilibrium phase concentrations $c_s$ and $c_l$. The overall solute concentration $c$ is given by Eq.~(\ref{eq19}). Si-9 at\% As is used as the representative system.}
    \label{supfig1}
\end{figure}

\subsection{Interfacial energy}
Following Cahn and Hilliard, we define the interfacial energy $\gamma$ as the difference per unit area of the interface between the actual free energy of the system and that it would have if the properties of the phases were continuous throughout. Assuming the dividing surface at $x = 0$ we have
\begin{eqnarray}\label{eq20}
    \gamma = \int_{-\infty}^0\Bigg[ Hf_{\mathrm{dw}}(\phi_o) + \frac{\sigma}{2}\Big(\frac{\mathrm{d}\phi_o}{\mathrm{d}x}\Big)^2 + f(c_s^e,c_l^e,\phi_o) + \lambda[c_o - \{c_s^e h(\phi_o) + c_l^e(1-h(\phi_o))\}] \nonumber \\
    - f(c_s^e,1)-\lambda[c_o-c_s^e] \Bigg]\mathrm{d}x + \int_{0}^{+\infty} \Bigg[Hf_{\mathrm{dw}}(\phi_o) + \frac{\sigma}{2}\Big(\frac{\mathrm{d}\phi_o}{\mathrm{d}x}\Big)^2 + f(c_s^e,c_l^e,\phi_o) \nonumber \\
    + \lambda[c_o - \{c_s^e h(\phi_o) + c_l^e(1-h(\phi_o))\}]- f(c_l^e,0)-\lambda[c_o-c_l^e] \Bigg]\mathrm{d}x .
\end{eqnarray}
From equilibrium condition of the phase-field i.e. $\frac{\delta \mathcal{\overline{F}}}{\delta \phi} = 0$ we have,
\begin{equation}\label{eq21}
    Hf_{\mathrm{dw}}^{\prime}(\phi_o) - \sigma \frac{\mathrm{d}^2\phi_o}{\mathrm{d}x^2} + \frac{\partial f(c_s,c_l,\phi)}{\partial \phi} - \lambda (c_s - c_l)h^{\prime}(\phi_o) = 0.
\end{equation}
Multiplying both sides by $\frac{\mathrm{d}\phi_o}{\mathrm{d}x}$ we have
\begin{equation}\label{eq22}
   Hf_{\mathrm{dw}}^{\prime}(\phi_o)\frac{\mathrm{d}\phi_o}{\mathrm{d}x} - \sigma \frac{\mathrm{d}^2\phi_o}{\mathrm{d}x^2}\frac{\mathrm{d}\phi_o}{\mathrm{d}x} + \frac{\partial f(c_s,c_l,\phi)}{\partial \phi}\frac{\mathrm{d}\phi_o}{\mathrm{d}x} - \lambda (c_s - c_l)h^{\prime}(\phi_o)\frac{\mathrm{d}\phi_o}{\mathrm{d}x} = 0.
\end{equation}
Using chain-rule we have
\begin{equation}\label{eq23}
    \frac{\mathrm{d}f}{\mathrm{d}x} = \frac{\partial f}{\partial c_s}\frac{\mathrm{d}c_s}{\mathrm{d}x} + \frac{\partial f}{\partial c_l}\frac{\mathrm{d}c_l}{\mathrm{d}x} + \frac{\partial f}{\partial \phi}\frac{\mathrm{d}\phi_o}{\mathrm{d}x} .
\end{equation}
Substituting for $\frac{\partial f}{\partial c_s}$ and $\frac{\partial f}{\partial c_l}$ we have
\begin{equation}\label{eq24}
    \frac{\mathrm{d}f}{\mathrm{d}x} = \lambda h(\phi_o)\frac{\mathrm{d}c_s}{\mathrm{d}x} + \lambda \{1-h(\phi_o)\}\frac{\mathrm{d}c_l}{\mathrm{d}x} + \frac{\partial f}{\partial \phi}\frac{\mathrm{d}\phi_o}{\mathrm{d}x} .
\end{equation}
The first two terms on the right hand side of the equation can be re-expressed as
\begin{equation}\label{eq25}
   \frac{\mathrm{d}f}{\mathrm{d}x} = \lambda \frac{\mathrm{d}}{\mathrm{d}x} [c_s h(\phi_o) + c_l\{1-h(\phi_o)\}] - \lambda (c_s - c_l)h^{\prime}(\phi_o) \frac{\mathrm{d}\phi_o}{\mathrm{d}x} + \frac{\partial f}{\partial \phi}\frac{\mathrm{d}\phi_o}{\mathrm{d}x}.
\end{equation}
Hence,
\begin{equation}\label{eq26}
   \frac{\partial f}{\partial \phi}\frac{\mathrm{d}\phi_o}{\mathrm{d}x} =  \frac{\mathrm{d}f}{\mathrm{d}x} - \lambda \frac{\mathrm{d}}{\mathrm{d}x} [c_s h(\phi_o) + c_l\{1-h(\phi_o)\}] + \lambda (c_s - c_l)h^{\prime}(\phi_o) \frac{\mathrm{d}\phi_o}{\mathrm{d}x}.
\end{equation}
Substituting the above expression back in Eq.~(\ref{eq22}),
\begin{equation}\label{eq27}
     Hf_{\mathrm{dw}}^{\prime}(\phi_o)\frac{\mathrm{d}\phi_o}{\mathrm{d}x} - \sigma \frac{\mathrm{d}^2\phi_o}{\mathrm{d}x^2}\frac{\mathrm{d}\phi_o}{\mathrm{d}x} + \frac{\mathrm{d}f}{\mathrm{d}x} - \lambda \frac{\mathrm{d}}{\mathrm{d}x} [c_s h(\phi_o) + c_l\{1-h(\phi_o)\}] = 0.
\end{equation}
Integrating within the limits $-\infty$ to $x$ we have
\begin{equation}\label{eq28}
    Hf_{\mathrm{dw}}(\phi_o)\Big|_{-\infty}^{x} - \frac{\sigma}{2}\Big(\frac{\mathrm{d}\phi_o}{\mathrm{d}x}\Big)^2 \Big|_{-\infty}^{x} + f(c_s,c_l,\phi_o)\Big|_{-\infty}^{x} - \lambda[c_sh(\phi_o)+c_l\{1-h(\phi_o)\}]\Big|_{-\infty}^{x} = 0.
\end{equation}
Substituting the limits and re-arranging 
\begin{equation}\label{eq29}
    \frac{\sigma}{2}\Big(\frac{\mathrm{d}\phi_o}{\mathrm{d}x}\Big)^2 = Hf_{\mathrm{dw}}(\phi_o) + f(c_s^e,c_l^e,\phi_o) - \lambda[c_s^eh(\phi_o) + c_l^e\{1-h(\phi_o)\}]-f(c_s^e,1) + \lambda c_s^e .
\end{equation}
The above expression is the statement of equipartition of energy which states that at equilibrium the bulk part and the gradient part contribute equally to the phase-field profile. Similarly integrating Eq.~(\ref{eq27}) within the limits $x$ to $+\infty$ we have
\begin{equation}\label{eq30}
    \frac{\sigma}{2}\Big(\frac{\mathrm{d}\phi_o}{\mathrm{d}x}\Big)^2 = Hf_{\mathrm{dw}}(\phi_o) + f(c_s^e,c_l^e,\phi_o) - \lambda[c_s^eh(\phi_o) + c_l^e\{1-h(\phi_o)\}]-f(c_l^e,0) + \lambda c_l^e .
\end{equation}
Substituting Eqs.~(\ref{eq29}) and (\ref{eq30}) in Eq.~(\ref{eq20}) we obtain
\begin{equation}\label{eq31}
    \gamma = \int_{-\infty}^{0}\sigma\Big(\frac{\mathrm{d}\phi_o}{\mathrm{d}x}\Big)^2 \mathrm{d}x +  \int_{0}^{+\infty}\sigma\Big(\frac{\mathrm{d}\phi_o}{\mathrm{d}x}\Big)^2 \mathrm{d}x,
\end{equation}
which can be combined to give
\begin{equation}\label{eq32}
    \gamma = \sigma\int_{-\infty}^{+\infty}\Big(\frac{\mathrm{d}\phi_o}{\mathrm{d}x}\Big)^2 \mathrm{d}x.
\end{equation}
Employing Eq.~(\ref{eq17}) the above expression can be re-written as
\begin{equation}\label{eq33}
    \gamma = \sqrt{2}WH \int_{0}^{1}\sqrt{f_{\mathrm{dw}}(\phi_o)}\mathrm{d}\phi_o.
\end{equation}
Substituting $f_{\mathrm{dw}}(\phi_o) = \phi_o^2(1-\phi_o)^2$ and integrating we obtain
\begin{equation}\label{eq34}
    \gamma = IWH
\end{equation}
where $I = 1/3\sqrt{2}$.

\section{Non-dimensionalization}
To perform the asymptotic analysis, we first non-dimensionalize the governing equations and make certain choices on the scale of the various terms. As an example we show the non-dimensionalization procedure for the form of the governing equations with $g(\phi) = h(\phi)$ of the model with assumption $\mu_s = \mu_l$. The non-dimenisonalization of the general model $\mu_s \neq \mu_l$ follows along similar lines. The dimensional form of the governing equations write as
\begin{align}\label{phi}
    \Bigg[\frac{1}{M_{\phi}} & + \frac{h(\phi)\{1-h(\phi)\}}{M_s\{1-h(\phi)\} + M_l h(\phi)}\frac{1}{|\nabla\phi|^2}\Bigg]\frac{\partial\phi}{\partial t} =  \sigma \nabla^2\phi -  H f_{dw}^{\prime}(\phi) -  \{f_s(c_s) - f_l(c_l)   \nonumber \\
    &- \mu(c_s - c_l)\}h^{\prime}(\phi) + \frac{(\tilde{c}_s-\tilde{c}_l)}{|\nabla\phi|}\frac{(M_s - M_l)h(\phi)\{1-h(\phi)\}}{M_s\{1-h(\phi)\} + M_l h(\phi)} \nabla\mu \cdot \frac{\nabla\phi}{|\nabla\phi|}
\end{align}
and
\begin{align}\label{conc}
    \frac{\partial c}{\partial t} = \nabla \cdot \Bigg[\frac{M_sM_l}{M_s\{1-h(\phi)\} + M_l h(\phi)}\nabla_n \mu & + \{M_sh(\phi) + M_l(1-h(\phi))\}\nabla_t\mu  \nonumber \\
    + & \frac{(\tilde{c}_s-\tilde{c}_l)}{|\nabla\phi|}\frac{(M_s-M_l)h(\phi)\{1-h(\phi)\}}{M_s\{1-h(\phi)\} + M_l h(\phi)}\frac{\nabla\phi}{|\nabla\phi|}\frac{\partial\phi}{\partial t}\Bigg]
\end{align}

To non-dimensionalize the above Eqs.~(\ref{phi}) and (\ref{conc}), we introduce dimensionless quantities (denoted by tilde) : $\mathbf{x} = d_o\tilde{\mathbf{x}}$, $t = \frac{d_o^2}{D_l}\tilde{t}$ and $E = H\tilde{E}$ where $d_o$ is the capillary length and $D_l$ is the diffusivity of the liquid phase. Furthermore, it is imperative to represent the mobilites in terms of the diffusivities as $M_s = D_s\frac{\partial c_s}{\partial \mu}$ and $M_l = D_l\frac{\partial c_l}{\partial \mu}$. Therefore, Eq.~({\ref{phi}}) writes as
\begin{align}\label{nondim1}
    \Big[ \frac{\tau D_l}{d_o^2} + \frac{X}{H}\frac{(\overline{c_s} - \overline{c_l})^2}{|\nabla\phi|^2}\tilde{\alpha_1}(\phi)\Big]\frac{\partial\phi}{\partial\tilde{t}} = & \frac{W^2}{d_o^2}\tilde{\nabla}^2\phi - f_{\mathrm{dw}}^{\prime}(\phi) -  \frac{X}{H}\{\tilde{f_s}(c_s) - \tilde{f_l}(c_l)  \nonumber \\
    & - \tilde{\mu}(c_s - c_l) \}h^{\prime}(\phi) +  \frac{X}{H}\frac{(\overline{c_s}-\overline{c_l})}{|\nabla\phi|}\tilde{a}(\phi))\tilde{\nabla}\tilde{\mu}\cdot\frac{\tilde{\nabla}\phi}{|\tilde{\nabla}\phi|},\nonumber \\
\end{align}
where $\tau = 1/M_{\phi}H$ and $X$is the scale of the driving force. We introduce non-dimensional parameter $\alpha = \frac{\tau D_l}{W^2}$, $\epsilon = \frac{W}{d_o}$ and $\Lambda = \frac{X}{H}$
The interfacial energy $\gamma$ and the interfacial width $W$ are related by the well height $H$ and gradient energy coefficient $\sigma$ as
\begin{equation}\label{nondim2}
    \gamma = a_1 \sqrt{\sigma H},
\end{equation}
and 
\begin{equation}\label{nondim3}
    W = \sqrt{\frac{\sigma}{H}},
\end{equation}
where $a_1$ is a constant related to the specific form of the chosen double well potential and is equal to $I$ as given by Eq.~(\ref{eq34}). Furthermore, it is to noted that the capillary length is set by the ratio of the surface energy to the driving force i.e. $d_o = \frac{\gamma}{X}$. Using Eqs.~(\ref{nondim2}) and (\ref{nondim3}), $d_o = a_1 W\frac{H}{X}$, implies $\Lambda = a_1\epsilon$. We observe that as $\epsilon \rightarrow 0$, $\Lambda \rightarrow 0$. Physically this corresponds to decreasing the interfacial width $W$ and simultaneously raising the well height $H$, while holding the interfacial energy and capillary length constant. Substituting the non-dimensional parameters and scaling, Eq.~(\ref{nondim1}) translates to
\begin{align}\label{nondim4}
    \Big[ \alpha\epsilon^2 + a_1\epsilon \frac{(\tilde{c}_s - \tilde{c}_l)^2}{|\tilde{\nabla}\phi|^2}\tilde{\alpha_1}(\phi)\Big]\frac{\partial \phi}{\partial \tilde{t}} = \epsilon^2\tilde{\nabla}^2\phi - f_{\mathrm{dw}}^{\prime}(\phi) - & a_1\epsilon\{\tilde{f_s}(c_s) - \tilde{f_l}(c_l) - \tilde{\mu}(c_s - c_l) \}h^{\prime}(\phi) \nonumber \\
   & + a_1\epsilon\frac{(\tilde{c}_s-\tilde{c}_l)}{|\tilde{\nabla}\phi|}\tilde{a}(\phi)\tilde{\nabla}\tilde{\mu}\cdot\frac{\tilde{\nabla}\phi}{|\tilde{\nabla}\phi|}, \nonumber \\
\end{align}
where
\begin{equation}\label{nondim5}
    \tilde{a}(\phi) = \frac{\chi\frac{D_s}{D_l}-1}{\chi\frac{D_s}{D_l}\{1-h(\phi)\} + h(\phi)}
\end{equation}
and
\begin{equation}\label{nondim6}
    \tilde{\alpha_1}(\phi) = \frac{h(\phi)\{1-h(\phi)\}}{\chi\frac{D_s}{D_l}\{1-h(\phi)\}+ h(\phi)}.
\end{equation}
$\chi$ is the thermodynamic factor equal to $\frac{\partial c_s}{\partial \mu}/\frac{\partial c_l}{\partial\mu}$.
Similarly, the non-dimensional form of concentration equation can be written as
\begin{equation}\label{nondim7}
    \frac{\partial c}{\partial \tilde{t}} = \tilde{\nabla} \cdot \Bigg[ \frac{\partial c_l}{\partial \mu}q_n(\phi)\tilde{\nabla}_n\tilde{\mu} + \frac{\partial c_l}{\partial \mu}q_t(\phi)\tilde{\nabla}_t\tilde{\mu} + \frac{(\tilde{c}_s - \tilde{c}_l)}{|\tilde{\nabla}\phi|}\tilde{a}(\phi)\frac{\tilde{\nabla}\phi}{|\tilde{\nabla}\phi|}\frac{\partial\phi}{\partial\tilde{t}}\Bigg]
\end{equation}
where
\begin{equation}
    q_n(\phi) = \frac{\chi \frac{D_s}{D_l}}{\chi \frac{D_s}{D_l}\{1-h(\phi)\} + h(\phi)}
\end{equation}
and
\begin{equation}
    q_t(\phi) = \chi \frac{D_s}{D_l} h(\phi) + 1 - h(\phi)
\end{equation}
For the sake of compactness of notation we henceforth remove the tildes. 

\section{Asymptotic analysis}
We perform the asymptotic analysis by expanding the field variables in terms of a small parameter which in the present case is $\epsilon$. The solution domain is divided into outer (farther from the interface) and inner (closer to the interface) regions where the field variables vary slowly and rapidly respectively. For instance in the inner region the field variables are expanded as
\begin{equation}\label{asy1}
    \phi = \phi_0 + \epsilon\phi_1 + \epsilon^2\phi_2 ,
\end{equation}
\begin{equation}\label{asy2}
    c = c_0 + \epsilon c_1 + \epsilon^2 c_2 ,
\end{equation}
\begin{equation}\label{asy3}
    c_s = c_{s,0} + \epsilon c_{s,1} + \epsilon^2 c_{s,2} ,
\end{equation}
\begin{equation}\label{asy4}
    c_l = c_{l,0} + \epsilon c_{l,1} + \epsilon^2 c_{l,2}.
\end{equation}
\begin{equation}\label{asy5}
    \mu = \mu_0 + \epsilon \mu_1 + \epsilon^2 \mu_2
\end{equation}
The superscript on $\epsilon$ denotes power, where as the subscript on the field variables denotes the order of the solution. The power series is substituted in the governing equations and terms are re-grouped and solved for each power of $\epsilon$. The solutions from the inner region are matched to the outer solution at each order.

In the outer region it can be shown that leading order $\phi$ assumes constant values of $1$ and $0$ in the solid and the liquid. The higher order corrections trivially turn out to be zero. Similarly the $c$ equation obeys Fick's law of diffusion at all orders. In the inner region, the equations are re-casted in a curvilinear co-ordinate system with the reference frame attached to the interface. $r$ and $s$ measures the direction normal and along the interface respectively. Furthermore, the distance in the normal direction of the interface is scaled as $\eta = r/\epsilon$.  We transform all the operators as follows,
\begin{equation}\label{asy6}
    \frac{\partial}{\partial t}  =  -\frac{V_n}{\epsilon}\frac{\partial}{\partial \eta} + \frac{\mathrm{d}}{\mathrm{d}t} - V_t\frac{\partial}{\partial s}
\end{equation}
\begin{equation}\label{asy7}
    \nabla = \frac{\hat{n}}{\epsilon}\frac{\partial}{\partial \eta} + \hat{s}\frac{\partial}{\partial s} - \epsilon\kappa \eta \hat{s}\frac{\partial}{\partial s}
\end{equation}
\begin{equation}\label{asy8}
         \nabla^2  =  \frac{1}{\epsilon^2}\frac{\partial^2}{\partial \eta} + \frac{\kappa}{\epsilon}\frac{\partial}{\partial \eta} - \kappa^2\eta\frac{\partial}{\partial \eta} + \frac{\partial^2}{\partial s^2}
\end{equation}
\begin{equation}\label{asy9}
   \nabla \cdot (q\nabla)  =  \frac{1}{\epsilon^2}\frac{\partial}{\partial \eta}(q\frac{\partial}{\partial \eta}) + \frac{1}{\epsilon}\kappa q\frac{\partial}{\partial\eta} - \kappa^2 q \eta\frac{\partial}{\partial \eta} + \frac{\partial}{\partial s}(q\frac{\partial}{\partial s}) 
\end{equation}
\begin{equation}\label{asy10}
    \nabla \cdot(q\nabla_n) = \frac{1}{\epsilon^2}\frac{\partial}{\partial \eta}(q\frac{\partial}{\partial \eta}) + \frac{1}{\epsilon}\kappa q\frac{\partial}{\partial\eta} - \kappa^2 q \eta\frac{\partial}{\partial \eta}
\end{equation}
\begin{equation}\label{asy11}
    \nabla \cdot(q\nabla_t) = \frac{\partial}{\partial s}(q\frac{\partial}{\partial s})
\end{equation}
\begin{equation}\label{asy12}
    \nabla \cdot \vec{f} = \frac{1}{\epsilon}\frac{\partial}{\partial \eta}(\hat{n}\cdot \vec{f}) + \frac{\partial}{\partial s}(\hat{s}\cdot \vec{f}) + \kappa \hat{n}\cdot \vec{f}
\end{equation}
\begin{equation}\label{asy13}
    \nabla \cdot (q\hat{n}) = \frac{1}{\epsilon}\frac{\partial q}{\partial \eta}.
\end{equation}
Using the above transformations, the governing equations in the inner region can be expressed as
\begin{eqnarray}\label{asy14}
 \frac{\partial^2\phi}{\partial\eta^2}- f^{\prime}_{\mathrm{dw}}(\phi) + \epsilon\Bigg[(\alpha V_n + \kappa)\frac{\partial\phi}{\partial\eta} - a_1\{f_s(c_s)-f_l(c_l)-\mu(c_s-c_l)\}h^{\prime}(\phi) \nonumber \\
 - a_1\frac{(\tilde{c}_s-\tilde{c}_l)}{|\frac{\partial\phi}{\partial\eta}|}a(\phi)\frac{\partial\mu}{\partial\eta}\Bigg] +
 \epsilon^2\Bigg[\frac{\partial^2\phi}{\partial s^2} - \kappa^2\eta\frac{\partial\phi}{\partial\eta} + a_1\frac{(\tilde{c}_s-\tilde{c}_l)^2}{|\frac{\partial\phi}{\partial\eta}|^2}\alpha_1(\phi)\frac{\partial\phi}{\partial\eta}V_n\Bigg] = \mathcal{O}(\epsilon^3), \nonumber \\
\end{eqnarray}
\begin{eqnarray}
 \frac{1}{\epsilon^2}\frac{\partial}{\partial\eta}\Big[\frac{\partial c_l}{\partial\mu}q_n(\phi)\frac{\partial\mu}{\partial\eta}\Big] + \frac{1}{\epsilon}\Bigg[V_n\frac{\partial c}{\partial\eta} + \frac{\partial}{\partial\eta}\Bigg\{ \frac{(\tilde{c}_s-\tilde{c}_l)}{|\frac{\partial\phi}{\partial\eta}|}a(\phi)V_n\frac{\partial\phi}{\partial\eta}\Bigg\}\Bigg] \nonumber \\
 + \epsilon^0\Bigg[\frac{\partial}{\partial s}\Bigg\{q_t(\phi)\frac{\partial\phi}{\partial s}\Bigg\} + \kappa V_n \frac{(\tilde{c}_s-\tilde{c}_l)}{|\frac{\partial\phi}{\partial\eta}|}a(\phi)\frac{\partial\phi}{\partial\eta} - \kappa^2\eta q_n(\phi)\frac{\partial\mu}{\partial\eta}\Bigg] = \mathcal{O}(\epsilon)
\end{eqnarray}
\subsection{$\mathcal{O}(\epsilon^0)$ in $\phi$ : Phase-field profile}
At the leading order $\mathcal{O}(\epsilon^0)$ in $\phi$ we have,
\begin{equation}\label{sl1}
    \frac{\partial^2{\phi_0}}{\partial \eta^2} - f^{\prime}_{\mathrm{dw}}({\phi_0}) = 0
\end{equation}
 Multiplying both sides by $\frac{\partial{\phi_0}}{\partial \eta}$ and integrating we obtain,
 \begin{equation}\label{sl2}
    \Big(\frac{\partial{\phi_0}}{\partial\eta}\Big)^2 = 2 f_{\mathrm{dw}}({\phi_0}) + C_1
 \end{equation}
 Using the following matching conditions
 \begin{equation}\label{sl3}
    \lim_{\eta \to \pm\infty} {\phi_0} = \phi_0|^{\pm} = 0,1 
 \end{equation}
 and 
 \begin{equation}\label{sl4}
  \lim_{\eta \to \pm\infty} \frac{\partial{\phi_0}}{\partial \eta} = 0   
 \end{equation}
we evaluate the integration constant $C_1 = 0$. Therefore, we have
\begin{equation}\label{sl5}
 \frac{\partial {\phi_0}}{\partial\eta} = \sqrt{2f_{\mathrm{dw}}({\phi_0})}
\end{equation}
Integrating the above Eq.~ then gives the leading order solution to be
\begin{equation}\label{sl6}
     {\phi_0}(\eta) = \frac{1}{2}\Bigg[1-\tanh\left(\frac{\eta}{\sqrt{2}}\right)\Bigg] .
\end{equation}

\subsection{$\mathcal{O}(1/\epsilon^2)$ in $c$ : Equality of diffusion potential}
At the leading order, equating $\mathcal{O}(1/\epsilon^2)$ in $c$ equation 
\begin{equation}\label{eqdp1}
   \frac{\partial}{\partial \eta}\Big[\frac{\partial c_{l,0}}{\partial \mu}q_n(\phi_0)\frac{\partial {\mu_0}}{\partial \eta} \Big] = 0 .
\end{equation}
Integrating once,
\begin{equation}\label{eqdp2}
    \frac{\partial c_{l,0}}{\partial \mu}q_n({\phi_0})\frac{\partial {\mu_0}}{\partial \eta} = A_1(s)
\end{equation}
Taking the limit $\eta \rightarrow \pm\infty$ and employing the matching conditions $\lim_{\eta \to \pm\infty} = \frac{\partial{\mu_0}}{\partial \eta} = 0$, we obtain $A_1(s) = 0$. Since $q_n(\phi_0)$ and $\frac{\partial c_{l,0}}{\partial\mu}$ are finite (non-zero) in the inner region, Eq.~(\ref{eqdp2}) reduces to
\begin{equation}\label{eqdp3}
   \frac{\partial {\mu_0}}{\partial \eta} = 0 .
\end{equation}
Integrating again we have
\begin{equation}\label{eqdp4}
    {\mu_0} = \overline{\mu}_0(s)
\end{equation}
which is a constant and independent of the normal co-ordinate $\eta$. From matching conditions, we have $\lim_{\eta \to \pm\infty} {\mu_0} = \mu_0\Big|^{\pm} = \overline{\mu}_0$. This implies,
\begin{equation}\label{eqdp5}
 \frac{\partial f_l(c_{l,0})}{\partial c_l} = \frac{\partial f_s(c_{s,0})}{\partial c_s} = \overline{\mu}_0
\end{equation}
i.e. the slopes of the tangents to the free energy curves of solid and liquid at the solid and liquid compositions are equal. In addition Eq.~(\ref{eqdp5}) also implies that lowest order concentration fields $c_{s,0}$ and $c_{l,0}$ are independent of $\eta$.
The value of the constant $\overline{\mu}_0(s)$ is fixed by considering the next-to-leading order $\phi$ equation. 

\subsection{$\mathcal{O}(\epsilon)$ in $\phi$ : Gibbs Thomson equation}
At $\mathcal{O}(\epsilon)$ we obtain,
\begin{equation}\label{gt1}
    -\alpha V_n\frac{\partial {\phi_0}}{\partial \eta} = \frac{\partial^2{\phi_1}}{\partial \eta^2} - f^{\prime\prime}_{\mathrm{dw}}({\phi_0}){\phi_1} + \kappa\frac{\partial {\phi_0}}{\partial \eta} - a_1[f_s(c_{s,0}) - f_l(c_{l,0}) - \mu_0(c_{s,0} - c_{l,0}) ]h^{\prime}(\phi_0)
\end{equation}
The above Eq.~is multiplied by $\frac{\partial{\phi_0}}{\partial \eta}$ on both sides, rearranged and integrated within the limits $\pm \infty$ to obtain
\begin{eqnarray}\label{gt2}
 -(\alpha V_n + \kappa)\int_{-\infty}^{+\infty}\Big( \frac{\partial{\phi_0}}{\partial \eta}\Big)^2 \mathrm{d}\eta = \int_{-\infty}^{+\infty}\frac{\partial^2{\phi_1}}{\partial \eta^2}\frac{\partial{\phi_0}}{\partial \eta}\mathrm{d}\eta - \int_{-\infty}^{+\infty}f^{\prime\prime}_{\mathrm{dw}}({\phi_0}){\phi_1}\frac{\partial {\phi_0}}{\partial \eta}\mathrm{d}\eta \nonumber \\
 - a_1[f_s(c_{s,0}) - f_l(c_{l,0}) - \mu_0(c_{s,0} - c_{l,0})] \int_{-\infty}^{+\infty} \frac{\partial h{\phi_0})}{\partial \phi}\frac{\partial {\phi_0}}{\partial \eta}\mathrm{d}\eta
\end{eqnarray}
Let $I_1 = \int_{-\infty}^{+\infty}\frac{\partial^2{\phi_1}}{\partial \eta^2}\frac{\partial {\phi_0}}{\partial \eta}\mathrm{d}\eta$ which can be simplified by integration by parts (with the second term as first function) as below,
\begin{equation}\label{gt3}
 I_1 = \int_{-\infty}^{+\infty}\frac{\partial^2{\phi_1}}{\partial \eta^2}\frac{\partial {\phi_0}}{\partial \eta}\mathrm{d}\eta = \frac{\partial {\phi_0}}{\partial \eta}\frac{\partial {\phi_1}}{\partial \eta} \Bigg|_{-\infty}^{+\infty} - \int_{-\infty}^{+\infty}\frac{\partial^2{\phi_0}}{\partial\eta^2}\frac{\partial {\phi_1}}{\partial \eta}\mathrm{d}\eta
\end{equation}
From the derivative matching condition, the first term goes to zero. The second integral can be further simplified by integration by parts (with the first term as the first function) as follows,
\begin{equation}\label{gt4}
I_1 = - \int_{-\infty}^{+\infty}\frac{\partial^2{\phi_0}}{\partial\eta^2}\frac{\partial {\phi_1}}{\partial \eta}\mathrm{d}\eta = -\Bigg[ \frac{\partial^2{\phi_0}}{\partial\eta^2}{\phi_1}\Bigg|_{-\infty}^{+\infty} - \int_{-\infty}^{+\infty}\frac{\partial^3{\phi_0}}{\partial\eta^3}{\phi_1}\mathrm{d}\eta\Bigg]
\end{equation}
Using Eq.~(\ref{sl1}) the first term can be re-written as,
\begin{equation}\label{gt5}
 I_1 = -\Bigg[ f^{\prime}_{\mathrm{dw}}({\phi_0}){\phi_1}\Bigg|_{-\infty}^{+\infty} - \int_{-\infty}^{+\infty}\frac{\partial^3{\phi_0}}{\partial\eta^3}{\phi_1}\mathrm{d}\eta\Bigg] = \int_{-\infty}^{+\infty}\frac{\partial^3{\phi_0}}{\partial\eta^3}{\phi_1}\mathrm{d}\eta
\end{equation}
We next simplify the second integral in Eq.~(\ref{gt2}) as follows,
\begin{equation}\label{gt6}
I_2 = \int_{-\infty}^{+\infty}f^{\prime\prime}_{\mathrm{dw}}({\phi_0}){\phi_1}\frac{\partial {\phi_0}}{\partial \eta}\mathrm{d}\eta = \int_{-\infty}^{+\infty}\frac{\partial}{\partial \eta}f^{\prime}_{\mathrm{dw}}({\phi_0}){\phi_1}\mathrm{d}\eta
\end{equation}
Therefore, the first two integrals in Eq.~(\ref{gt2}) simplifies as,
\begin{equation}\label{gt7}
 I_1 - I_2 = \int_{-\infty}^{+\infty} \Bigg[\frac{\partial^3{\phi_0}}{\partial\eta^3}{\phi_1} - \frac{\partial}{\partial \eta}f^{\prime}_{\mathrm{dw}}({\phi_0}){\phi_1}\Bigg] \mathrm{d}\eta = \int_{-\infty}^{+\infty} \frac{\partial}{\partial \eta}\Bigg[\frac{\partial^2{\phi_0}}{\partial\eta^2} - f^{\prime}_{\mathrm{dw}}({\phi_0}) \Bigg]{\phi_1}\mathrm{d}\eta = 0
\end{equation}
where we have utilized Eq.~(\ref{sl1}). Therefore, Eq.~(\ref{gt2}) reduces to
\begin{equation}\label{gt8}
  -(\alpha V_n + \kappa)\int_{-\infty}^{+\infty}\Big( \frac{\partial {\phi_0}}{\partial \eta}\Big)^2 \mathrm{d}\eta = - a_1[f_s(c_{s,0}) - f_l(c_{l,0}) - \mu_0(c_{s,0} - c_{l,0})] h({\phi_0})\Big|_{-\infty}^{+\infty}
\end{equation}
and finally to a form
\begin{equation}\label{gt9}
   -(\alpha V_n + \kappa)\int_{-\infty}^{+\infty}\Big( \frac{\partial {\phi_0}}{\partial \eta}\Big)^2 \mathrm{d}\eta =  a_1[f_s(c_{s,0}) - f_l(c_{l,0}) - \mu_0(c_{s,0} - c_{l,0})].
\end{equation}
which on re-arrangement yields
\begin{equation}\label{gt10}
    \mu_0 = \overline{\mu}_0(s) = \frac{f_s(c_{s,0})-f_l(c_{l,0})}{c_{s,0}-c_{l,0}} + \frac{\alpha V_n + \kappa}{c_{s,0}-c_{l,0}}.
\end{equation}
Furthermore, re-arranging the above Eq.~ we obtain,
\begin{equation}\label{gt11}
 \omega_{s,0} - \omega_{l,0} = -(\alpha V_n + \kappa),
\end{equation}
where $\omega_{s,0} = f_s(c_{s,0}) - \mu_0 c_{s,0}$ and $\omega_{l,0} = f_l(c_{l,0}) - \mu_0 c_{l,0}$ are the grand-potentials of the solid and the liquid respectively. Eqs.~(\ref{eqdp5}) and (\ref{gt11}) implies a parallel tangent construction whereby the slopes of the tangent to the free energy at solid and liquid compositions are equal but are displaced by an amount proportional to the velocity and curvature of the interface.

\subsection{$\mathcal{O}(1/\epsilon)$ in $c$ : Solute trapping}
The $\mathcal{O}(1/\epsilon)$ in $c$ equation can be equated to be
\begin{equation}\label{st1}
    -V_n\frac{\partial c_0}{\partial\eta} = \frac{\partial}{\partial\eta}\Big[\frac{\partial c_{l,0}}{\partial\mu} q_n(\phi_0)\frac{\partial\mu_1}{\partial\eta}\Big] - \frac{\partial}{\partial\eta}\Big[(\tilde{c}_{s,0}-\tilde{c}_{l,0})a(\phi_0)V_n\Big]
\end{equation}
Integrating once we have
\begin{equation}\label{st2}
    -V_n c_0 = \frac{\partial c_{l,0}}{\partial\mu} q_n(\phi_0)\frac{\partial\mu_1}{\partial\eta} - (\tilde{c}_{s,0}-\tilde{c}_{l,0})a(\phi_0)V_n + A_2(s).
\end{equation}
To evaluate the integration constant $A_2(s)$ we take the limit $\eta \rightarrow -\infty$ to obtain
\begin{equation}\label{st3}
    A_2(s) = -V_n c_{s,0} - \frac{\partial c_{l,0}}{\partial\mu} q_n^-\frac{\partial\mu_0}{\partial r}\Big|^-,
\end{equation}
where we have employed the fact that $\lim_{\eta \rightarrow -\infty} a(\phi_0) = 0$. Substituting Eq.~(\ref{st3}) into Eq.~(\ref{st2}) and rearranging we obtain
\begin{equation}\label{st4}
    -V_n(c_0 - c_{s,0})= \frac{\partial c_{l,0}}{\partial\mu}q_n(\phi_0)\frac{\partial\mu_1}{\partial\eta} - (\tilde{c}_{s,0}-\tilde{c}_{l,0})a(\phi_0)V_n - \frac{\partial c_{l,0}}{\partial\mu}q_n^-\frac{\partial\mu_0}{\partial r}\Big|^-.
\end{equation}
The above expression can be re-expressed in terms of first order diffusion potential gradient as
\begin{equation}\label{st5}
    \frac{\partial\mu_1}{\partial\eta} = -\frac{V_n}{\partial c_{l,0}/\partial\mu}\frac{(c_0 - c_{s,0})}{q_n(\phi_0)} + (\tilde{c}_{s,0}-\tilde{c}_{l,0})\frac{V_n}{\partial c_{l,0}/\partial\mu}\frac{a(\phi_0)}{q_n(\phi_0)} + q_n^-\frac{\partial\mu_0}{\partial r}\Big|^- \frac{1}{q_n(\phi_0)}.
\end{equation}
Integrating both sides,
\begin{align}\label{st6}
    \mu_1 = -\frac{V_n}{\partial c_{l,0}/\partial\mu}\int_{0}^{\eta}\frac{(c_0 - c_{s,0})}{q_n(\phi_0)}\mathrm{d}\eta &+ \frac{V_n}{\partial c_{l,0}/\partial\mu}\int_{0}^{\eta}(\tilde{c}_{s,0}-\tilde{c}_{l,0})\frac{a(\phi_0)}{q_n(\phi_0)}\mathrm{d}\eta \nonumber \\
    &+ q_n^-\frac{\partial\mu_0}{\partial r}\Big|^- \int_{0}^{\eta}\frac{1}{q_n(\phi_0)}\mathrm{d}\eta + \overline{\mu}_1(s),
\end{align}
where $\overline{\mu}_1(s)$ is the constant of integration. We split the Eq.~(\ref{st6}) into two parts \\
(1) For $\eta > 0$
\begin{align}\label{st7}
    \mu_1 = & - \frac{V_n}{\partial c_{l,0}/\partial\mu}\int_{0}^{\eta}\Bigg[\frac{(c_0 - c_{s,0})}{q_n(\phi_0)} - \frac{(c_{l,0} - c_{s,0})}{q_n^+}\Bigg]\mathrm{d}\eta - \eta \frac{V_n}{\partial c_{l,0}/\partial\mu} \frac{(c_{l,0} - c_{s,0})}{q_n^+} \nonumber \\
   & + \frac{V_n}{\partial c_{l,0}/\partial\mu}\int_{0}^{\eta}(\tilde{c}_{s,0}-\tilde{c}_{l,0})\frac{a(\phi_0)}{q_n(\phi_0)}\mathrm{d}\eta 
    + q_n^-\frac{\partial\mu_0}{\partial r}\Big|^- \int_{0}^{\eta}\Bigg[\frac{1}{q_n(\phi_0)}-\frac{1}{q_n^+}\Bigg]\mathrm{d}\eta \nonumber \\
    & +\eta q_n^-\frac{\partial\mu_0}{\partial r}\Big|^-\frac{1}{q_n^+} + \overline{\mu}_1(s) 
\end{align}
and
(2) For $\eta < 0$
\begin{align}\label{st8}
 \mu_1 = & -\frac{V_n}{\partial c_{l,0}/\partial\mu}\int_{0}^{\eta}\frac{(c_0 - c_{s,0})}{q_n(\phi_0)}\mathrm{d}\eta + \frac{V_n}{\partial c_{l,0}/\partial\mu}\int_{0}^{\eta}(\tilde{c}_{s,0}-\tilde{c}_{l,0})\frac{a(\phi_0)}{q_n(\phi_0)}\mathrm{d}\eta \nonumber \\
 & + q_n^-\frac{\partial\mu_0}{\partial r}\Big|^- \int_{0}^{\eta}\Bigg[\frac{1}{q_n(\phi_0)}-\frac{1}{q_n^-}\Bigg]\mathrm{d}\eta 
 + \eta q_n^-\frac{\partial\mu_0}{\partial r}\Big|^- \frac{1}{q_n^+} + \overline{\mu}_1(s).
\end{align}
Taking far field limit $\eta \rightarrow +\infty$ in Eq.~(\ref{st7}) and employing the matching condition we have
\begin{align}\label{st9}
 \mu_1|^+ + \eta\frac{\partial\mu_0}{\partial r}\Bigg|^+ = & -\frac{V_n}{\partial c_{l,0}/\partial\mu} \int_{0}^{+\infty}\Bigg[\frac{(c_0 - c_{s,0})}{q_n(\phi_0)} - \frac{(c_{l,0} - c_{s,0}}{q_n^+}\Bigg]\mathrm{d}\eta - \eta \frac{V_n}{\partial c_{l,0}/\partial\mu} \frac{(c_{l,0} - c_{s,0})}{q_n^+}  \nonumber \\
 & + \frac{V_n}{\partial c_{l,0}/\partial\mu}\int_{0}^{+\infty}(\tilde{c}_{s,0}-\tilde{c}_{l,0})\frac{a(\phi_0)}{q_n(\phi_0)}\mathrm{d}\eta 
    + q_n^-\frac{\partial\mu_0}{\partial r}\Big|^- \int_{0}^{+\infty}\Bigg[\frac{1}{q_n(\phi_0)}-\frac{1}{q_n^+}\Bigg]\mathrm{d}\eta \nonumber \\
    & +\eta q_n^-\frac{\partial\mu_0}{\partial r}\Big|^-\frac{1}{q_n^+} + \overline{\mu}_1(s) \nonumber \\
\end{align}
Comparing coefficients of $\eta$ on both sides
\begin{equation}\label{st10}
    \frac{\partial\mu_0}{\partial r}\Bigg|^+ = -\frac{V_n}{\partial c_{l,0}/\partial\mu}\frac{(c_{l,0}-c_{s,0})}{q_n^+} + q_n^-\frac{\partial\mu_0}{\partial r}\Big|^-\frac{1}{q_n^+},
\end{equation}
which on re-arranging gives the flux conservation or the Stefan's condition at the interface at the lowest order
\begin{equation}\label{st11}
     q_n^+\frac{\partial\mu_0}{\partial r}\Bigg|^+ - q_n^-\frac{\partial\mu_0}{\partial r}\Big|^- = -\frac{V_n}{\partial c_{l,0}/\partial\mu}(c_{l,0}-c_{s,0})  .
\end{equation}
Similarly comparing coefficients of $\eta^0$ in Eq.~(\ref{st9}) we have
\begin{align}\label{st12}
 \mu_1|^+ =& -\frac{V_n}{\partial c_{l,0}/\partial\mu} \int_{0}^{+\infty}\Bigg[\frac{(c_0 - c_{s,0})}{q_n(\phi_0)} - \frac{(c_{l,0} - c_{s,0}}{q_n^+}\Bigg]\mathrm{d}\eta + \frac{V_n}{\partial c_{l,0}/\partial\mu}\int_{0}^{+\infty}(\tilde{c}_{s,0}-\tilde{c}_{l,0})\frac{a(\phi_0)}{q_n(\phi_0)}\mathrm{d}\eta \nonumber \\
    &+ q_n^-\frac{\partial\mu_0}{\partial r}\Big|^- \int_{0}^{+\infty}\Bigg[\frac{1}{q_n(\phi_0)}-\frac{1}{q_n^+}\Bigg]\mathrm{d}\eta  + \overline{\mu}_1(s) \nonumber \\
\end{align}
Introducing the lowest order solute profile $c_0 = c_{s,0}h(\phi_0) + c_{l,0}\{1-h(\phi_0)\}$ and expanding $\tilde{c}_{s,0} - \tilde{c}_{l,0} = c_{s,0} - c_{l,0} + S_sh(\phi_0) + S_l\{1-h(\phi_0)\}$ in Eq.~(\ref{st12}) and rearranging we obtain
\begin{align}\label{st13}
 \mu_1|^+ =& -\frac{V_n}{\partial c_{l,0}/\partial\mu}(c_{s,0}-c_{l,0})\int_{0}^{+\infty}\{\overline{p}(\phi_0) - \overline{p}(\phi_0|^+)\}\mathrm{d}\eta \nonumber \\
 &+ \frac{V_n}{\partial c_{l,0}/\partial\mu}\int_{0}^{+\infty}[S_sh(\phi_0)+S_l\{1-h(\phi_0)\}]\frac{a(\phi_0)}{q_n(\phi_0)}\mathrm{d}\eta \nonumber \\
 &+ q_n^-\frac{\partial\mu_0}{\partial r}\Big|^- \int_{0}^{+\infty}\Bigg[\frac{1}{q_n(\phi_0)}-\frac{1}{q_n^+}\Bigg]\mathrm{d}\eta  + \overline{\mu}_1(s),
\end{align}
where,
\begin{equation}\label{st14}
    \overline{p}(\phi_0) = \frac{h(\phi_0)-1-a(\phi_0)}{q_n(\phi_0)}
\end{equation}
and we have utilized the relation $\overline{p}(\phi_0|^+) = 1/q_n^+= 1$. Taking the far field limit $\eta \rightarrow -\infty$ of Eq.~(\ref{st8}) and comparing the coefficients $\eta^0$ we obtain
\begin{align}\label{st15}
 \mu_1|^- = & -\frac{V_n}{\partial c_{l,0}/\partial\mu}(c_{s,0}-c_{l,0})\int_{0}^{-\infty}\{\overline{p}(\phi_0)-\overline{p}(\phi_0|^-)\}\mathrm{d}\eta \nonumber \\
 & + \frac{V_n}{\partial c_{l,0}/\partial\mu}\int_{0}^{-\infty}[S_sh(\phi_0)+S_l\{1-h(\phi_0)\}]\frac{a(\phi_0)}{q_n(\phi_0)}\mathrm{d}\eta \nonumber \\
 & + q_n^-\frac{\partial\mu_0}{\partial r}\Big|^- \int_{0}^{-\infty}\Bigg[\frac{1}{q_n(\phi_0)}-\frac{1}{q_n^-}\Bigg]\mathrm{d}\eta  + \overline{\mu}_1(s). 
\end{align}
Subtracting Eq.~(\ref{st13}) from Eq.~(\ref{st15}) we have
\begin{align}\label{st16}
 \mu_1|^- - \mu_1|^+ = & -\frac{V_n}{\partial c_{l,0}/\partial\mu}(c_{s,0}-c_{l,0})(F_1^- - F_1^+) + \frac{V_n}{\partial c_{l,0}/\partial\mu}(F_2^- - F_2^+) \nonumber \\
  &+ q_n^-\frac{\partial\mu_0}{\partial r}\Bigg|^- (G^- - G^+)
\end{align}
where ,
\begin{equation}\label{st17}
    F_1^{\pm} = \int_{0}^{\pm\infty}\{\overline{p}(\phi_0) - \overline{p}(\phi_0|^{\pm})\}\mathrm{d}\eta ,
\end{equation}
\begin{equation}\label{st18}
    F_2^{\pm} = \int_{0}^{\pm\infty}[S_sh(\phi_0)+S_l\{1-h(\phi_0)\}]\frac{a(\phi_0)}{q_n(\phi_0)}\mathrm{d}\eta,
\end{equation}
and
\begin{equation}\label{st19}
    G^{\pm} = \int_{0}^{\pm\infty}\Big[\frac{1}{q_n(\phi_0)}-\frac{1}{q_n^{\pm}}\Big]\mathrm{d}\eta .
\end{equation}
The third term in Eq.~(\ref{st16}) proportional to the solute $q_n^-\frac{\partial\mu_0}{\partial r}\Big|^-$ is the Kapitza jump and results in diffusion potential jump even for a stationary interface. Since there is little evidence that this term plays an important role in solute trapping, it can be eliminated by a judicious choice of diffusivity interpolation function $q_n(\phi_0)$ that will result in $G^+ = G^-$. The inverse interpolation given by
\begin{equation}\label{st20}
    \frac{1}{q_n(\phi_0)} = \frac{h(\phi_0)}{\chi\frac{D_s}{D_l}} + 1-h(\phi_0)
\end{equation}
eliminates the term. Thus the diffusion potential jump can be controlled by the second term in Eq.~(\ref{st16}) which depends upon the source terms. 

Perhaps the most simplest choice is given by the choice $S_s = S_l = -A(c_{l,0} - c_{s,0})$. Substituting back in Eq.~(\ref{st16}) we obtain
\begin{equation}\label{st23}
    \mu_1|^- - \mu_1|^+ = -\frac{V_n}{\partial c_{l,0}/\partial\mu}(c_{l,0}-c_{s,0})(F^- - F^+).
\end{equation}
where,
\begin{equation}
    F_2^{\pm} = \int_{0}^{\pm}\{\tilde{p}(\phi_0) - \tilde{p}(\phi_0|^{\pm})\}\mathrm{d}\eta
\end{equation}
with
\begin{equation}
    \tilde{p}(\phi_0) = \frac{h(\phi_0)-1-(1-A)a(\phi_0)}{q_n(\phi_0)}
\end{equation}
The overall jump in diffusion potential upto first order therefore writes as
\begin{equation}\label{st28}
    \mu|^- - \mu|^+ = (\mu_0 + \epsilon \mu_1|^-) - (\mu_0 + \epsilon \mu_1|^+) = \epsilon(\mu_1|^- - \mu_1|^+)
\end{equation}
The overall jump expression writes as
\begin{equation}\label{st30}
    \mu|^- - \mu|^+ = \epsilon \frac{V_n}{\partial c_{l,0}/\partial\mu}(c_{l,0}-c_{s,0})(F^- - F^+).
\end{equation}
Reverting to dimensional quantities, for instance Eq.~(\ref{st30})
\begin{equation}\label{st31}
    \mu|^- - \mu|^+ = \frac{WV}{D_l\partial c_l/\partial\mu}(c_{l,0}-c_{s,0})(F^- - F^+)
\end{equation}
it is easy to see that the diffusion potential is proportional to the product of interface width $W$, velocity $V$ and the trapping parameter $A$.

For the second set of interpolation functions discussed in the main article we have
\begin{equation}\label{alt_diff}
    \frac{1}{q_n(\phi_0)} = \frac{h_n(\phi_0)}{\chi\frac{D_s}{D_l}} + 1-h_n(\phi_0)
\end{equation}
which satisfies the constraint $G^+ = G^-$. The diffusion potential jump is given by Eq.~(\ref{st31}) with 
\begin{equation}
    \tilde{p}(\phi_0) = \frac{p(\phi) - 1 - (1-A)a(\phi_0)}{q_n(\phi_0)}
\end{equation}
where,
\begin{equation}
    p(\phi_0) = h(\phi_0) + a\left(\phi_0 - \frac{1}{2}\right)\phi_0^2(1-\phi_0)^2
\end{equation}
and 
\begin{equation}
    a(\phi_0) = \frac{\left(\chi\frac{D_s}{D_l}-b\right)p(\phi_0)\{1-p(\phi_0)\}}{\chi\frac{D_s}{D_l}\{1-p(\phi_0)\}+bp(\phi_0)}
\end{equation}

\subsection{$\mathcal{O}(\epsilon^2)$ in $\phi$ : Modified Gibbs-Thomson Equation}
The $\mathcal{O}(\epsilon^2)$ terms in $\phi$ equation writes as
\begin{align}\label{gb1}
 -\alpha V_n\frac{\partial\phi_1}{\partial \eta} -& a_1 {\alpha_1}(\phi_o)\frac{(\tilde{c}_{s,0}-\tilde{c}_{l,0})^2}{\frac{\partial c_{l,0}}{\partial\mu}}\frac{V_n}{(\frac{\partial\phi_0}{\partial\eta})^2}\frac{\partial\phi_0}{\partial\eta}  =  \frac{\partial^2\phi_2}{\partial\eta^2} + \kappa\frac{\partial\phi_1}{\partial\eta} - \eta\kappa^2\frac{\partial\phi_0}{\partial\eta} - f_{\mathrm{dw}}^{\prime\prime}(\phi_0)\phi_2 \nonumber \\
 &- f_{\mathrm{dw}}^{\prime\prime\prime}(\phi_0)\frac{\phi_1^2}{2} -  a_1[f_s(c_{s,0}) - f_l(c_{l,0}) - \mu_0(c_{s,0} - c_{l,0})]h^{\prime\prime}(\phi_0)\phi_1 \nonumber \\
 & + a_1\mu_1(c_{s,0}-c_{l,0})h^{\prime}(\phi_0) 
  -a_1(\tilde{c}_{s,0}-\tilde{c}_{l,0})a(\phi_0)\frac{\partial\mu_1}{\partial\eta}\frac{1}{|\frac{\partial\phi_0}{\partial\eta}|} 
\end{align}
Multiplying both sides by $\frac{\partial\phi_0}{\partial\eta}$, rearranging and integrating within the limits $-\infty$ to $+\infty$ we obtain
\begin{align}\label{gb2}
 -(\alpha V_n + \kappa)\int_{-\infty}^{+\infty}\frac{\partial\phi_1}{\partial\eta}\frac{\partial\phi_0}{\partial\eta}\mathrm{d}\eta & - a_1\frac{V_n}{\frac{\partial c_{l,0}}{\partial\mu}}\int_{-\infty}^{+\infty}(\tilde{c}_{s,0}-\tilde{c}_{l,0})^2{\alpha_1}(\phi_0)\mathrm{d}\eta =  
  \int_{-\infty}^{+\infty}\frac{\partial^2\phi_2}{\partial\eta}\frac{\partial\phi_0}{\partial\eta} \nonumber \\
 & -\kappa^2\int_{-\infty}^{+\infty}\eta\Big(\frac{\partial\phi_0}{\partial\eta}\Big)^2 \mathrm{d}\eta -\int_{-\infty}^{+\infty}f_{\mathrm{dw}}^{\prime\prime}(\phi_0)\phi_2\frac{\partial\phi_0}{\partial\eta}\mathrm{d}\eta \nonumber \\
  &- \int_{-\infty}^{+\infty}f_{\mathrm{dw}}^{\prime\prime\prime}(\phi_0)\frac{\phi_1^2}{2}\frac{\partial\phi_0}{\partial\eta}\mathrm{d}\eta -  a_1[f_s(c_{s,0}) - f_l(c_{l,0}) - \mu_0(c_{s,0} - c_{l,0})] \nonumber \\
 &\times \int_{-\infty}^{+\infty}h^{\prime\prime}(\phi_0)\phi_1\frac{\partial\phi_0}{\partial\eta}\mathrm{d}\eta + a_1(c_{s,0}-c_{l,0})\int_{-\infty}^{+\infty}\mu_1h^{\prime}(\phi_0) \frac{\partial\phi_0}{\partial\eta}\mathrm{d}\eta \nonumber \\
 & + a_1(\tilde{c}_{s,0}-\tilde{c}_{l,0})\int_{-\infty}^{+\infty}a(\phi_0)\frac{\partial\mu_1}{\partial\eta}\mathrm{d}\eta 
\end{align}

Let $I_1 = \int_{-\infty}^{+\infty}\frac{\partial ^2{\phi_2}}{\partial\eta^2}\frac{\partial {\phi_0}}{\partial \eta} \mathrm{d}\eta$ which can be simplified by integration by parts (with the second term as the first function) as below,
\begin{equation}\label{gb3}
 I_1 = \int_{-\infty}^{+\infty}\frac{\partial^2{\phi_2}}{\partial \eta^2}\frac{\partial {\phi_0}}{\partial \eta}\mathrm{d}\eta = \frac{\partial{\phi_0}}{\partial \eta}\frac{\partial {\phi_2}}{\partial \eta} \Bigg|_{-\infty}^{+\infty} - \int_{-\infty}^{+\infty}\frac{\partial^2{\phi_0}}{\partial\eta^2}\frac{\partial {\phi_2}}{\partial \eta}\mathrm{d}\eta
\end{equation}
From the derivative matching condition, the first term goes to zero. The second integral can be further simplified by integration by parts (with the first term as the first function) as follows,
\begin{equation}\label{gb4}
I_1 = - \int_{-\infty}^{+\infty}\frac{\partial^2{\phi_0}}{\partial\eta^2}\frac{\partial {\phi_2}}{\partial \eta}\mathrm{d}\eta = -\Bigg[ \frac{\partial^2{\phi_0}}{\partial\eta^2}{\phi_2}\Bigg|_{-\infty}^{+\infty} - \int_{-\infty}^{+\infty}\frac{\partial^3{\phi_0}}{\partial\eta^3}{\phi_2}\mathrm{d}\eta\Bigg]
\end{equation}
Using Eq.~(\ref{sl1}) the first term can be re-written as,
\begin{equation}\label{gb5}
 I_1 = -\Bigg[ f^{\prime}_{\mathrm{dw}}({\phi_0}){\phi_2}\Bigg|_{-\infty}^{+\infty} - \int_{-\infty}^{+\infty}\frac{\partial^3{\phi_0}}{\partial\eta^3}{\phi_2}\mathrm{d}\eta\Bigg] = \int_{-\infty}^{+\infty}\frac{\partial^3{\phi_0}}{\partial\eta^3}{\phi_2}\mathrm{d}\eta .
\end{equation}
We next simplify the third integral on the RHS of Eq.~(\ref{gb2}) as follows,
\begin{equation}\label{gb6}
I_2 = \int_{-\infty}^{+\infty}f^{\prime\prime}_{\mathrm{dw}}({\phi_0}){\phi_2}\frac{\partial {\phi_0}}{\partial \eta}\mathrm{d}\eta = \int_{-\infty}^{+\infty}\frac{\partial}{\partial \eta}f^{\prime}_{\mathrm{dw}}({\phi_0}){\phi_2}\mathrm{d}\eta
\end{equation}
Therefore, the first and the third integrals on the right hand side in Eq.~(\ref{gb2}) simplifies as,
\begin{equation}\label{gb7}
 I_1 - I_2 = \int_{-\infty}^{+\infty} \Bigg[\frac{\partial^3{\phi_0}}{\partial\eta^3}{\phi_2} - \frac{\partial}{\partial \eta}f^{\prime}_{\mathrm{dw}}({\phi_0}){\phi_2}\Bigg] \mathrm{d}\eta = \int_{-\infty}^{+\infty} \frac{\partial}{\partial \eta}\Bigg[\frac{\partial^2{\phi_0}}{\partial\eta^2} - f^{\prime}_{\mathrm{dw}}({\phi_0}) \Bigg]{\phi_2} \mathrm{d}\eta = 0
\end{equation}
where we have utilized Eq.~(\ref{sl1}). To simplify the remaining terms of Eq.~(\ref{gb2}) we utilize the symmetry properties of the involved functions. For this purpose we first examine a re-arranged form of Eq.~(\ref{gt1})
\begin{equation}\label{gb8}
    L{\phi^1} = -(\alpha V_n + \kappa)\frac{\partial {\phi_0}}{\partial \eta} + a_1[f_s(c_{s,0}) - f_{l}(c_{l,0})-\mu_0(c_{s,0}-c_{l,0})]h^{\prime}({\phi_0})
\end{equation}
where the operator $L = \Big[\frac{\partial^2}{\partial \eta^2} - f^{\prime\prime}_{\mathrm{dw}}({\phi_0})\Big]$. From the leading order solution, ${\phi_0}$ is a sum of a constant and an odd function. Therefore it's derivative $\frac{\partial \hat{\phi^0}}{\partial \eta}$ is an even function. Similarly, for $h(\hat{\phi^0})$ we have
\begin{equation}\label{gb9}
    h(\hat{\phi^0}(\eta)) = 1 - h(\hat{\phi^0}(-\eta))
\end{equation}
Differentiating with respect to $\eta$,
\begin{equation}\label{gb10}
    \frac{\partial h({\phi_0}(\eta))}{\partial \eta} = -\frac{\partial h({\phi_0}(-\eta))}{\partial \eta}
\end{equation}
Employing chain-rule of differentiation,
\begin{equation}\label{gb11}
    \frac{\partial h({\phi_0}(\eta))}{\partial {\phi_0}}\frac{\partial {\phi_0}}{\partial \eta} = -\frac{\partial h({\phi_0}(-\eta))}{\partial \phi_0(-\eta)}\frac{\partial {\phi_0(-\eta)}}{\partial (-\eta)}\frac{\partial (-\eta)}{\partial \eta} .
\end{equation}
Using even property of $\frac{\partial\phi_0}{\partial\eta}$ we have
\begin{equation}\label{gb12}
   \frac{\partial h({\phi_0}(\eta))}{\partial {\phi_0}} =  \frac{\partial h({\phi_0}(-\eta))}{\partial \phi_0(-\eta)}
\end{equation}
which proves $h^{\prime}(\phi_0)$ is an even function. Since a constant multiplied by an even function yields an even function both terms on the RHS of Eq.~(\ref{gb8}) turn out to be even functions. Therefore, the LHS of Eq.~(\ref{gb8}) should also be even. This is only possible if ${\phi_1}$ is even, so that first derivative $\frac{\partial{\phi_1}}{\partial \eta}$ is odd and the second derivative $\frac{\partial ^2{\phi_1}}{\partial \eta^2}$ is even. As for the second term on the LHS, we next show that $f^{\prime\prime}_{\mathrm{dw}}$ is even. Assuming $f_{dw}({\phi_0}) = {{\phi_0}}^2{(1-{\phi_0})}^2$, $f^{\prime\prime}_{\mathrm{dw}}({\phi_0}) = 12{{\phi_0}}^2 - 12{\phi_0} + 2$. Therefore,
\begin{equation}\label{gb13}
    f^{\prime\prime}_{\mathrm{dw}}({\phi_0}(-\eta)) = 12{{\phi_0}(-\eta)}^2 - 12{\phi_0}(-\eta) + 2 . 
\end{equation}
Using anti-symmetric property of $\phi_0$ profile,
\begin{equation}\label{gb14}
   f^{\prime\prime}_{\mathrm{dw}}({\phi_0}(-\eta)) = 12{(1-{\phi_0}(\eta))}^2 - 12(1-{\phi_0}(\eta)) + 2 = 12{{\phi_0}(\eta)}^2 - 12{\phi_0}(\eta) + 2 = f^{\prime\prime}_{\mathrm{dw}}({\phi_0}(\eta))
\end{equation}
Therefore, $f^{\prime\prime}_{\mathrm{dw}}({\phi_0}){\phi_1}$ is also even.

Returning back to Eq.~(\ref{gb2}), we again examine the nature of the integrands in each term
\begin{equation}\label{gb15}
 -(\alpha V_n + \kappa) \int_{-\infty}^{+\infty}\frac{\partial {\phi_1}}{\partial \eta}\frac{\partial {\phi_0}}{\partial \eta} \mathrm{d}\eta = -(\alpha V_n + \kappa)\int_{-\infty}^{+\infty} (\mathrm{Odd} \times \mathrm{Even})  = -(\alpha V_n + \kappa)\int_{-\infty}^{+\infty} \mathrm{Odd} = 0
\end{equation}

For the fourth integral on the RHS in Eq.~(\ref{gb2}) we need to determine the property of $f^{\prime\prime\prime}_{\mathrm{dw}}({\phi_0})$. If we assume $f_{\mathrm{dw}}({\phi_0}) = {{\phi_0}}^2{(1-{\phi_0})}^2$, $f^{\prime\prime\prime}_{\mathrm{dw}}({\phi_0}) = 24{\phi_0} - 12$. Thus,
\begin{equation}\label{gb16}
 f^{\prime\prime\prime}_{\mathrm{dw}}({\phi_0}(-\eta)) = 24{\phi_0}(-\eta) - 12
\end{equation}
Using anti-symmetric property of $\phi_0$ profile,
\begin{equation}\label{gb17}
 f^{\prime\prime\prime}_{\mathrm{dw}}({\phi_0}(-\eta)) = 24(1 - {\phi_0}(\eta)) - 12 = 12 - 24{\phi_0}(\eta) = -f^{\prime\prime\prime}_{\mathrm{dw}}({\phi_0}(\eta))
\end{equation}
which proves $f^{\prime\prime\prime}_{\mathrm{dw}}({\phi_0})$ is odd. Therefore, the fourth integral on the RHS in Eq.~(\ref{gb2}) reads as
\begin{equation}\label{gb18}
  \int_{-\infty}^{+\infty} f^{\prime\prime\prime}_{\mathrm{dw}}({\phi_0})\frac{{\phi_1}^2}{2}\frac{\partial {\phi_0}}{\partial \eta} \mathrm{d}\eta = \int_{-\infty}^{+\infty} (\mathrm{Odd} \times \mathrm{Even} \times \mathrm{Even}) = \int_{-\infty}^{+\infty} \mathrm{Odd} = 0.
\end{equation}
 We next examine the fifth integral on the RHS in Eq.~(\ref{gb2}),
\begin{equation}\label{gb19}
   \int_{-\infty}^{+\infty}h^{\prime\prime}({\phi_0}){\phi_1} \frac{\partial{\phi_0}}{\partial \eta} \mathrm{d}\eta = \int_{-\infty}^{+\infty} (\mathrm{Odd} \times \mathrm{Even} \times \mathrm{Even}) = \int_{-\infty}^{+\infty} \mathrm{Odd} = 0
\end{equation}
where we have utilized the fact that since $h^{\prime}({\phi_0})$ is even, as proved in Eq.~(\ref{gb12}), $h^{\prime\prime}({\phi_0})$ is odd. Evaluating the second integral on the RHS in Eq.~(\ref{gb2}) ,
\begin{equation}\label{gb20}
   \int_{-\infty}^{+\infty}\kappa^2\eta\Bigg(\frac{\partial {\phi_0}}{\partial \eta}\Bigg)^2 \mathrm{d}\eta = \int_{-\infty}^{+\infty} (\mathrm{Odd} \times \mathrm{Even}) = \int_{-\infty}^{+\infty} \mathrm{Odd} = 0 .
\end{equation}
The remaining non-zero terms in Eq.~(\ref{gb2}) writes as
\begin{align}\label{gb21}
   \frac{V_n}{\partial c_{l,0}/\partial\mu} \int_{-\infty}^{+\infty}(\tilde{c}_{s,0}-\tilde{c}_{l,0})^2{\alpha_1}(\phi_0) = & (c_{s,0}-c_{l,0})\int_{-\infty}^{+\infty}\mu_1 h^{\prime}(\phi_0)\frac{\partial\phi_0}{\partial\eta} \mathrm{d}\eta \nonumber \\ 
   & +\int_{-\infty}^{+\infty}(\tilde{c}_{s,0}-\tilde{c}_{l,0})a(\phi_0)\frac{\partial\mu_1}{\partial\eta}\mathrm{d}\eta
\end{align}
The expressions for $\mu_1$ and $\frac{\partial\mu_1}{\partial\eta}$ incorporating the choice of source terms $S_s=S_l= -A(c_{s,0}-c_{l,0})$ can be written as
\begin{equation}\label{gb22}
    \mu_1 = -\frac{V_n}{\partial c_{l,0}/\partial\mu}(c_{s,0}-c_{l,0})\int_{0}^{\eta}\frac{h(\phi_0)-1-(1-A)a(\phi_0)}{q_n(\phi)}\mathrm{d}\eta + q_n^-\frac{\partial\mu_0}{\partial r}\Bigg|^-\int_{0}^{\eta}\frac{1}{q_n(\phi_0)}\mathrm{d}\eta + \overline{\mu_1}(s)
\end{equation}
and
\begin{equation}\label{gb23}
\frac{\partial\mu_1}{\partial \eta} = -\frac{V_n}{\partial c_{l,0}/\partial\mu}(c_{s,0}-c_{l,0})\frac{h(\phi_0)-1-(1-A)a(\phi)}{q_n(\phi_0)} +  q_n^-\frac{\partial\mu_0}{\partial r}\Bigg|^-\frac{1}{q_n(\phi_0)}   
\end{equation}
Introducing the notation 
\begin{equation}\label{gb24}
    \tilde{p}(\phi_0)= \frac{h(\phi_0)-1-(1-A)a(\phi_0)}{q_n(\phi)},
\end{equation}
and substituting Eqs.~(\ref{gb22}) and (\ref{gb23}) in Eq.~(\ref{gb21}) we obtain
\begin{align}\label{gb25}
 -(1-A)^2(c_{s,0}-c_{l,0})^2\frac{V_n}{\partial c_{l,0}/\partial\mu}&\int_{-\infty}^{+\infty}{\alpha_1}(\phi_0)\mathrm{d}\eta = \nonumber \\
 & -\frac{V_n}{\partial c_{l,0}/\partial\mu}(c_{s,0}-c_{l,0})^2\int_{-\infty}^{+\infty}\Big[\int_{0}^{\eta}\tilde{p}(\phi_0)\mathrm{d}\eta^{\prime}\Big]h^{\prime}(\phi_0)\frac{\partial\phi_0}{\partial\eta}\mathrm{d}\eta \nonumber \\
 & + (c_{s,0}-c_{l,0})q_n^-\frac{\partial\mu_0}{\partial r}\Big|^- \int_{-\infty}^{+\infty}\Big[\int_{0}^{\eta}\frac{1}{q_n(\phi_0)}\mathrm{d}\eta^{\prime}\Big]h^{\prime}(\phi_0)\frac{\partial\phi_0}{\partial\eta}\mathrm{d}\eta\nonumber \\
 & + (c_{s,0}-c_{l,0})\overline{\mu}_1(s)\int_{-\infty}^{+\infty} h^{\prime}(\phi_0)\frac{\partial\phi_0}{\partial\eta}\mathrm{d}\eta \nonumber \\
 &- (1-A)(c_{s,0}-c_{l,0})^2\frac{V_n}{\partial c_{l,0}/\partial\mu}\int_{-\infty}^{+\infty}a(\phi_0)\tilde{p}(\phi_0)\mathrm{d}\eta \nonumber \\
 & +(1-A)(c_{s,0}-c_{l,0})q_n^-\frac{\partial\mu_0}{\partial r}\Big|^- \int_{-\infty}^{+\infty} \frac{a(\phi_0)}{q_n(\phi_0)}\mathrm{d}\eta \nonumber \\
\end{align}
Solving for $\overline{\mu}_1(s)$ we obtain
\begin{eqnarray}\label{gb26}
 \overline{\mu}_1(s) = \frac{V_n}{\partial c_{l,0}/\partial\mu}(c_{s,0}-c_{l,0})\Big[(1-A)^2 N - M - (1-A)P\Big] + q_n^-\frac{\partial\mu_0}{\partial r}\Big|^-\Big[Q + (1-A)H\Big].\nonumber \\
\end{eqnarray}
where,
\begin{equation}\label{gb27}
    N = \int_{-\infty}^{+\infty}{\alpha_1}(\phi_0)\mathrm{d}\eta ,
\end{equation}
\begin{equation}\label{gb28}
    M = \int_{-\infty}^{+\infty}\Big[\int_{0}^{\eta}\tilde{p}(\phi_0)\mathrm{d}\eta^{\prime}\Big]h^{\prime}(\phi_0)\frac{\partial\phi_0}{\partial\eta}\mathrm{d}\eta ,
\end{equation}
\begin{equation}\label{gb29}
    P = \int_{-\infty}^{+\infty}a(\phi_0)\tilde{p}(\phi_0)\mathrm{d}\eta ,
\end{equation}
\begin{equation}\label{gb30}
    Q = \int_{-\infty}^{+\infty}\Big[\int_{0}^{\eta}\frac{1}{q_n(\phi_0)}\mathrm{d}\eta^{\prime}\Big]h^{\prime}(\phi_0)\frac{\partial\phi_0}{\partial\eta}\mathrm{d}\eta
\end{equation}
\begin{equation}\label{gb31}
    H = \int_{-\infty}^{+\infty} \frac{a(\phi_0)}{q_n(\phi_0)}\mathrm{d}\eta .
\end{equation}
Therefore, the first order correction to the diffusion potential writes as
\begin{equation}\label{gb32}
 \mu_1|^{\pm} =  -\frac{V_n}{\partial c_{l,0}/\partial\mu}(c_{s,0}-c_{l,0})\Big[F^{\pm} + M - (1-A)^2 N  + (1-A)P\Big] + q_n^-\frac{\partial\mu_0}{\partial r}\Big|^-\Big[G^{\pm} + Q + (1-A)H\Big]. 
\end{equation}
Since the choice of $q_n(\phi_0)$ ensures $G^+ = G^-$ and it can be further shown that $G+Q+H = 0$, Eq.~(\ref{gb32}) can be further simplified as
\begin{equation}\label{gb33}
   \mu_1|^{\pm} =  -\frac{V_n}{\partial c_{l,0}/\partial\mu}(c_{s,0}-c_{l,0})\Big[F^{\pm} + M - (1-A)^2 N  + (1-A)P\Big] - q_n^-\frac{\partial\mu_0}{\partial r}\Big|^- AH. 
\end{equation}
It is to be noted that $H$ can also be expressed as $H = F^+ - F^-$. The diffusion potential upto first order can then be written as
\begin{equation}\label{gb34}
    \mu|^{\pm} = \mu_0 + \epsilon\mu_1|^{\pm}
\end{equation}
Substituting Eqs.~(\ref{gt10}) and (\ref{gb33}) in the above equation we obtain
\begin{align}\label{gb35}
     \mu|^{\pm} = \frac{f_s(c_{s,0}) - f_l(c_{l,0})}{c_{s,0}-c_{l,0}} + \frac{\kappa}{c_{s,0}-c_{l,0}} & + \frac{V_n}{c_{s,0}-c_{l,0}}\Bigg\{ \alpha - \epsilon\frac{(c_{s,0}-c_{l,0})^2}{\partial c_{l,0}/\partial\mu}\Big[F^{\pm} + M - (1-A)^2 N \nonumber \\
     & + (1-A)P\Big]\Bigg\}- \epsilon q_n^-\frac{\partial\mu_0}{\partial r}\Big|^- AH 
\end{align}
For the second set of interpolation functions, the integrals $F$, $M$, $N$, $P$ and $H$ can be redefined.

\subsection{$\mathcal{O}$($\epsilon^0$) in $c$ : Modified Mass Conservation Equation}
The $\mathcal{O}(\epsilon^0)$ terms in $c$ equation writes as
\begin{align}\label{mc1}
 -V_n\frac{\partial c_1}{\partial \eta}  = & \frac{\partial}{\partial \eta}\Big[\frac{\partial c_{l,0}}{\partial \mu} q_n(\phi_0)\frac{\partial \mu_2}{\partial \eta}\Big] + \frac{\partial}{\partial\eta}\Big[\frac{\partial c_{l,0}}{\partial \mu}q_n^{\prime}(\phi_0)\phi_1\frac{\partial\mu_1}{\partial\eta}\Big] + \kappa\frac{\partial c_{l,0}}{\partial \mu} q_n(\phi_0)\frac{\partial \mu_1}{\partial\eta}  \nonumber \\
 & + \kappa\frac{\partial c_{l,0}}{\partial \mu} q_n^{\prime}(\phi_0)\phi_1\frac{\partial\mu_0}{\partial\eta}c- \kappa^2\eta \frac{\partial c_{l,0}}{\partial \mu}q_n(\phi_0)\frac{\partial\mu_0}{\partial\eta} 
 + \frac{\partial}{\partial s}\Big[\frac{\partial c_{l,0}}{\partial \mu}q_t(\phi_0)\frac{\partial\mu_0}{\partial s}\Big]  \nonumber \\
 & + \frac{\partial}{\partial\eta}\Big[a^{\prime}(\phi_0)\phi_1 (\tilde{c}_{s,0} - \tilde{c}_{l,0})V_n\Big] + \frac{\partial}{\partial\eta}\Big[a(\phi_0)(\tilde{c}_{s,0}-\tilde{c}_{l,0})\frac{V_n}{\frac{\partial\phi_0}{\partial\eta}}\frac{\partial\phi_1}{\partial\eta}\Big] \nonumber \\
 & +\frac{\partial}{\partial\eta}\Big[a(\phi_0)(\tilde{c}_{s,1}-\tilde{c}_{l,1})V_n\Big]  
\end{align}
Using the leading order solution $\frac{\partial\mu_0}{\partial\eta} = 0$ and integrating the above equation we obtain
\begin{align}\label{mc2}
 -V_n c_1 = & \frac{\partial c_{l,0}}{\partial \mu}q_n(\phi_0)\frac{\partial \mu_2}{\partial \eta} + \frac{\partial c_{l,0}}{\partial \mu}q_n^{\prime}(\phi_0)\phi_1\frac{\partial\mu_1}{\partial\eta} + \frac{\partial c_{l,0}}{\partial \mu}\kappa \int_{0}^{\eta} q_n(\phi_0)\frac{\partial \mu_1}{\partial\eta}\mathrm{d}\eta  \nonumber \\
 & + \frac{\partial c_{l,0}}{\partial \mu}\frac{\partial ^2 \mu_0}{\partial s^2}\int_{0}^{\eta}q_t(\phi_0)\mathrm{d}\eta + a^{\prime}(\phi_0)\phi_1 (\tilde{c}_{s,0} - \tilde{c}_{l,0})V_n + a(\phi_0)(\tilde{c}_{s,0}-\tilde{c}_{l,0})\frac{V_n}{\frac{\partial\phi_0}{\partial\eta}}\frac{\partial\phi_1}{\partial\eta}  \nonumber \\
 & + a(\phi_0)(\tilde{c}_{s,1}-\tilde{c}_{l,1})V_n + \kappa V_n \int_{0}^{\eta} a(\phi_0)(\tilde{c}_{s,0}-\tilde{c}_{l,0}) + A_3(s)
\end{align}
Substituting the value of $\frac{\partial \mu_1}{\partial\eta}$ from the next-to-leading order solution in the third term of the right hand side of the above expression
\begin{align}\label{mc3}
 -V_n c_1 = & \frac{\partial c_{l,0}}{\partial \mu}q_n(\phi_0)\frac{\partial \mu_2}{\partial \eta} + \frac{\partial c_{l,0}}{\partial \mu}q_n^{\prime}(\phi_0)\phi_1\frac{\partial\mu_1}{\partial\eta} - V_n \kappa\int_{0}^{\eta} (c_o - c_{s,0})\mathrm{d}\eta  \nonumber \\
 & - V_n \kappa \int _{0}^{\eta} (\tilde{c}_{s,0} - \tilde{c}_{l,0})a(\phi_0)\mathrm{d}\eta + \eta \kappa \frac{\partial c_{l,0}}{\partial \mu}q_n^- \frac{\partial \mu_0}{\partial r}\Big|^- + \frac{\partial c_{l,0}}{\partial \mu}\frac{\partial ^2 \mu_0}{\partial s^2}\int_{0}^{\eta}q_t(\phi_0)\mathrm{d}\eta   \nonumber \\
 & + a^{\prime}(\phi_0)\phi_1 (\tilde{c}_{s,0} - \tilde{c}_{l,0})V_n + a(\phi_0)(\tilde{c}_{s,0}-\tilde{c}_{l,0})\frac{V_n}{\frac{\partial\phi_0}{\partial\eta}}\frac{\partial\phi_1}{\partial\eta} \nonumber \\
  & + a(\phi_0)(\tilde{c}_{s,1}-\tilde{c}_{l,1})V_n + \kappa V_n \int_{0}^{\eta} a(\phi_0)(\tilde{c}_{s,0}-\tilde{c}_{l,0}) + A_3(s)
\end{align}
Cancelling the fourth and the tenth term on the right side of the above expression we have
\begin{align}\label{mc4}
 -V_n c_1 = & \frac{\partial c_{l,0}}{\partial \mu}q_n(\phi_0)\frac{\partial \mu_2}{\partial \eta} + \frac{\partial c_{l,0}}{\partial \mu}q_n^{\prime}(\phi_0)\phi_1\frac{\partial\mu_1}{\partial\eta} - V_n \kappa\int_{0}^{\eta} (c_o - c_{s,0})\mathrm{d}\eta + \eta \kappa \frac{\partial c_{l,0}}{\partial \mu}q_n^- \frac{\partial \mu_0}{\partial r}\Big|^- \nonumber \\
 & + \frac{\partial c_{l,0}}{\partial \mu}\frac{\partial ^2 \mu_0}{\partial s^2}\int_{0}^{\eta}q_t(\phi_0)\mathrm{d}\eta 
 + a^{\prime}(\phi_0)\phi_1 (\tilde{c}_{s,0} - \tilde{c}_{l,0})V_n + a(\phi_0)(\tilde{c}_{s,0}-\tilde{c}_{l,0})\frac{V_n}{\frac{\partial\phi_0}{\partial\eta}}\frac{\partial\phi_1}{\partial\eta} \nonumber \\
 & + a(\phi_0)(\tilde{c}_{s,1}-\tilde{c}_{l,1})V_n + A_3(s) 
\end{align}
Splitting the above equation into two parts one for $\eta > 0$ and the other $\eta < 0$.
\\
(1) For $\eta > 0$
\begin{align}\label{mc5}
 -V_n c_1 = & \frac{\partial c_{l,0}}{\partial \mu}q_n(\phi_0)\frac{\partial \mu_2}{\partial \eta} + \frac{\partial c_{l,0}}{\partial \mu}q_n^{\prime}(\phi_0)\phi_1\frac{\partial\mu_1}{\partial\eta} - V_n \kappa \int_{0}^{\eta}\Big[ (c_0 - c_{s,0}) - (c_{l,0}-c_{s,0})\Big]\mathrm{d}\eta  \nonumber \\
  & - \eta V_n \kappa (c_{l,0}-c_{s,0}) + \eta \kappa \frac{\partial c_{l,0}}{\partial \mu}q_n^- \frac{\partial \mu_0}{\partial r}\Big|^- + \frac{\partial^2\mu_0}{\partial s^2}\int_{0}^{\eta}\Big[q_t(\phi_0) - q_t(\phi_o|^+)\Big]\mathrm{d}\eta \nonumber \\
 & + \eta \frac{\partial c_{l,0}}{\partial \mu}\frac{\partial^2\mu_0}{\partial s^2}q_t(\phi_o|^+) + a^{\prime}(\phi_0)\phi_1 (\tilde{c}_{s,0} - \tilde{c}_{l,0})V_n + a(\phi_0)(\tilde{c}_{s,0}-\tilde{c}_{l,0})\frac{V_n}{\frac{\partial\phi_0}{\partial\eta}}\frac{\partial\phi_1}{\partial\eta} \nonumber \\
 & + a(\phi_0)(\tilde{c}_{s,1}-\tilde{c}_{l,1})V_n + A_3(s)
\end{align}
(2) For $\eta < 0$
\begin{align}\label{mc6}
 -V_n c_1 = & \frac{\partial c_{l,0}}{\partial \mu}q_n(\phi_0)\frac{\partial \mu_2}{\partial \eta} + \frac{\partial c_{l,0}}{\partial \mu}q_n^{\prime}(\phi_0)\phi_1\frac{\partial\mu_1}{\partial\eta} - V_n \kappa \int_{0}^{\eta} (c_0 - c_{s,0})\mathrm{d}\eta + \eta \kappa \frac{\partial c_{l,0}}{\partial \mu}q_n^- \frac{\partial \mu_0}{\partial r}\Big|^- \nonumber \\
 & + \frac{\partial c_{l,0}}{\partial \mu}\frac{\partial^2\mu_0}{\partial s^2}\int_{0}^{\eta}\Big[q_t(\phi_0) - q_t(\phi_o|^+)\Big]\mathrm{d}\eta 
 +  \eta \frac{\partial c_{l,0}}{\partial \mu}\frac{\partial^2\mu_0}{\partial s^2}q_t(\phi_o|^+)  \nonumber \\
 &+ a^{\prime}(\phi_0)\phi_1 (\tilde{c}_{s,0} - \tilde{c}_{l,0})V_n + a(\phi_0)(\tilde{c}_{s,0}-\tilde{c}_{l,0})\frac{V_n}{\frac{\partial\phi_0}{\partial\eta}}\frac{\partial\phi_1}{\partial\eta} + a(\phi_0)(\tilde{c}_{s,1}-\tilde{c}_{l,1})V_n + A_3(s) \nonumber \\
\end{align}
Taking the far field limit $\eta \rightarrow +\infty$ in Eq.~(\ref{mc5}) and employing the matching conditions we have
\begin{equation}\label{mc7}
    \lim_{\eta\rightarrow +\infty} - V_n c_1 = -V_n \Big( c_1|^+ + \eta\frac{\partial c_0}{\partial r}\Bigg|^+\Big)
\end{equation}
\begin{equation}\label{mc8}
    \lim_{\eta\rightarrow +\infty}q_n(\phi_0)\frac{\partial \mu_2}{\partial \eta}  = q_n^+ \Bigg[\frac{\partial \mu_1}{\partial r}\Bigg|^+ + \eta \frac{\partial ^2 \mu_0}{\partial r^2}\Bigg|^+\Bigg]
\end{equation}
In addition we have $\lim_{\eta\rightarrow +\infty} q_n^{\prime}(\phi_0), a(\phi_0), a^{\prime}(\phi_0) = 0$ such that the second, seventh, eighth and ninth term in Eq.~(\ref{mc5}) is equal to zero. Substituting back in Eq.~(\ref{mc5})
\begin{align}\label{mc9}
 -V_n \Big( c_1|^+ + \eta\frac{\partial c_0}{\partial r}\Bigg|^+\Big) = & \frac{\partial c_{l,0}}{\partial \mu}q_n^+ \Bigg[\frac{\partial \mu_1}{\partial r}\Bigg|^+ + \eta \frac{\partial ^2 \mu_0}{\partial r^2}\Bigg|^+\Bigg] - V_n \kappa \int_{0}^{+\infty}\Big[ (c_0 - c_{s,0}) - (c_{l,0}-c_{s,0})\Big]\mathrm{d}\eta \nonumber \\
 - & \eta V_n \kappa (c_{l,0}-c_{s,0}) + \eta \kappa \frac{\partial c_{l,0}}{\partial \mu}q_n^- \frac{\partial \mu_0}{\partial r}\Big|^- \nonumber \\
 &  + \frac{\partial c_{l,0}}{\partial \mu}\frac{\partial^2\mu_0}{\partial s^2}\int_{0}^{+\infty}\Big[q_t(\phi_0) - q_t(\phi_o|^+)\Big]\mathrm{d}\eta + \frac{\partial c_{l,0}}{\partial \mu}\eta \frac{\partial^2\mu_0}{\partial s^2}q_t(\phi_o|^+) \nonumber \\
 &+ A_3(s)
\end{align}
Comparing the coefficients of $\eta^0$ on both sides we obtain
\begin{align}\label{mc10}
 -V_n c_1|^+ = & \frac{\partial c_{l,0}}{\partial \mu}q_n^+ \frac{\partial \mu_1}{\partial r}\Bigg|^+ - V_n \kappa\int_{0}^{+\infty}\Big[ (c_0 - c_{s,0}) - (c_{l,0}-c_{s,0})\Big]\mathrm{d}\eta \nonumber \\
 & + \frac{\partial c_{l,0}}{\partial \mu}\frac{\partial^2\mu_0}{\partial s^2}\int_{0}^{+\infty}\Big[q_t(\phi_0) - q_t(\phi_o|^+)\Big]\mathrm{d}\eta + A_3(s)
\end{align}
Substituting the lowest order concentration profile i.e. $c_0 = c_{s,0}h(\phi_0) + c_{l,0}\{1-h(\phi_0)\}$ and re-arranging the equation we have
\begin{align}\label{mc11}
 q_n^+ \frac{\partial \mu_1}{\partial r}\Bigg|^+ = & -\frac{V_n}{\partial c_{l,0}/\partial \mu} c_1|^+ - \frac{V_n}{\partial c_{l,0}/\partial\mu} \kappa (c_{l,0}-c_{s,0})\int_{0}^{+\infty}h(\phi_0)\mathrm{d}\eta \nonumber \\
 & - \frac{\partial^2\mu_0}{\partial s^2}\int_{0}^{+\infty}\Big[q_t(\phi_0) - q_t(\phi_o|^+)\Big]\mathrm{d}\eta + A_3(s) 
\end{align}
Similarly taking the limit $\eta \rightarrow -\infty$ in Eq.~(\ref{mc6}) and comparing the coefficients of $\eta^0$ we obtain
\begin{align}\label{mc12}
 q_n^- \frac{\partial \mu_1}{\partial r}\Bigg|^- = & -\frac{V_n}{\partial c_{l,0}/\partial\mu} c_1|^- - \frac{V_n}{\partial c_{l,0}/\partial \mu} \kappa (c_{l,0}-c_{s,0})\int_{0}^{-\infty}\{1-h(\phi_0)\}\mathrm{d}\eta \nonumber \\
 & - \frac{\partial^2\mu_0}{\partial s^2}\int_{0}^{+\infty}\Big[q_t(\phi_0) - q_t(\phi_o|^-)\Big]\mathrm{d}\eta +A_3(s) 
\end{align}
Subtracting Eq.~(\ref{mc12}) from Eq.~(\ref{mc11}) we have
\begin{eqnarray}\label{mc13}
 q_n^+ \frac{\partial \mu_1}{\partial r}\Bigg|^+ - q_n^- \frac{\partial \mu_1}{\partial r}\Bigg|^- = -\frac{V_n}{\partial c_{l,0}/\partial\mu}(c_{l,1}-c_{s,1}) - \frac{V_n}{\partial c_{l,0}/\partial\mu} \kappa (H^+ - H^-) - \frac{\partial^2\mu_0}{\partial s^2} (S^+ - S^-) \nonumber \\
\end{eqnarray}
where,
\begin{equation}\label{mc14}
    H^{\pm} = \int_{0}^{\pm\infty} [h(\phi_0) - h(\phi_0|^{\pm})]\mathrm{d}\eta,
\end{equation}
and 
\begin{equation}\label{mc15}
    S^{\pm} = \int_{0}^{\pm\infty}[q_t(\phi_0) - q_t(\phi_0|^{\pm})]\mathrm{d}\eta.
\end{equation}
The second and the third term on the right hand side of Eq.~(\ref{mc13}) corresponds to the correction to the mass conservation because of the presence of interface stretching and surface diffusion. From Eqs.~(\ref{mc14}) and (\ref{mc15}) both these effects can be eliminated if $h(\phi_0)$ and $q_t(\phi_0)$ are anti-symmetric. Therefore, the first order mass conservation reads as
\begin{equation}\label{mc16}
    q_n^+ \frac{\partial \mu_1}{\partial r}\Bigg|^+ - q_n^- \frac{\partial \mu_1}{\partial r}\Bigg|^- = -\frac{V_n}{\partial c_{l,0}/\partial\mu}(c_{l,1}-c_{s,1}) .
\end{equation}
The overall mass conservation reads as
\begin{align}
 q_n^+\frac{\partial\mu}{\partial r}\Bigg|^+ - q_n^-\frac{\partial\mu}{\partial r}\Bigg|^- = & q_n^+ \frac{\partial}{\partial r}(\mu_0 + \epsilon\mu_1|^+) - q_n^- \frac{\partial}{\partial r}(\mu_0 + \epsilon\mu_1^-) \nonumber \\
  = & \Bigg(q_n^+\frac{\partial\mu_0}{\partial r}\Big|^+ - q_n^-\frac{\partial\mu_0}{\partial r}\Big|^-\Bigg) + \epsilon \Bigg(q_n^+\frac{\partial\mu_1}{\partial r}\Big|^+ - q_n^-\frac{\partial\mu_1}{\partial r}\Big|^-\Bigg)
\end{align}
Substituting Eqs.~(\ref{st11}) and (\ref{mc16}) we obtain
\begin{align}
 q_n^+\frac{\partial\mu}{\partial r}\Bigg|^+ - q_n^-\frac{\partial\mu}{\partial r}\Bigg|^- = & -\frac{V_n}{\partial c_{l,0}/\partial\mu}(c_{l,0}-c_{s,0}) - \epsilon \frac{V_n}{\partial c_{l,0}/\partial\mu}(c_{l,1}-c_{s,1}) \nonumber \\
  = & -\frac{V_n}{\partial c_{l,0}/\partial\mu}\Big[(c_{l,0} + \epsilon c_{l,1}) - (c_{s,0}+\epsilon c_{l,1})\Big]
\end{align}

\section{Asymptotics of the model with $\mu_s \neq \mu_l$}
We proceed similarly as in the preceeding section and work out the asymptotics by assuming $g(\phi) = h(\phi)$.
\subsection{$\mathcal{O}(\epsilon^0)$ in $\phi$ : Equilibrium profile}
The leading order terms in $\phi$ equation writes as
\begin{equation}\label{eq2_1}
    \frac{\chi\frac{D_s}{D_l}\{1-h(\phi_0)\}}{\chi\frac{D_s}{D_l}\{1-h(\phi_0)\} + h(\phi_0)}\Big[\frac{\partial^2\phi_0}{\partial\eta^2} - f_{\mathrm{dw}}^{\prime}(\phi_0)\Big] + \frac{h(\phi_0)}{\chi\frac{D_s}{D_l}\{1-h(\phi_0)\} + h(\phi_0)}\Big[\frac{\partial^2\phi_0}{\partial\eta^2} - f_{\mathrm{dw}}^{\prime}(\phi_0)\Big] = 0.
\end{equation}
where $\chi = \partial c_s/\partial \mu_s\Big/\partial c_l/\partial\mu_l$. The above equation can be simplified and integrated to obtain the equilibrium phase-field profile as
\begin{equation}
    \phi_0(\eta) = \frac{1}{2}\Bigg[1-\tanh\left(\frac{\eta}{\sqrt{2}}\right)\Bigg]
\end{equation}
\subsection{$\mathcal{O}(1/\epsilon^2)$ in $c$ : Equality of diffusion potential}
At the leading order in the $c$ equation we have
\begin{equation}\label{dp2_1}
    \frac{\partial}{\partial \eta}\Bigg[\frac{\partial c_{l,0}}{\partial\mu_l}q_n(\phi_0)\frac{\partial}{\partial\eta}\Big\{\mu_{s,0}h(\phi_0)+\mu_{l,0}\{1-h(\phi_0)\}\Big\}\Bigg] = 0.
\end{equation}
Integrating once,
\begin{equation}\label{dp2_2}
    \frac{\partial c_{l,0}}{\partial\mu_l}q_n(\phi_0)\frac{\partial}{\partial\eta}\Big\{\mu_{s,0}h(\phi_0)+\mu_{l,0}\{1-h(\phi_0)\}\Big\} = A_1(s)
\end{equation}
Taking the limit $\eta \rightarrow \pm\infty$ and employing the matching condition $\lim_{\eta \rightarrow -\infty}\frac{\partial\mu_{s,0}}{\partial\eta} = 0$ and $\lim_{\eta \rightarrow +\infty}\frac{\partial\mu_{l,0}}{\partial\eta} = 0$ we obtain $A_1(s) = 0$. Since $q_n(\phi_0)$ and $\frac{\partial c_{l,0}}{\partial \mu_l}$ are finite in the inner region, Eq.~(\ref{dp2_2}) reduces to
\begin{equation}\label{dp2_3}
   \frac{\partial}{\partial\eta}\Big\{\mu_{s,0}h(\phi_0)+\mu_{l,0}\{1-h(\phi_0)\}\Big\} = 0 .
\end{equation}
Integrating again we have
\begin{equation}\label{dp2_4}
    \mu_{s,0}h(\phi_0)+\mu_{l,0}\{1-h(\phi_0)\} = \overline{\mu}_0(s),
\end{equation}
where $\overline{\mu}_0(s)$ is a constant of integration only dependent on the arc length. Taking $\lim_{\eta \rightarrow -\infty}$
\begin{equation}\label{dp2_5}
    \lim_{\eta \rightarrow -\infty} \mu_{s,0} = \overline{\mu}_0(s),
\end{equation}
which implies
\begin{equation}\label{dp2_6}
    \mu_{s,0}|^- = \frac{\partial f_s(c_{s,0}|^-)}{\partial c_s} =  \overline{\mu}_0(s).
\end{equation}
Similarly taking the $\lim_{\eta \rightarrow +\infty}$ we obtain
\begin{equation}\label{dp2_7}
    \mu_{l,0}|^+ = \frac{\partial f_l(c_{l,0}|^+)}{\partial c_l} =  \overline{\mu}_0(s).
\end{equation}
From Eqs.~(\ref{dp2_6}) and (\ref{dp2_7}) we have
\begin{equation}\label{dp2_8}
    \mu_{s,0}|^- = \mu_{l,0}|^+
\end{equation}
which implies the equality of diffusion potential at the interface which is similar to the lowest order solution of the model in which pointwise equality of the diffusion potential (i.e. $\mu_s \ \mu_l$) was assumed.

\subsection{$\mathcal{O}(\epsilon)$ in $\phi$}
\begin{align}\label{gb2_1}
    -\alpha V_n\frac{\partial\phi_0}{\partial\eta} & =  \frac{\partial^2\phi_1}{\partial\eta^2} + \kappa\frac{\partial\phi_0}{\partial\eta} - f_{\mathrm{dw}}^{\prime\prime}(\phi_0)\phi_1 - a_1 \frac{\chi\frac{D_s}{D_l}\{1-h(\phi_0)\}}{\chi\frac{D_s}{D_l}\{1-h(\phi_0)\} + h(\phi_0)}\Bigg[ f_s(c_{s,0}) - f_l(c_{l,0}) \nonumber \\
   & - \mu_{s,0}(c_{s,0}-c_{l,0}) - S_l\{1-h(\phi_0)\}(\mu_{s,0}-\mu_{l,0})\Bigg]h^{\prime}(\phi_0) - a_1\frac{h(\phi_0)}{\chi\frac{D_s}{D_l}\{1-h(\phi_0)\} + h(\phi_0)} \times \nonumber \\
   & \Bigg[f_s(c_{s,0}) - f_l(c_{l,0})- \mu_{l,0}(c_{s,0}-c_{l,0}) - S_s h(\phi_0)(\mu_{l,0}-\mu_{s,0})\Bigg]h^{\prime}(\phi_0) \nonumber \\
   & - a_1 \frac{(\tilde{c}_{s,0}-\tilde{c}_{l,0})}{|\frac{\partial\phi_0}{\partial\eta}|}\frac{h(\phi_0)\{1-h(\phi_0)\}}{\chi\frac{D_s}{D_l}\{1-h(\phi_0)\} + h(\phi_0)}\Big( \chi\frac{D_s}{D_l}\frac{\partial\mu_{s,0}}{\partial\eta} - \frac{\partial\mu_{l,0}}{\partial\eta}\Big).
\end{align}
Re-arranging the above equation
\begin{align}\label{gb2_2}
    -(\alpha V_n + \kappa)\frac{\partial\phi_0}{\partial\eta} = & \frac{\partial^2\phi_1}{\partial\eta^2} - f_{\mathrm{dw}}^{\prime\prime}(\phi_0)\phi_1 - a_1[f_s(c_{s,0}) - f_l(c_{l,0})]h^{\prime}(\phi_0) \nonumber \\
   & + a_1 \frac{\chi\frac{D_s}{D_l}\{1-h(\phi_0)\}}{\chi\frac{D_s}{D_l}\{1-h(\phi_0)\}+h(\phi_0)}\mu_{s,0}(c_{s,0}-c_{l,0})h^{\prime}(\phi_0) \nonumber \\
    & + a_1\frac{h(\phi_0)}{\chi\frac{D_s}{D_l}\{1-h(\phi_0)\}+h(\phi_0)}\mu_{l,0}(c_{s,0}-c_{l,0})h^{\prime}(\phi_0) \nonumber \\
    & - a_1 \frac{(c_{s,0}-c_{l,0})h(\phi_0)\{1-h(\phi_0)\}}{|\frac{\partial\phi_0}{\partial\eta}|\Big[\chi\frac{D_s}{D_l}\{1-h(\phi_0)\} + h(\phi_0)\Big]}\chi\frac{D_s}{D_l}\frac{\partial\mu_{s,0}}{\partial \eta} \nonumber \\
    & + a_1 \frac{(c_{s,0}-c_{l,0})h(\phi_0)\{1-h(\phi_0)\}}{|\frac{\partial\phi_0}{\partial\eta}|\Big[\chi\frac{D_s}{D_l}\{1-h(\phi_0)\} + h(\phi_0)\Big]}\frac{\partial\mu_{l,0}}{\partial \eta} \nonumber \\
    & + a_1 \frac{\chi\frac{D_s}{D_l}\{1-h(\phi_0)\}}{\chi\frac{D_s}{D_l}\{1-h(\phi_0)\}+h(\phi_0)}S_l\{1-h(\phi_0)\}(\mu_{s,0}-\mu_{l,0})h^{\prime}(\phi_0) \nonumber \\
    & + a_1\frac{h(\phi_0)}{\chi\frac{D_s}{D_l}\{1-h(\phi_0)\}+h(\phi_0)}S_sh(\phi_0)(\mu_{l,0}-\mu_{s,0})h^{\prime}(\phi_0) \nonumber \\
    & -a_1 \frac{S_sh(\phi_0)}{|\frac{\partial\phi_0}{\partial\eta}|}\frac{h(\phi_0)\{1-h(\phi_0)\}}{\chi\frac{D_s}{D_l}\{1-h(\phi_0)\} + h(\phi_0)} \chi\frac{D_s}{D_l}\frac{\partial\mu_{s,0}}{\partial\eta} \nonumber \\
    & -a_1 \frac{S_l\{1-h(\phi_0)\}}{|\frac{\partial\phi_0}{\partial\eta}|}\frac{h(\phi_0)\{1-h(\phi_0)\}}{\chi\frac{D_s}{D_l}\{1-h(\phi_0)\} + h(\phi_0)} \chi\frac{D_s}{D_l}\frac{\partial\mu_{s,0}}{\partial\eta} \nonumber \\
    & + a_1 \frac{S_sh(\phi_0)}{|\frac{\partial\phi_0}{\partial\eta}|}\frac{h(\phi_0)\{1-h(\phi_0)\}}{\chi\frac{D_s}{D_l}\{1-h(\phi_0)\} + h(\phi_0)} \frac{\partial\mu_{l,0}}{\partial\eta} \nonumber \\
    & + a_1 \frac{S_l\{1-h(\phi_0)\}}{|\frac{\partial\phi_0}{\partial\eta}|}\frac{h(\phi_0)\{1-h(\phi_0)\}}{\chi\frac{D_s}{D_l}\{1-h(\phi_0)\} + h(\phi_0)} \frac{\partial\mu_{l,0}}{\partial\eta}
\end{align}
We next multiply both sides by $\frac{\partial\phi_0}{\partial\eta}$ and integrate within the limits $-\infty$ to $+\infty$. As shown before, the first two terms on the RHS can be simplified as
\begin{equation}\label{gb2_3}
   1 + 2 = \int_{-\infty}^{+\infty}\frac{\partial^2\phi_1}{\partial\eta^2}\frac{\partial\phi_0}{\partial\eta}\mathrm{d}\eta - \int_{-\infty}^{+\infty}f_{\mathrm{dw}}^{\prime\prime}(\phi_0)\phi_1\frac{\partial\phi_0}{\partial\eta}\mathrm{d}\eta = 0.
\end{equation}
The third term on the RHS can be simplified by considering
\begin{align}\label{gb2_4}
    \frac{\mathrm{d}f}{\partial\eta}(c_{s,0},c_{l,0},\phi_0) = \frac{\partial f}{\partial c_s}\frac{\partial c_{s,0}}{\partial\eta} + \frac{\partial f}{\partial c_l}\frac{\partial c_{l,0}}{\partial\eta} + \frac{\partial f}{\partial \phi}\frac{\partial \phi_0}{\partial\eta}.
\end{align}
Substituting the partial derivatives of the free energy we obtain
\begin{align}\label{gb2_5}
   \frac{\mathrm{d}f}{\partial\eta}(c_{s,0},c_{l,0},\phi_0) = \mu_{s,0}h(\phi_0)\frac{\partial c_{s,0}}{\partial\eta} + \mu_{l,0}\{1-h(\phi_0)\}\frac{\partial c_{l,0}}{\partial\eta} + \Big\{f_s(c_{s,0}) - f_l(c_{l,0})\Big\}h^{\prime}(\phi_0)\frac{\partial \phi_0}{\partial\eta}.
\end{align}
Integrating both sides within the limits $\pm\infty$ the third term on the RHS of Eq.~(\ref{gb2_2}) (after multiplication with $\frac{\partial\phi_0}{\partial\eta}$) can be written as
\begin{align}\label{gb2_6}
    -a_1\int_{-\infty}^{+\infty}\Big\{f_s(c_{s,0}) - f_l(c_{l,0})\Big\}h^{\prime}(\phi_0)\frac{\partial \phi_0}{\partial\eta}\mathrm{d}\eta = & -a_1\int_{-\infty}^{+\infty}\frac{\mathrm{d}f}{\mathrm{d}\eta}\mathrm{d}\eta + a_1\int_{-\infty}^{+\infty}\mu_{s,0}h(\phi_0)\frac{\partial c_{s,0}}{\partial\eta}\mathrm{d}\eta \nonumber \\
    & + a_1\int_{-\infty}^{+\infty}\mu_{l,0}\{1-h(\phi_0)\}\frac{\partial c_{l,0}}{\partial\eta}\mathrm{d}\eta
\end{align}
The second and the third integral on the RHS can be simplified via integration by parts with $\mu_{s,0}h(\phi_0)$ and $\mu_{l,0}\{1-h(\phi_0)\}$ as the first functions to yield
\begin{align}\label{gb2_7}
    -a_1\int_{-\infty}^{+\infty}\Big\{f_s(c_{s,0}) - f_l(c_{l,0})\Big\}h^{\prime}(\phi_0)\frac{\partial \phi_0}{\partial\eta}\mathrm{d}\eta = & a_1 (\omega_s - \omega_l) - a_1\int_{-\infty}^{+\infty}\frac{\partial}{\partial\eta}[\mu_{s,0}h(\phi_0)]c_{s,0}\mathrm{d}\eta \nonumber \\
    & - a_1\int_{-\infty}^{+\infty}\frac{\partial}{\partial\eta}[\mu_{l,0}\{1-h(\phi_0)\}]c_{l,0}\mathrm{d}\eta
\end{align}
where, 
\begin{equation}
    \omega_s - \omega_l = f_s(c_{s,0}|^-,1) - \mu_{s,0}|^-c_{s,0}|^- f_l(c_{l,0}|^+) + \mu_{l,0}^+c_{l,0}^+
\end{equation}
which can be further simplified using Eq.~(\ref{dp2_8}) to yield
\begin{equation}
    \omega_s - \omega_l = f_s(c_{s,0}|^-,1) - f_l(c_{l,0}|^+,0) - \mu_{l,0}^+(c_{s,0}|^-- c_{l,0}^+)
\end{equation}
Combining the fourth and sixth term on the RHS of Eq.~(\ref{gb2_2}) we obtain
\begin{align}\label{gb2_8}
    &a_1\int_{-\infty}^{+\infty}\frac{\chi\frac{D_s}{D_l}\{1-h(\phi_0)\}}{\chi\frac{D_s}{D_l}\{1-h(\phi_0)\} + h(\phi_0)}\Bigg\{ \mu_{s,0}(c_{s,0}-c_{l,0})h^{\prime}(\phi_0)\frac{\partial\phi_0}{\partial\eta} + (c_{s,0}-c_{l,0})h(\phi_0)\frac{\partial\mu_{s,0}}{\partial\eta}\Bigg\}\mathrm{d}\eta \nonumber \\
    & = a_1 \int_{-\infty}^{+\infty}\frac{\chi\frac{D_s}{D_l}\{1-h(\phi_0)\}}{\chi\frac{D_s}{D_l}\{1-h(\phi_0)\} + h(\phi_0)} (c_{s,0}-c_{l,0})\frac{\partial}{\partial\eta}\Big[\mu_{s,0}h(\phi_0)\Big]\mathrm{d}\eta
\end{align}
Similarly, combining fifth and seventh term on the RHS of Eq.~(\ref{gb2_2}) we obtain
\begin{align}\label{gb2_9}
    &a_1\int_{-\infty}^{+\infty}\frac{h(\phi_0)}{\chi\frac{D_s}{D_l}\{1-h(\phi_0)\} + h(\phi_0)}\Bigg\{ \mu_{l,0}(c_{s,0}-c_{l,0})h^{\prime}(\phi_0)\frac{\partial\phi_0}{\partial\eta} - (c_{s,0}-c_{l,0})\{1-h(\phi_0)\}\frac{\partial\mu_{l,0}}{\partial\eta}\Bigg\}\mathrm{d}\eta \nonumber \\
    & = -a_1 \int_{-\infty}^{+\infty}\frac{h(\phi_0)}{\chi\frac{D_s}{D_l}\{1-h(\phi_0)\} + h(\phi_0)} (c_{s,0}-c_{l,0})\frac{\partial}{\partial\eta}\Big[\mu_{l,0}\{1-h(\phi_0)\}\Big]\mathrm{d}\eta 
\end{align}
Therefore using Eqs.~(\ref{gb2_7}),(\ref{gb2_8}) and (\ref{gb2_9}) the third, fourth, fifth, sixth and seventh term on the RHS of Eq.~(\ref{gb2_2}) writes as
\begin{equation}\label{gb2_10}
    3 + 4 + 5 + 6 +7 = a_1(\omega_s - \omega_l)
\end{equation}
Combining eighth and eleventh terms on the RHS of Eq.~(\ref{gb2_2})
\begin{align}\label{gb2_11}
  8+11 =&  a_1\int_{-\infty}^{+\infty}\frac{S_l\{1-h(\phi_0)\}\chi\frac{D_s}{D_l}\{1-h(\phi_0)\}}{\chi\frac{D_s}{D_l}\{1-h(\phi_0)\}+h(\phi_0)}\frac{\partial}{\partial\eta}\Big[\mu_{s,0}h(\phi_0)\Big]\mathrm{d}\eta \nonumber \\
  & -a_1\int_{-\infty}^{+\infty}\frac{S_l\{1-h(\phi_0)\}\chi\frac{D_s}{D_l}\{1-h(\phi_0)\}}{\chi\frac{D_s}{D_l}\{1-h(\phi_0)\}+h(\phi_0)}\mu_{l,0}h^{\prime}(\phi_0)\frac{\partial\phi_0}{\partial\eta}\mathrm{d}\eta
\end{align}
Similarly combining ninth and twelfth terms on the RHS of Eq.~(\ref{gb2_2}) we have
\begin{align}\label{gb2_12}
  9+12 =& -a_1\int_{-\infty}^{+\infty}\frac{S_sh(\phi_0) h(\phi_0)}{\chi\frac{D_s}{D_l}\{1-h(\phi_0)\}+h(\phi_0)}\frac{\partial}{\partial\eta}\Big[\mu_{l,0}\{1-h(\phi_0)\}\Big]\mathrm{d}\eta \nonumber \\
  & -a_1\int_{-\infty}^{+\infty}\frac{S_sh(\phi_0)h(\phi_0)}{\chi\frac{D_s}{D_l}\{1-h(\phi_0)\}+h(\phi_0)}\mu_{s,0}h^{\prime}(\phi_0)\frac{\partial\phi_0}{\partial\eta}\mathrm{d}\eta
\end{align}
Using the results of Eqs.~(\ref{gb2_10}), (\ref{gb2_11}) and (\ref{gb2_12}) and adding the tenth and thirteenth terms in Eq.~(\ref{gb2_2}) we obtain
\begin{align}\label{gb2_13}
    -(\alpha V_n + \kappa)\int_{-\infty}^{+\infty}\Big(\frac{\partial\phi_0}{\partial\eta}\Big)^2\mathrm{d}\eta = & a_1(\omega_s - \omega_l) + a_1\int_{-\infty}^{+\infty}S_s h(\phi_0)\frac{\partial \mu_{s,0}}{\partial\eta}h(\phi_0)\mathrm{d}\eta \nonumber \\
    & -a_1 \int_{-\infty}^{+\infty} S_l\{1-h(\phi_0)\}\frac{\partial\mu_{l,0}}{\partial\eta}\{1-h(\phi_0)\}\mathrm{d}\eta.
\end{align}
The above equation can also be expressed in the following form
\begin{align}
   -(\alpha V_n + \kappa)\int_{-\infty}^{+\infty}&\Big(\frac{\partial\phi_0}{\partial\eta}\Big)^2\mathrm{d}\eta =  a_1(\omega_s - \omega_l) - a_1\int_{-\infty}^{+\infty}[S_sh(\phi_0)+S_l\{1-h(\phi_0)\}] \times \nonumber \\
   & \frac{\partial\mu_{l,0}}{\partial\eta}\{1-h(\phi_0)\}\mathrm{d}\eta -a_1\int_{-\infty}^{+\infty} S_sh(\phi_0)(\mu_{s,0}-\mu_{l,0})h^{\prime}(\phi_0)\frac{\partial\phi_0}{\partial\eta}\mathrm{d}\eta
\end{align}
The Gibbs-Thomson condition at the lowest order depends upon the choice of the source terms.

\subsection{$\mathcal{O}(1/\epsilon)$ in $c$ : Solute trapping}
The $\mathcal{O}(1/\epsilon)$ in $c$ equation can be equated as
\begin{align}\label{sl2_1}
    -V_n\frac{\partial c_0}{\partial\eta} = & \frac{\partial}{\partial\eta}\Bigg[\frac{\partial c_{l,0}}{\partial\mu_l} q_n(\phi_0)\frac{\partial}{\partial\eta}\Big\{\mu_{s,1}h(\phi_0) + \mu_{l,1}\{1-h(\phi_0)\}\Big\} \nonumber \\
    & + \frac{\partial c_{l,0}}{\partial\mu_l}q_n(\phi_0)\frac{\partial}{\partial\eta}\Big\{(\mu_{s,0} - \mu_{l,0})h^{\prime}(\phi_0)\phi_1\Big\} \nonumber \\
     & +\frac{\partial c_{l,0}}{\partial\mu_l}q_n^{\prime}(\phi_0)\phi_1\frac{\partial}{\partial\eta}\Big\{\mu_{s,0}h(\phi_0) + \mu_{l,0}\{1-h(\phi_0)\}\Big\}\Bigg] \nonumber \\
     &- \frac{\partial}{\partial\eta}\Bigg[(\tilde{c}_{s,0} - \tilde{c}_{l,0}) a(\phi_0)V_n\Bigg].
\end{align}
From the leading order solution, since, $\mu_{s,0}h(\phi_0) + \mu_{l,0}\{1-h(\phi_0)\}$ is independent of $\eta$, the third term on the RHS of the above equation is equal to zero. Integrating the above equation we obtain
\begin{align}\label{sl2_2}
    -V_n c_0 = & \frac{\partial c_{l,0}}{\partial\mu_l}q_n(\phi_0)\frac{\partial}{\partial\eta}\Big\{\mu_{s,1}h(\phi_0) + \mu_{l,1}\{1-h(\phi_0)\}\Big\} \nonumber \\
    & + \frac{\partial c_{l,0}}{\partial\mu_l}q_n(\phi_0)\frac{\partial}{\partial\eta}\Big\{(\mu_{s,0} - \mu_{l,0})h^{\prime}(\phi_0)\phi_1\Big\} 
    - (\tilde{c}_{s,0} - \tilde{c}_{l,0}) a(\phi_0)V_n + A_2(s).
\end{align}
Taking the limit $\eta \rightarrow -\infty$ we evaluate the constant of integration $A_2(s)$ as
\begin{equation}\label{sl2_3}
    A_2(s) = -V_n c_{s,0} - \frac{\partial c_{l,0}}{\partial\mu_l}q_n^-\frac{\partial\mu_{s,0}}{\partial r}\Bigg|^-.
\end{equation}
Substituting for $A_2(s)$ from the above equation to Eq.~(\ref{sl2_2}) and rearranging we obtain
\begin{align}\label{sl2_4}
    \frac{\partial}{\partial\eta}\Big[\mu_{s,1}h(\phi_0) + \mu_{l,1}\{1-h(\phi_0)\}\Big] = & -\frac{V_n}{\partial c_{l,0}/\partial\mu_l}\frac{(c_0 - c_{s,0})}{q_n(\phi_0)} - \frac{\partial}{\partial\eta}\Big[(\mu_{s,0}-\mu_{l,0})h^{\prime}(\phi_0)\phi_1\Big]\nonumber \\
    & +(\tilde{c}_{s,0} - \tilde{c}_{l,0})\frac{V_n}{\partial c_{l,0}/\partial\mu_l}\frac{a(\phi_0)}{q_n(\phi_0)} + q_n^-\frac{\partial\mu_{s,0}}{\partial r}\Bigg|^-\frac{1}{q_n(\phi_0)}.
\end{align}
Integrating the above equation
\begin{align}\label{sl2_5}
    \mu_{s,1}h(\phi_0) + \mu_{l,1}\{1-h(\phi_0)\} = & -V_n\int_0^{\eta}\frac{1}{\partial c_{l,0}/\partial\mu_l}\frac{(c_0 - c_{s,0})}{q_n(\phi_0)}\mathrm{d}\eta - (\mu_{s,0}-\mu_{l,0})h^{\prime}(\phi_0)\phi_1 \nonumber \\
    & + V_n\int_{0}^{\eta}\frac{V_n}{\partial c_{l,0}/\partial\mu_l}(\tilde{c}_{s,0} - \tilde{c}_{l,0})\frac{a(\phi_0)}{q_n(\phi_0)}\mathrm{d}\eta \nonumber \\
    & + q_n^-\frac{\partial\mu_{s,0}}{\partial r}\Bigg|^-\int_0^{\eta}\frac{1}{q_n(\phi_0)}\mathrm{d}\eta + \overline{\mu}_1(s),
\end{align}
where $\overline{\mu}_1(s)$ is the constant of integration.  The above equation can be split into two parts as follows \\
(1) For $\eta >0$ 
\begin{align}\label{sl2_6}
    \mu_{s,1}h(\phi_0) + \mu_{l,1}\{1-h(\phi_0)\} = & -V_n \int_0^{\eta}\frac{1}{\partial c_{l,0}/\partial\mu_l}\Bigg[\frac{(c_0 - c_{s,0})}{q_n(\phi_0)} - \frac{c_{l,0} - c_{s,0}}{q_n^+}\Bigg]\mathrm{d}\eta \nonumber \\
    & - \eta \frac{V_n}{\partial c_{l,0}/\partial\mu_l} \frac{c_{l,0} - c_{s,0}}{q_n^+}  + (\mu_{s,0}-\mu_{l,0})h^{\prime}(\phi_0)\phi_1 \nonumber \\
    & + V_n\int_{0}^{\eta}\frac{1}{\partial c_{l,0}/\partial\mu_l}(\tilde{c}_{s,0} - \tilde{c}_{l,0})\frac{a(\phi_0)}{q_n(\phi_0)}\mathrm{d}\eta \nonumber \\
    & + q_n^-\frac{\partial\mu_{s,0}}{\partial r}\Bigg|^-\int_0^{\eta}\Bigg[\frac{1}{q_n(\phi_0)} - \frac{1}{q_n^+}\Bigg]\mathrm{d}\eta \nonumber \\
    & +\eta q_n^-\frac{\partial\mu_{s,0}}{\partial r}\Bigg|^-\frac{1}{q_n^+} + \overline{\mu}_1(s)
\end{align}
(2) For $\eta < 0$
\begin{align}\label{sl2_7}
    \mu_{s,1}h(\phi_0) + \mu_{l,1}\{1-h(\phi_0)\} = & -V_n \int_0^{\eta}\frac{1}{\partial c_{l,0}/\partial\mu_l}\frac{(c_0 - c_{s,0})}{q_n(\phi_0)} \mathrm{d}\eta  + (\mu_{s,0}-\mu_{l,0})h^{\prime}(\phi_0)\phi_1 \nonumber \\ 
    & + V_n \int_{0}^{\eta}\frac{1}{\partial c_{l,0}/\partial\mu_l}(\tilde{c}_{s,0} - \tilde{c}_{l,0})\frac{a(\phi_0)}{q_n(\phi_0)}\mathrm{d}\eta \nonumber \\
    & + q_n^-\frac{\partial\mu_{s,0}}{\partial r}\Bigg|^-\int_0^{\eta}\Bigg[\frac{1}{q_n(\phi_0)} - \frac{1}{q_n^-}\Bigg]\mathrm{d}\eta \nonumber \\
    & +\eta q_n^-\frac{\partial\mu_{s,0}}{\partial r}\Bigg|^-\frac{1}{q_n^-} + \overline{\mu}_1(s)
\end{align}
Taking the far field limit $\eta \rightarrow +\infty$ in Eq.~(\ref{sl2_6}) and employing the matching condition we have
\begin{align}\label{sl2_8}
    \mu_{l,1}|^+ + \eta \frac{\partial \mu_{l,0}}{\partial r}\Bigg|^+ = & -V_n \int_0^{+\infty}\frac{1}{\partial c_{l,0}/\partial\mu_l}\Bigg[\frac{(c_0 - c_{s,0})}{q_n(\phi_0)} - \frac{c_{l,0} - c_{s,0}}{q_n^+}\Bigg]\mathrm{d}\eta - \eta \frac{V_n}{\partial c_{l,0}/\partial\mu_l} \frac{c_{l,0} - c_{s,0}}{q_n^+} \nonumber \\
    & + V_n \int_{0}^{+\infty}\frac{1}{\partial c_{l,0}/\partial\mu_l}(\tilde{c}_{s,0} - \tilde{c}_{l,0})\frac{a(\phi_0)}{q_n(\phi_0)}\mathrm{d}\eta \nonumber \\
    & + q_n^-\frac{\partial\mu_{s,0}}{\partial r}\Bigg|^-\int_0^{+\infty}\Bigg[\frac{1}{q_n(\phi_0)} - \frac{1}{q_n^+}\Bigg]\mathrm{d}\eta 
    +\eta q_n^-\frac{\partial\mu_{s,0}}{\partial r}\Bigg|^-\frac{1}{q_n^+} + \overline{\mu}_1(s),
\end{align}
where, we have utilized the fact that $\lim_{\eta \rightarrow +\infty} (\mu_{s,0}-\mu_{l,0})h^{\prime}(\phi_0)\phi_1 = 0$.
Comparing coefficients of $\eta$ on both sides we have
\begin{equation}\label{sl2_9}
    \frac{\partial \mu_{l,0}}{\partial r}\Bigg|^+ = -\frac{V_n}{\partial c_{l,0}/\partial\mu_l} \frac{c_{l,0} - c_{s,0}}{q_n^+} + q_n^-\frac{\partial\mu_{s,0}}{\partial r}\Bigg|^-\frac{1}{q_n^+} ,
\end{equation}
which on re-arranging gives the flux conservation or the Stefan's condition at the interface at the lowest order as
\begin{equation}\label{sl2_10}
   q_n^+ \frac{\partial \mu_{l,0}}{\partial r}\Bigg|^+ - q_n^- \frac{\partial \mu_{s,0}}{\partial r}\Bigg|^- = -\frac{V_n}{\partial c_{l,0}/\partial\mu_l} (c_{l,0} - c_{s,0}) .
\end{equation}
Similarly, comparing the coefficients of $\eta^0$ in Eq.~(\ref{sl2_8}) we have
\begin{align}\label{sl2_11}
   \mu_{l,1}|^+ =  & -V_n\int_0^{+\infty}\frac{1}{\partial c_{l,0}/\partial\mu_l}\Bigg[\frac{(c_0 - c_{s,0})}{q_n(\phi_0)} - \frac{c_{l,0} - c_{s,0}}{q_n^+}\Bigg]\mathrm{d}\eta  \nonumber \\
   & +  V_n \int_{0}^{+\infty}\frac{1}{\partial c_{l,0}/\partial\mu_l}(\tilde{c}_{s,0} - \tilde{c}_{l,0})\frac{a(\phi_0)}{q_n(\phi_0)}\mathrm{d}\eta \nonumber \\
   & + q_n^-\frac{\partial\mu_{s,0}}{\partial r}\Bigg|^-\int_0^{+\infty}\Bigg[\frac{1}{q_n(\phi_0)} - \frac{1}{q_n^+}\Bigg]\mathrm{d}\eta + \overline{\mu}_1(s).
\end{align}
Introducing the lowest order solute profile $c_0 = c_{s,0}h(\phi_0) + c_{l,0}\{1-h(\phi_0)\}$ and expanding $\tilde{c}_{s,0} - \tilde{c}_{l,0} = c_{s,0} -c_{l,0} + S_sh(\phi_0) + S_l\{1-h(\phi_0)\}$ in above equation and rearranging we obtain
\begin{align}\label{sl2_12}
    \mu_{l,1}|^+ =&  - V_n \int_{0}^{+\infty}\frac{(c_{s,0}-c_{l,0})}{\partial c_{l,0}/\partial\mu_l} \{\overline{p}(\phi_0) - \overline{p}(\phi_0|^+)\}\mathrm{d}\eta \nonumber \\
    & + V_n \int_{0}^{+\infty}\frac{1}{\partial c_{l,0}/\partial \mu_l} [S_sh(\phi_0) + S_l\{1-h(\phi_0)\}]\frac{a(\phi_0)}{q_n(\phi_0)}\mathrm{d}\eta \nonumber \\
    & + q_n^-\frac{\partial\mu_{s,0}}{\partial r}\Bigg|^-\int_0^{+\infty}\Bigg[\frac{1}{q_n(\phi_0)} - \frac{1}{q_n^+}\Bigg]\mathrm{d}\eta + \overline{\mu}_1(s),
\end{align}
where,
\begin{equation}\label{sl2_13}
    \overline{p}(\phi_0) = \frac{h(\phi_0) - 1 -a(\phi_0)}{q_n(\phi_0)}
\end{equation}
and we have utilized the relation $\overline{p}(\phi_0|^+) = 1/q_n^+ = 1$. Similarly, taking the far field $\eta \rightarrow -\infty$ of Eq.~(\ref{sl2_7}) and comparing the coefficients of $\eta^0$ we obtain,
\begin{align}\label{sl2_14}
    \mu_{s,1}|^- =& -V_n\int_{0}^{-\infty}\frac{(c_{s,0}-c_{l,0})}{\partial c_{l,0}/\partial\mu_l} \{p(\phi_0) - p(\phi_0|^-)\}\mathrm{d}\eta \nonumber \\
    & + V_n \int_{0}^{-\infty}\frac{1}{\partial c_{l,0}/\partial \mu_l} [S_sh(\phi_0) + S_l\{1-h(\phi_0)\}]\frac{a(\phi_0)}{q_n(\phi_0)}\mathrm{d}\eta \nonumber \\
    & + q_n^-\frac{\partial\mu_{s,0}}{\partial r}\Bigg|^-\int_0^{-\infty}\Bigg[\frac{1}{q_n(\phi_0)} - \frac{1}{q_n^+}\Bigg]\mathrm{d}\eta + \overline{\mu}_1(s).
\end{align}
Subtracting Eq.~(\ref{sl2_12}) from Eq.~(\ref{sl2_14}),
\begin{equation}\label{sl2_15}
    \mu_s|^- - \mu_l|^+ = -V_n(F_1^- - F_1^+) + V_n(F_2^- - F_2^+) + q_n^-\frac{\partial\mu_{s,0}}{\partial r}\Bigg|^-(G^- - G^+),
\end{equation}
where,
\begin{equation}\label{sl2_16}
    F_1^{\pm} = \int_{0}^{\pm\infty}\frac{(c_{s,0}-c_{l,0})}{\partial c_{l,0}/\partial\mu_l}\{\overline{p}(\phi_0) - \overline{p}(\phi_0|^{\pm})\}\mathrm{d}\eta,
\end{equation}
\begin{equation}\label{sl2_17}
  F_2^{\pm} = \int_{0}^{\pm\infty}\frac{[S_sh(\phi_0) + S_l\{1-h(\phi_0)\}]}{\partial c_{l,0}/\partial\mu_l}\frac{a(\phi_0)}{q_n(\phi_0)}\mathrm{d}\eta,    
\end{equation}
and
\begin{equation}\label{sl2_18}
    G^{\pm} = \int_0^{\pm\infty}\Bigg[\frac{1}{q_n(\phi_0)} - \frac{1}{q_n^{\pm}}\Bigg]\mathrm{d}\eta .
\end{equation}
With the choice of the source terms $S_s = S_l = A(c_{l,0}-c_{s,0})$ as deduced earlier, the diffusion potential jump can simplified to be
\begin{equation}\label{sl2_19}
    \mu_s|^- - \mu_l|^+ = V_n(F^- - F^+) + q_n^-\frac{\partial\mu_{s,0}}{\partial r}\Bigg|^-(G^- - G^+),
\end{equation}
where
\begin{equation}
    F^{\pm} = \int_0^{\pm}\frac{(c_{l,0}-c_{s,0})}{\partial c_{l,0}/\partial \mu_l}\{\tilde{p}(\phi_0) - \tilde{p}(\phi_0|^{\pm})\}\mathrm{d}\eta
\end{equation}
and 
\begin{equation}
    \tilde{p}(\phi_0) = \frac{h(\phi_0)-1-(1-A)a(\phi_0)}{q_n(\phi_0)}
\end{equation}
which is similar to the Eq.~(\ref{st16}) derived for the case where we had assumed $\mu_s = \mu_l$ with the exception that lowest order phase concentrations (and the thermodynamic factors $\partial c_{s,0}/\partial \mu_s$ and $\partial c_{l,0}/\partial \mu_l$) are dependent on the normal co-ordinate system $\eta$ i.e. $c_{s,0} = c_{s,0}(\eta)$, $c_{l,0} = c_{l,0}(\eta)$ and cannot be taken out of the integral to obtain closed form expressions. However, it is clear that the parameter $A$ can still be tuned to control partitioning.

As discussed in the next section, a numerical comparison of the two models suggest that the steady state diffusion potential in both the models (Fig.(\ref{supfig3})) are almost identical. A comparison of the velocity-dependent partition coefficient (Fig.(\ref{supfig2})) suggests that the difference between the two models depend on the value of $A$. The agreement between the predictions of the two models is better at higher values of $A$ within the entire velocity range considered. Even with lower values of $A$ (for instance $A=0.1$) the agreement is good until velocities of about $0.1$ m/s. A good agreement between both the models motivates us to utilize the same interpolation functions to negate the Kapitza jump (i.e. satisfy $G^- = G^+$).

The next order $\phi$ and $c$ equations are analytically not tractable. Hence no information regarding the correction to the Gibbs-Thomson and mass-conservation conditions are available.

\section{Comparative study of both the models }
\begin{sidewaysfigure}
    \centering
    \includegraphics[scale = 0.34]{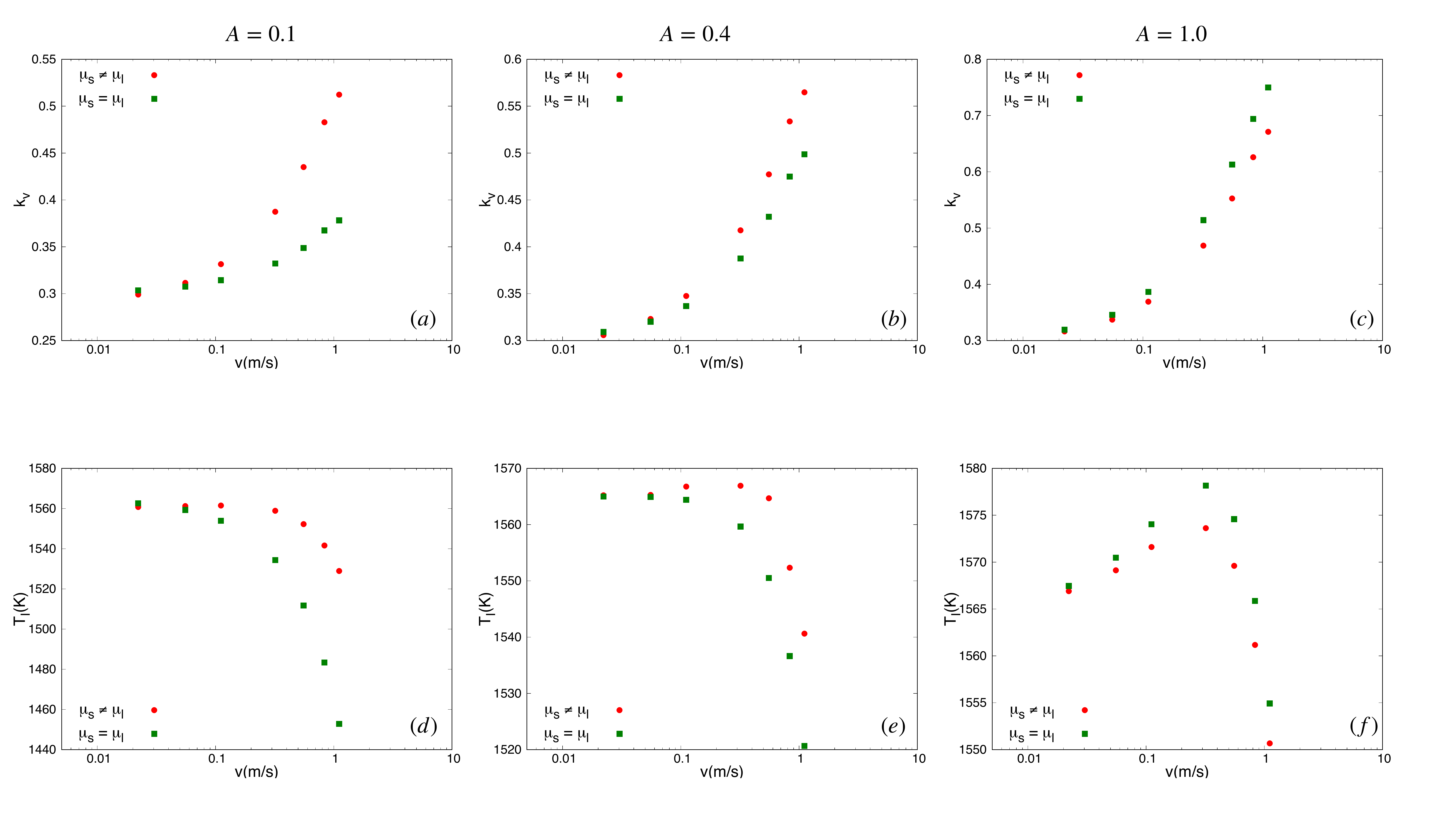}
    \caption{A comparison of the partition coefficient and interface temperature for the models with $\mu_s \neq \mu_l$ and with the assumption $\mu_s = \mu_l$ for different values of model parameter $A$. The interface width $W$ is chosen as $13.5$ nm. The assumption of $\mu_s = \mu_l$ is observed to be most valid at larger values of $AW$.}
    \label{supfig2}
\end{sidewaysfigure}

A comparative study of the partition coefficient and interface temperature of the general model (which assumes $\mu_s \neq \mu_l$) with the model where $\mu_s = \mu_l$ is imposed are presented in Fig.\ref{supfig2}. To study the effect of the model parameter $A$ on the assumption $\mu_s = \mu_l$, we select three values of $A = 0.1, 0.4,1.0$. The interface width has been fixed to a value of $W = 13.5$ nm in all the simulations.

For velocities $v < 0.1$ m/s, both the models essentially give similar results for partition coefficient and interface temperature across all values of $A$. At larger velocities, the difference is prominent for $A = 0.1$ for both the partition coefficient and interface temperature. The difference between the results of the two models decrease with increasing $A$. For $A = 1.0$ and at $v = 1$m/s the difference in $k_v$ is about $10\%$ and $T_I$ is less than $1\%$. Since for solute trapping we mostly work in the regime of higher values of $A$, the asymptotic expression derived for the model $\mu_s = \mu_l$ acts as a good initial guess.

\begin{sidewaysfigure}
    \centering
    \includegraphics[scale = 0.34]{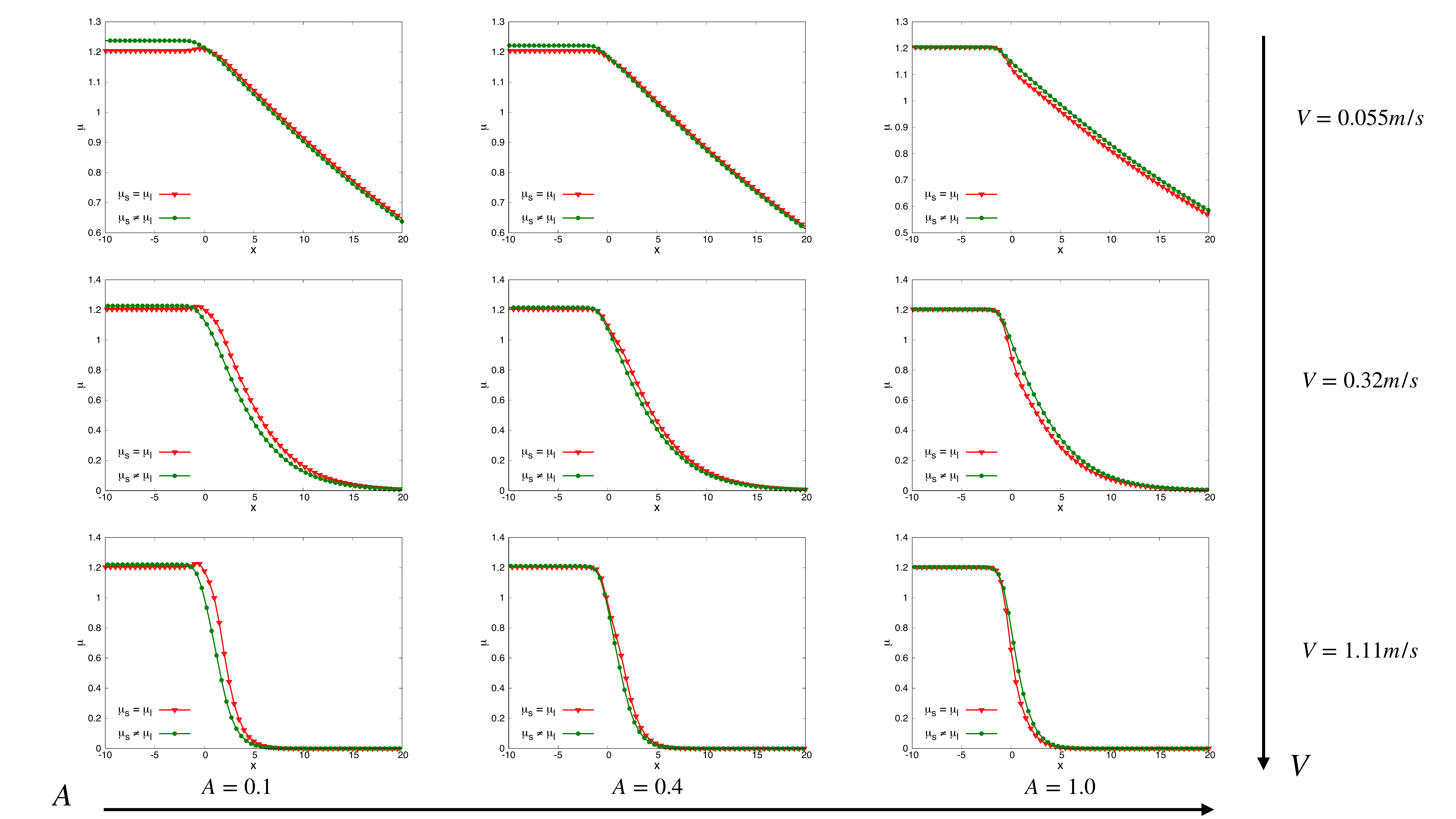}
    \caption{A comparison of the steady state diffusion potential for the models with $\mu_s \neq \mu_l$ and with the assumption of $\mu_s = \mu_l$ for different values of model parameter $A$ and velocities. For the general model ($\mu_s \neq \mu_l$) the diffusion potential is defined as $\mu = \mu_s h(\phi) + \mu_l\{1-h(\phi)\}$.}
    \label{supfig3}
\end{sidewaysfigure}

The steady state diffusion potentials across the interface from both the models are compared in Fig.\ref{supfig3}. The diffusion potential for the model $\mu_s \neq \mu_l$ is defined as $\mu = \mu_s h(\phi) + \mu_l\{1-h(\phi)\}$. The steady state diffusion potential from both the models are within few percent of each other.

